\documentclass[longauth]{aa}  

\usepackage{graphicx}
\usepackage{txfonts}
\usepackage[colorlinks=true,urlcolor=blue,linkcolor=blue,citecolor=blue]{hyperref}
\usepackage{threeparttable}
\usepackage{booktabs}
\usepackage{longtable}
\usepackage{ragged2e}

\usepackage{tikz,xcolor,hyperref}

\definecolor{lime}{HTML}{A6CE39}
\DeclareRobustCommand{\orcidicon}{%
	\hspace{-1.5mm}
	\begin{tikzpicture}
	\draw[lime, fill=lime] (0,0) 
	circle [radius=0.16] 
	node[white] {{\fontfamily{qag}\selectfont \tiny ID}};
	\draw[white, fill=white] (-0.0625,0.095) 
	circle [radius=0.007];
	\end{tikzpicture}
	\hspace{-2.5mm}
}

\foreach \x in {A, ..., Z}{%
	\expandafter\xdef\csname orcid\x\endcsname{\noexpand\href{https://orcid.org/\csname orcidauthor\x\endcsname}{\noexpand\orcidicon}}
}

\foreach \x in {a, ..., z}{%
	\expandafter\xdef\csname orcid\x\endcsname{\noexpand\href{https://orcid.org/\csname orcidauthor\x\endcsname}{\noexpand\orcidicon}}
}

\foreach \x in {a, ..., z}{%
	\expandafter\xdef\csname orcid\x\x\endcsname{\noexpand\href{https://orcid.org/\csname orcidauthor\x\x\endcsname}{\noexpand\orcidicon}}
}

\begin{document} 

   \title{Two temperate super-Earths transiting a nearby late-type M dwarf\thanks{The photometric and radial velocity data used in this work are available in electronic form at the CDS via anonymous ftp to \url{cdsarc.u-strasbg.fr} (130.79.128.5) or via \url{http://cdsweb.u-strasbg.fr/cgi-bin/qcat?J/A+A/}}}
   \titlerunning{Two temperate super-Earths transiting LP~890-9}

   \author{
   L.~Delrez\orcidA{}\inst{\ref{aru_liege},\ref{star_liege}}
   \and C.A.~Murray\orcida{}\inst{\ref{cavendish},\ref{boulder}}
   \and F.J.~Pozuelos\orcidC{}\inst{\ref{aru_liege},\ref{star_liege},\ref{iaa}}
   \and N.~Narita\orcidH{}\inst{\ref{komaba},\ref{astrobio_mitaka},\ref{iac}}
   \and E.~Ducrot\orcidb{}\inst{\ref{cea},\thanks{Paris Region Fellow, Marie Sklodowska-Curie Action}}
   \and M.~Timmermans\inst{\ref{aru_liege}}
   \and \\ N.~Watanabe\orcidI{}\inst{\ref{tokyo_uni}}
   \and A.J.~Burgasser\orcidX{}\inst{\ref{san_diego}}
   \and T.~Hirano\orcidJ{}\inst{\ref{astrobio_mitaka},\ref{naoj}}
   \and B.V.~Rackham\orcidG{}\inst{\ref{miteaps},\ref{mitkavli},\thanks{51 Pegasi b Fellow}}
   \and K.G.~Stassun\orcidc{}\inst{\ref{vanderbilt}}
   \and \\ V.~Van~Grootel\orcidd{}\inst{\ref{star_liege}}
   \and C.~Aganze\orcidg{}\inst{\ref{san_diego}}
   \and M.~Cointepas\inst{\ref{grenoble},\ref{geneve}}
   \and S.~Howell\orcidZ{}\inst{\ref{nasa_ames}}
   \and L.~Kaltenegger\orcidh{}\inst{\ref{cornell}}
   \and P.~Niraula\orcidi{}\inst{\ref{miteaps}}
   \and D.~Sebastian\orcidj{}\inst{\ref{birmingham}}
   \and J.M.~Almenara\inst{\ref{grenoble}}
   \and K.~Barkaoui\orcidk{}\inst{\ref{aru_liege},\ref{miteaps},\ref{iac}}
   \and T.A.~Baycroft\inst{\ref{birmingham}}
   \and X.~Bonfils\orcidl{}\inst{\ref{grenoble}}
   \and F.~Bouchy\orcidm{}\inst{\ref{geneve}}
   \and A.~Burdanov\orcidn{}\inst{\ref{miteaps}}
   \and D.A.~Caldwell\inst{\ref{nasa_ames},\ref{seti}}
   \and D.~Charbonneau\orcido{}\inst{\ref{cfa}}
   \and D.R.~Ciardi\orcidp{}\inst{\ref{caltech}}
   \and K.A.~Collins\inst{\ref{cfa}}
   \and T.~Daylan\orcidq{}\inst{\ref{mitkavli},\ref{mitphysics}, \ref{princeton}}
   \and B.-O.~Demory\orcidr{}\inst{\ref{unibe}}
   \and J.~de~Wit\inst{\ref{miteaps}}
   \and G.~Dransfield\orcids{}\inst{\ref{birmingham}} 
   \and S.B.~Fajardo-Acosta\inst{\ref{caltech}}
   \and M.~Fausnaugh\orcidB{}\inst{\ref{mitkavli}}
   \and A.~Fukui\orcidK{}\inst{\ref{komaba},\ref{iac}}
   \and E.~Furlan\orcidt{}\inst{\ref{caltech}}
   \and L.J.~Garcia\orcidu{}\inst{\ref{aru_liege}}
   \and C.L.~Gnilka\inst{\ref{nasa_ames}}
   \and Y.~G\'omez~Maqueo~Chew\orcidf{}\inst{\ref{ciudad}}
   \and M.A.~G\'omez-Mu\~noz\orcidv{}\inst{\ref{uname}}  
   \and \\ M.N.~G\"unther\orcidw{}\inst{\ref{estec},\thanks{ESA Research Fellow}}
   \and H.~Harakawa\inst{\ref{subaru}}
   \and K.~Heng\orcidx{}\inst{\ref{unibe},\ref{warwick}}
   \and M.J.~Hooton\orcidaa{}\inst{\ref{cavendish}}
   \and Y.~Hori\orcidL{}\inst{\ref{astrobio_mitaka},\ref{naoj}}
   \and M.~Ikoma\orcidM{}\inst{\ref{naoj}}
   \and \\ E.~Jehin\orcidbb{}\inst{\ref{star_liege}}
   \and J.M.~Jenkins\orcidcc{}\inst{\ref{nasa_ames}}
   \and T.~Kagetani\orcidN{}\inst{\ref{tokyo_uni}}
   \and K.~Kawauchi\orcidO{}\inst{\ref{iac},\ref{la_laguna}}
   \and T.~Kimura\inst{\ref{tokyo_uni2}}
   \and T.~Kodama\orcidP{}\inst{\ref{komaba}}
   \and T.~Kotani\inst{\ref{astrobio_mitaka},\ref{naoj},\ref{sokendai}}   
   \and V.~Krishnamurthy\orcidQ{}\inst{\ref{astrobio_mitaka},\ref{naoj}}
   \and T.~Kudo\orcidR{}\inst{\ref{subaru}}
   \and V.~Kunovac\orciddd{}\inst{\ref{lowell},\ref{birmingham}}
   \and N.~Kusakabe\orcidF{}\inst{\ref{astrobio_mitaka},\ref{naoj}}
   \and D.W.~Latham\orcidee{}\inst{\ref{cfa}}   
   \and C.~Littlefield\inst{\ref{bay_area},\ref{nasa_ames}}
   \and J.~McCormac\inst{\ref{warwick}}
   \and C.~Melis\orcidY{}\inst{\ref{san_diego}}
   \and M.~Mori\orcidT{}\inst{\ref{tokyo_uni3}}
   \and F.~Murgas\orcidS{}\inst{\ref{iac},\ref{la_laguna}}
   \and E.~Palle\orcidU{}\inst{\ref{iac},\ref{la_laguna}}
   \and P.P.~Pedersen\inst{\ref{cavendish}}
   \and D.~Queloz\orcidz{}\inst{\ref{cavendish}}   
   \and \\ G.~Ricker\orcidff{}\inst{\ref{mitkavli},\ref{mitphysics}}  
   \and L.~Sabin\orcidgg{}\inst{\ref{uname}} 
   \and N.~Schanche\inst{\ref{unibe}}   
   \and U.~Schroffenegger\inst{\ref{unibe}}   
   \and S.~Seager\inst{\ref{miteaps},\ref{mitkavli},\ref{mitphysics},\ref{mitaeroastro}} 
   \and B.~Shiao\inst{\ref{stsci}}
   \and \\ S.~Sohy\inst{\ref{star_liege}}
   \and M.R.~Standing\orcide{}\inst{\ref{birmingham}}
   \and M.~Tamura\orcidV{}\inst{\ref{tokyo_uni3},\ref{astrobio_mitaka},\ref{naoj}}
   \and C.A.~Theissen\orcidhh{}\inst{\ref{san_diego},\thanks{NASA Sagan Fellow}}
   \and S.J.~Thompson\orcidD{}\inst{\ref{cavendish}}   
   \and \\ A.H.M.J.~Triaud\orcidii{}\inst{\ref{birmingham}}   
   \and R.~Vanderspek\orcidjj{}\inst{\ref{mitkavli}}  
   \and S.~Vievard\orcidkk{}\inst{\ref{astrobio_mitaka},\ref{subaru}}
   \and R.D.~Wells\orcidll{}\inst{\ref{unibe}}
   \and J.N.~Winn\inst{\ref{princeton}}   
   \and Y.~Zou\orcidW{}\inst{\ref{tokyo_uni}}  
   \and S.~Z\'u\~niga-Fern\'andez\orcidE{}\inst{\ref{aru_liege}}
   \and M.~Gillon\orcidy{}\inst{\ref{aru_liege}}
   }
   
   \institute{
   Astrobiology Research Unit, Universit\'e de Li\`ege, All\'ee du 6 Ao\^ut 19C, B-4000 Li\`ege, Belgium \label{aru_liege}
   \and Space Sciences, Technologies and Astrophysics Research (STAR) Institute, Universit\'e de Li\`ege, All\'ee du 6 Ao\^ut 19C, B-4000 Li\`ege, Belgium \label{star_liege}
   \and Cavendish Laboratory, JJ Thomson Avenue, Cambridge, CB3 0HE, UK \label{cavendish}
   \and University of Colorado Boulder, Boulder, CO 80309, USA \label{boulder}
   \and Instituto de Astrof\'isica de Andaluc\'ia (IAA-CSIC), Glorieta de la Astronom\'ia s/n, 18008 Granada, Spain \label{iaa}
   \and Komaba Institute for Science, The University of Tokyo, 3-8-1 Komaba, Meguro, Tokyo 153-8902, Japan \label{komaba}
   \and Astrobiology Center, 2-21-1 Osawa, Mitaka, Tokyo 181-8588, Japan \label{astrobio_mitaka}
   \and Instituto de Astrof\'{i}sica de Canarias (IAC), 38205 La Laguna, Tenerife, Spain \label{iac}
   \and AIM, CEA, CNRS, Universit\'e Paris-Saclay, Universit\'e de Paris, F-91191 Gif-sur-Yvette, France \label{cea}
   \and Department of Multi-Disciplinary Sciences, Graduate School of Arts and Sciences, The University of Tokyo, 3-8-1 Komaba, Meguro, Tokyo 153-8902, Japan \label{tokyo_uni}
   \and Center for Astrophysics and Space Sciences, UC San Diego, UCSD Mail Code 0424, 9500 Gilman Drive, La Jolla, CA 92093-0424, USA \label{san_diego}
   \and National Astronomical Observatory of Japan, 2-21-1 Osawa, Mitaka, Tokyo 181-8588, Japan \label{naoj}
   \and Department of Earth, Atmospheric and Planetary Science, Massachusetts Institute of Technology, 77 Massachusetts Avenue, Cambridge, MA 02139, USA \label{miteaps}
   \and Kavli Institute for Astrophysics and Space Research, Massachusetts Institute of Technology, Cambridge, MA 02139, USA \label{mitkavli}
   \and Department of Physics \& Astronomy, Vanderbilt University, 6301 Stevenson Center Ln., Nashville, TN 37235, USA \label{vanderbilt}
   \and Univ. Grenoble Alpes, CNRS, IPAG, 38000 Grenoble, France \label{grenoble}
   \and Observatoire de Genève, Département d’Astronomie, Université de Genève, Chemin Pegasi 51b, 1290 Versoix, Switzerland \label{geneve}
   \and NASA Ames Research Center, Moffett Field, CA 94035, USA \label{nasa_ames}
   \and Department of Astronomy and Carl Sagan Institute, Cornell University, 302 Space Sciences Building, Ithaca, NY 14853, USA \label{cornell}
   \and School of Physics \& Astronomy, University of Birmingham, Edgbaston, Birmingham B15 2TT, UK \label{birmingham}
   \and SETI Institute, Mountain View, CA 94043, USA \label{seti}
   \and Center for Astrophysics | Harvard \& Smithsonian, 60 Garden Street, Cambridge, MA, 02138, USA \label{cfa}
   \and Caltech/IPAC, Mail Code 100-22, 1200 E. California Blvd., Pasadena, CA 91125, USA \label{caltech}
   \and Department of Physics, Massachusetts Institute of Technology, Cambridge, MA 02139, USA \label{mitphysics}
   \and Department of Astrophysical Sciences, Princeton University, 4 Ivy Lane, Princeton, NJ 08544, USA \label{princeton} 
   \and Center for Space and Habitability, University of Bern, Gesellschaftsstrasse 6, 3012, Bern, Switzerland \label{unibe}
   \and Universidad Nacional Aut\'onoma de M\'exico, Instituto de Astronom\'ia, AP 70-264, Ciudad de M\'exico,  04510, M\'exico \label{ciudad}   
   \and Universidad Nacional Aut\'onoma de M\'exico, Instituto de Astronom\'ia, AP 106, Ensenada 22800, BC, M\'exico \label{uname} 
   \and European Space Agency (ESA), European Space Research and Technology Centre (ESTEC), Keplerlaan 1, 2201 AZ Noordwijk, Netherlands \label{estec}
   \and Subaru Telescope, 650 N. Aohoku Place, Hilo, HI 96720, USA \label{subaru}
   \and Department of Physics, University of Warwick, Gibbet Hill Road, Coventry CV4 7AL, United Kingdom \label{warwick}
   \and Departamento de Astrof\'{i}sica, Universidad de La Laguna (ULL), 38206 La Laguna, Tenerife, Spain \label{la_laguna}
   \and Department of Earth and Planetary Science, Graduate School of Science, The University of Tokyo, 7-3-1 Hongo, Bunkyo-ku, Tokyo 113-0033, Japan \label{tokyo_uni2}
   \and Department of Astronomy, School of Science, The Graduate University for Advanced Studies (SOKENDAI), 2-21-1 Osawa, Mitaka, Tokyo, Japan \label{sokendai}
   \and Lowell Observatory, 1400 W. Mars Hill Rd., Flagstaff, AZ 86001, USA \label{lowell}
   \and Bay Area Environmental Research Institute, Moffett Field, CA 94035, USA \label{bay_area}
   \and Department of Astronomy, Graduate School of Science, The University of Tokyo, 7-3-1 Hongo, Bunkyo-ku, Tokyo 113-0033, Japan \label{tokyo_uni3}
   \and Department of Aeronautics and Astronautics, MIT, 77 Massachusetts Avenue, Cambridge, MA 02139, USA \label{mitaeroastro}   
   \and{Space Telescope Science Institute, 3700 San Martin Drive, Baltimore, MD, 21218, USA} \label{stsci}
   }

  \date{Received ...; accepted ...}

  \abstract
   {In the age of JWST, temperate terrestrial exoplanets transiting nearby late-type M dwarfs provide unique opportunities for characterising their atmospheres, as well as searching for biosignature gases. In this context, the benchmark TRAPPIST-1 planetary system has garnered the interest of a broad scientific community.}
   {We report here the discovery and validation of two temperate super-Earths transiting LP~890-9 (TOI-4306, SPECULOOS-2), a relatively low-activity nearby \hbox{(32 pc)} M6V star. The inner planet, LP~890-9\,b, was first detected by TESS (and identified as TOI-4306.01) based on four sectors of data. Intensive photometric monitoring of the system with the SPECULOOS Southern Observatory then led to the discovery of a second outer transiting planet, LP~890-9\,c (also identified as SPECULOOS-2\,c), previously undetected by TESS. The orbital period of this second planet was later confirmed by MuSCAT3 follow-up observations.}
   {We first inferred the properties of the host star by analyzing its Lick/Kast optical and IRTF/SpeX near-infrared spectra, as well as its broadband spectral energy distribution, and \textit{Gaia} parallax. We then derived the properties of the two planets by modelling multi-colour transit photometry from TESS, SPECULOOS-South, MuSCAT3, ExTrA, TRAPPIST-South, and SAINT-EX. Archival imaging, Gemini-South/Zorro high-resolution imaging, and Subaru/IRD radial velocities also support our planetary interpretation.}
   {With a mass of $0.118 \pm 0.002$ $M_\odot$, a radius of $0.1556 \pm 0.0086$ $R_\odot$, and an effective temperature of $2850 \pm 75$ K, \hbox{LP~890-9} is the second-coolest star found to host planets, after TRAPPIST-1. The inner planet has an orbital period of 2.73 d, a radius of $1.320_{-0.027}^{+0.053}$ $R_\oplus$, and receives an incident stellar flux of $4.09 \pm 0.12$ $S_\oplus$. The outer planet has a similar size of $1.367_{-0.039}^{+0.055}$ $R_\oplus$ and an orbital period of 8.46 d. With an incident stellar flux of $0.906 \pm 0.026$ $S_\oplus$, it is located within the conservative habitable zone, very close to its inner limit (runaway greenhouse). Although the masses of the two planets remain to be measured, we estimated their potential for atmospheric characterisation via transmission spectroscopy using a mass-radius relationship and found that, after the TRAPPIST-1 planets, LP~890-9\,c is the second-most favourable habitable-zone terrestrial planet known so far (assuming for this comparison a similar atmosphere for all planets).}
   {The discovery of this remarkable system offers another rare opportunity to study temperate terrestrial planets around our smallest and coolest neighbours.}

   \keywords{Planets and satellites: detection -- Stars: individual: LP~890-9 -- Stars: individual: TIC 44898913 -- Stars: individual: TOI-4306 -- Stars: individual: SPECULOOS-2 -- Techniques: photometric}

   \maketitle
%
\section{Introduction}

One main goal of modern astronomy is the identification and atmospheric characterisation of temperate terrestrial exoplanets, to understand how frequently and under which conditions life may exist around other stars. Terrestrial planets transiting nearby late-type M dwarfs are key in this endeavor. Indeed, for a given planet (fixed radius, mass, and equilibrium temperature), the signal-to-noise ratio (S/N) of atmospheric spectral features probed by eclipse (transit or occultation) spectroscopy increases for smaller and cooler host stars (e.g. \citealt{2009ApJ...698..519K}, \citealt{2013Sci...342.1473D}). As a consequence, fewer eclipse observations need to be co-added to achieve a significant atmospheric detection. In this regard, the low luminosity of the latest-type M stars is also a considerable advantage, as it results in more frequent planetary eclipses for a given stellar irradiation, meaning that it takes much less time to obtain a certain amount of in-eclipse observations. The S/N of atmospheric spectral features also scales inversely with the distance from the Earth to the host star, making the nearest late-type M dwarfs the best targets to search for transiting temperate terrestrial planets whose atmospheres could be characterised with current or next-generation facilities.

These considerations motivated the development of the SPECULOOS project (Search for habitable Planets EClipsing ULtra-cOOl Stars; \citealt{2018NatAs...2..344G}, \citealt{2018haex.bookE.130B}, \citealt{2018SPIE10700E..1ID}), an exoplanet transit survey targeting a volume-limited (40 pc) sample of about 1700 late-type dwarfs with spectral type M6 and later -- mostly stars, but also $\sim$5\% brown dwarfs \citep{2021A&A...645A.100S} -- using a network of 1m-class robotic telescopes. While SPECULOOS started its scientific operations officially in 2019 \citep{2018Msngr.174....2J}, it was initiated in 2011 as a prototype survey \citep{2013EPJWC..4703001G} targeting fifty of the brightest southern late-type M dwarfs with the TRAPPIST-South telescope. This prototype survey led to the discovery of the \hbox{TRAPPIST-1} system\footnote{Internally, the SPECULOOS team also refers to this discovery as SPECULOOS-1, since it was the first official detection as part of the SPECULOOS project.}, consisting of seven temperate Earth-sized planets transiting a nearby (12 pc) M8V star \citep{2016Natur.533..221G,2017Natur.542..456G,2017NatAs...1E.129L}. The discovery of this benchmark system has caused a flurry of theoretical and observational follow-up studies, so that the TRAPPIST-1 planets are today the best-studied terrestrial planets outside our Solar System \citep{2021PSJ.....2....1A}. They are also the most favourable targets found so far in the temperate terrestrial regime for atmospheric characterisation with the recently-launched JWST (see, e.g., \citealt{2019AJ....158...27L,2020arXiv200204798G}) and are already approved for substantial ($\sim$200 hr) Cycle 1 Guaranteed Time Observations and General Observers programs. 

Besides TRAPPIST-1, only two other transiting systems are currently known around M dwarfs with spectral type M5 or later: \hbox{LHS 3844} \citep{2019ApJ...871L..24V} and LP 791-18 \citep{2019ApJ...883L..16C}. LHS 3844\,b is an ultra-short-period super-Earth \hbox{(1.3 $R_\oplus$)} that orbits its M5V star every 11 hours. LP 791-18 is an M6V star transited by a super-Earth (1.1 $R_\oplus$) and a sub-Neptune (2.3 $R_\oplus$) with respective orbital periods of 0.95 and 4.99 days. Both systems were detected by the \textit{Transiting Exoplanet Survey Satellite} (TESS, \citealt{2015JATIS...1a4003R}). While TESS is efficient at detecting small transiting planets around early-to-mid-type M dwarfs (e.g. \citealt{2019NatAs...3.1099G,2019AJ....158...32K,2020A&A...642A..49D,wells2021,nicole2022,2022MNRAS.514.4120G}), its detection potential drops sharply for objects later than $\sim$M5 \citep{2015ApJ...809...77S,2018ApJS..239....2B,2021A&A...645A.100S}, due to their faintness in TESS's bandpass. The two systems mentioned above demonstrate that TESS is able to detect some short-period and/or $\gtrsim$2 $R_\oplus$ planets around some bright late-type M stars. However, possible additional transiting planets in the same systems with longer orbital periods and/or smaller radii may be missed due to the limited photometric precision. With its near-infrared-optimised cameras and custom `$I+z$' filters (transmittance $>$90\% from 750 to $\sim$1100 nm), SPECULOOS is specifically designed to achieve high photometric precision on late-type/ultra-cool dwarf stars and can detect those additional transiting planets, thus allowing an effective synergy between the two surveys.

In this paper, we present a discovery that leverages that synergy: the detection of two temperate super-Earths transiting the nearby M6-type dwarf star LP~890-9. The inner planet was first detected by TESS based on four sectors of data. The announcement of this planet candidate (identified by TESS as \hbox{TOI-4306.01}) triggered intensive photometric monitoring of the target with the SPECULOOS Southern Observatory, which led to the discovery of a second longer-period transiting planet (identified by SPECULOOS as SPECULOOS-2\,c), previously undetected by TESS. The orbital period of this second planet was later confirmed by MuSCAT3 follow-up observations. We validated the planetary nature of the system through follow-up observations from several ground-based facilities, including high-precision multi-colour photometry, spectroscopy, high-angular-resolution imaging, and archival images. 

The paper is structured as follows. In Sect. \ref{sec:obs}, we describe the contributing facilities and datasets used in the validation and characterisation of the system. We infer the properties of the host star in Sect. \ref{sec:star} and validate the planetary nature of the two transit signals in Sect. \ref{sec:vetting}. In Sect. \ref{sec:phot}, we present our detailed analysis of the photometry, including a global transit analysis to derive the system's properties (Sect. \ref{sec:analysis}), a search for additional transiting planets in the TESS and SPECULOOS-South data and an assessment of their detection limits (Sect. \ref{sec:search}), as well as a study of the stellar variability (Sect. \ref{sec:activity}). Sect. \ref{sec:rv_analysis} describes our analysis of Subaru/IRD radial velocities and derivation of preliminary mass constraints. In Sect. \ref{sec:dynamics}, we present a dynamical analysis of the system, including studies of its tidal evolution and long-term stability. Finally, we discuss the system's properties, potential habitability, and prospects for follow-up in Sect. \ref{sec:discussion}, before concluding in Sect. \ref{sec:conclusion}. 

\section{Observations} \label{sec:obs}

In this section, we present all the observations of \hbox{LP~890-9} obtained with TESS and ground-based follow-up facilities. 

\begin{figure*}
    \centering
    \includegraphics[width=0.4\textwidth]{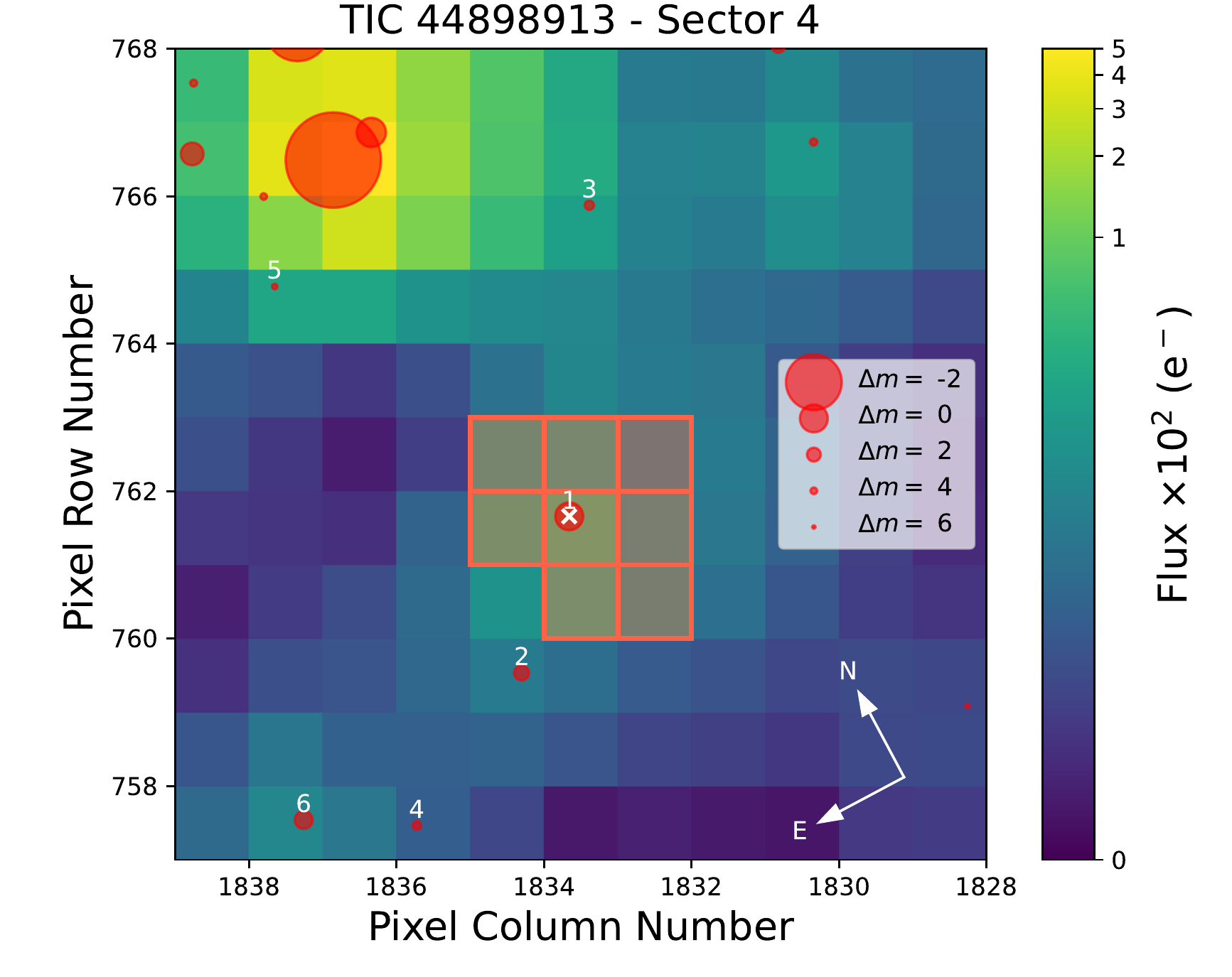}
    \includegraphics[width=0.4\textwidth]{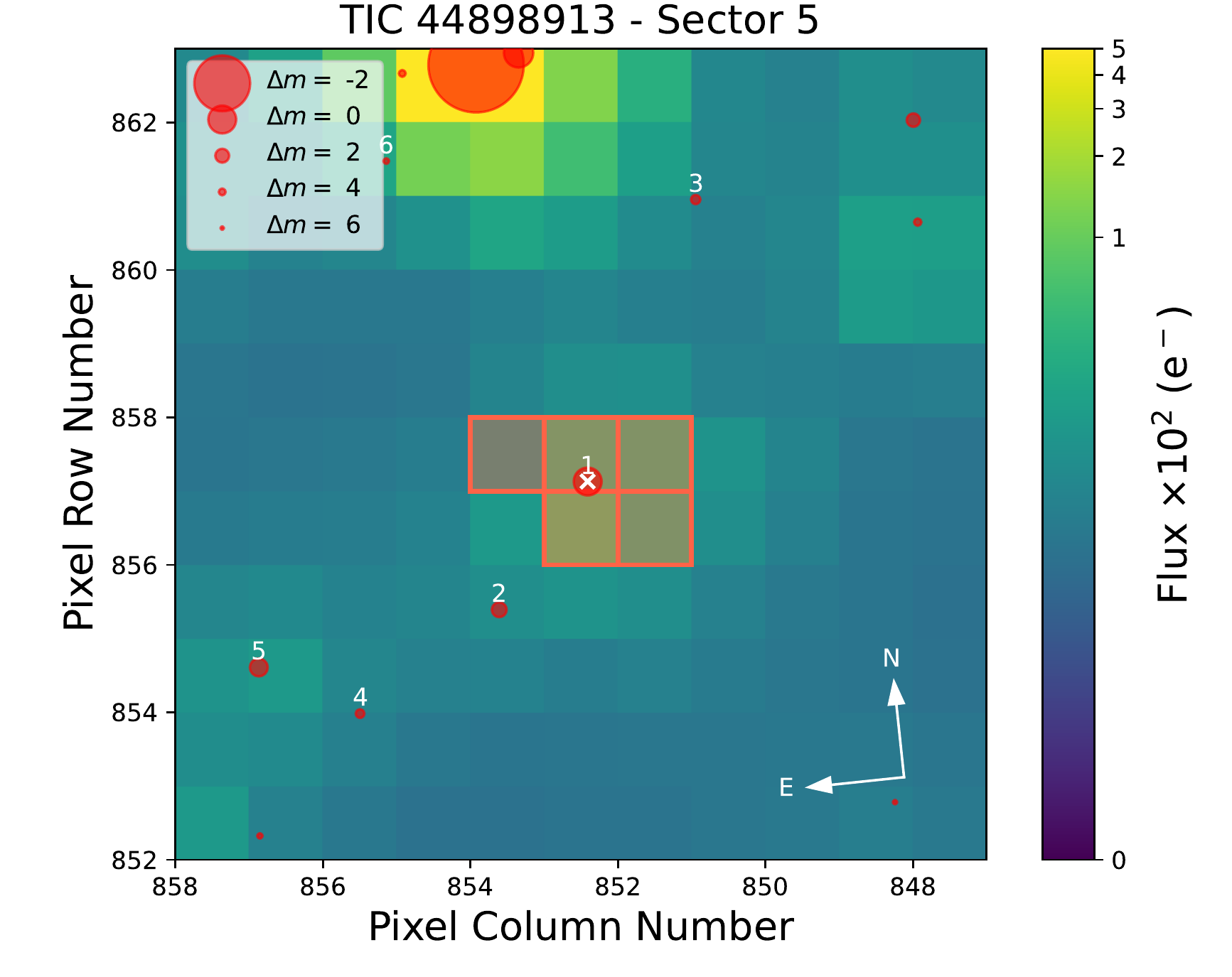}
    \includegraphics[width=0.4\textwidth]{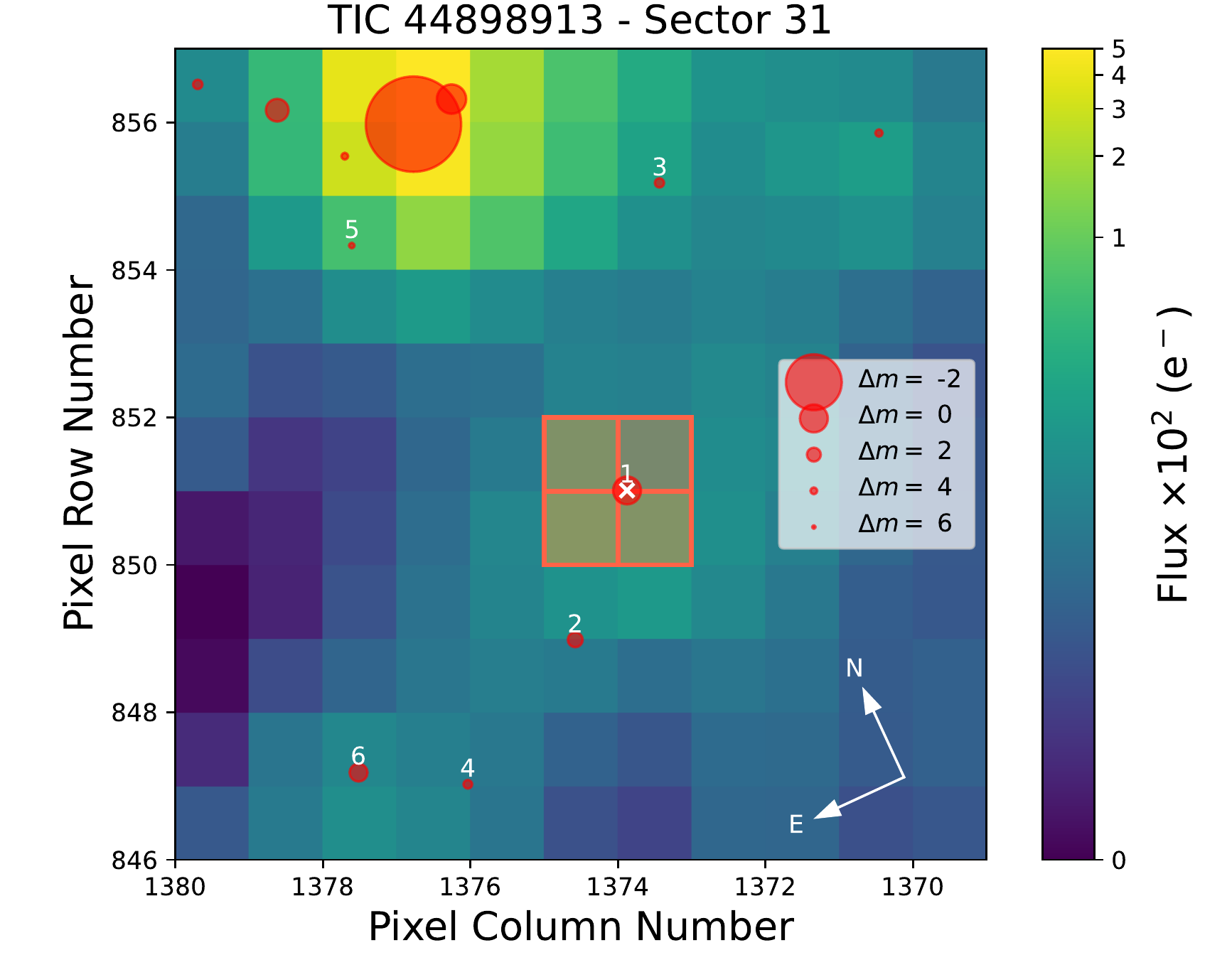}
    \includegraphics[width=0.4\textwidth]{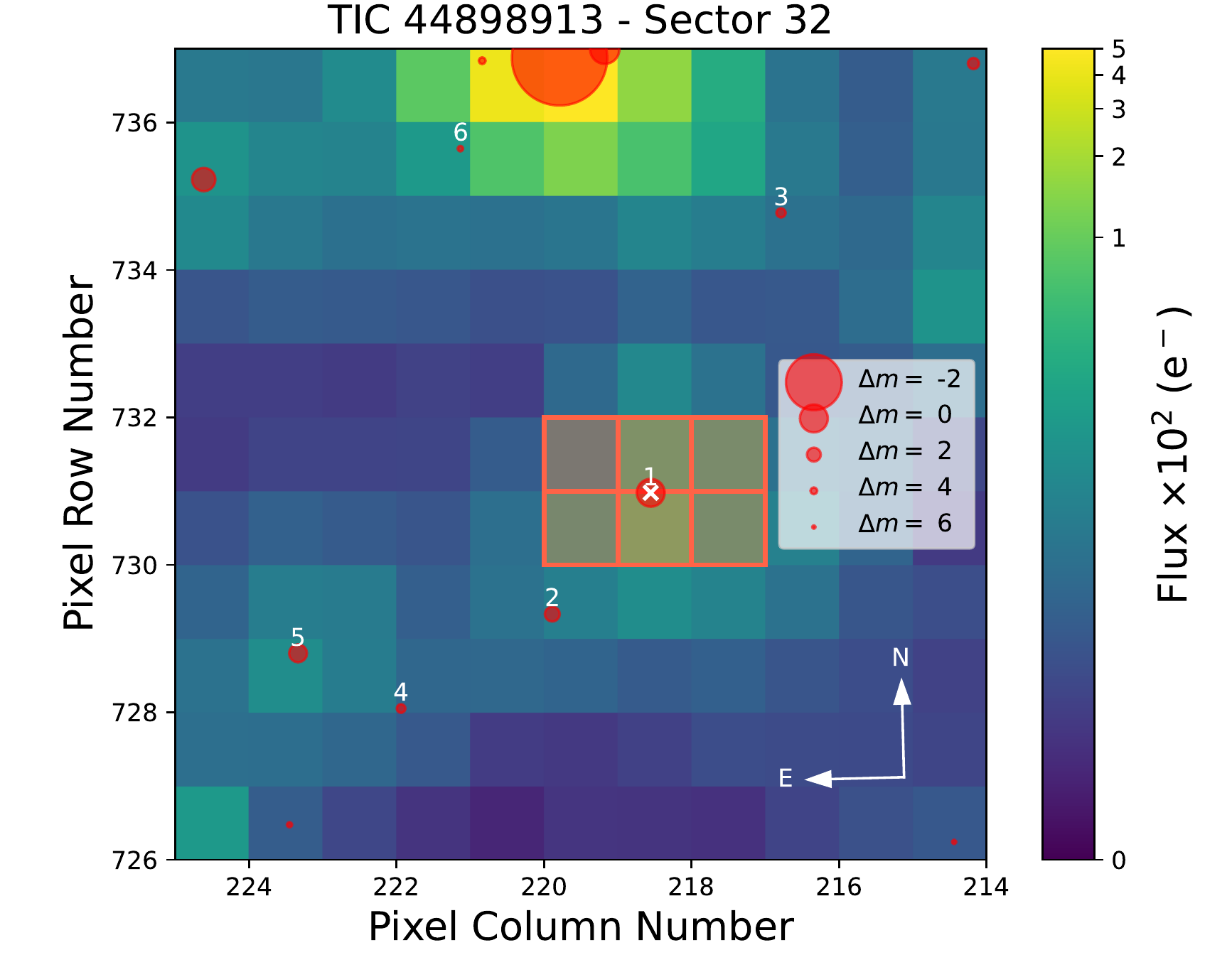}
    \caption{TESS target pixel files of LP~890-9 for sectors 4 (upper left), 5 (upper right), 31 (lower left), and 32 (lower right). The pixel scale is 21$\arcsec$ per pixel. The electron counts are colour-coded. The target is indicated by a white cross. Nearby sources identified in {\it Gaia\/} DR2 \citep{2018A&A...616A...1G}, up to 6 magnitudes in contrast with LP~890-9, are shown as red circles. The symbol size is proportional to the magnitude contrast with the target. The apertures used by the SPOC pipeline to extract the photometry are shown as red shaded regions. These plots were created with \texttt{tpfplotter} \citep{2020A&A...635A.128A}.}
    \label{fig:tpf}
\end{figure*}

\subsection{TESS photometry}
\label{obs:tess}

LP~890-9 (TIC 44898913) is part of the TESS Cool Dwarf catalogue \citep{2018AJ....155..180M}. It was identified as an attractive target for a transit search with TESS, and thus included in the TESS Candidate Target List \citep{2018AJ....156..102S}. The star was observed at a two-minute cadence during sectors 4-5 (18 October $-$ 11 December 2018) of the primary mission and again two years later during sectors 31-32 (21 October $-$ 17 December 2020) of the extended mission. The observations were acquired with CCDs 1 (sectors 5 and 32) and 2 (sectors 4 and 31) on Camera 2. The data were processed with the TESS Science Processing Operations Center (SPOC) pipeline \citep{2016SPIE.9913E..3EJ} at NASA Ames Research Center and searched for periodic transit signals \citep{2002ApJ...575..493J,2010ApJ...713L..87J,Jenkins2020}. While the analysis of the first two sectors did not identify any convincing signal, the addition of two more sectors during the extended mission revealed a $\sim$0.7\%-deep transit-like signature at a period of 2.73 days, with a Multiple Event Statistic (MES) of 8.3. The TESS Science Office reviewed the SPOC Data Validation Report \citep{2018PASP..130f4502T,2019PASP..131b4506L} for this signal and announced the planet candidate TOI-4306.01 on 21 July 2021 \citep{2021ApJS..254...39G}. Fig. \ref{fig:tpf} shows the target pixel files and photometric apertures used by SPOC for each of the four sectors, with the locations of nearby \textit{Gaia} DR2 sources, up to 6 magnitudes in contrast with LP~890-9. None of them lie within the SPOC photometric apertures.

\begin{figure*}
    \centering
    \includegraphics[width=0.85\textwidth]{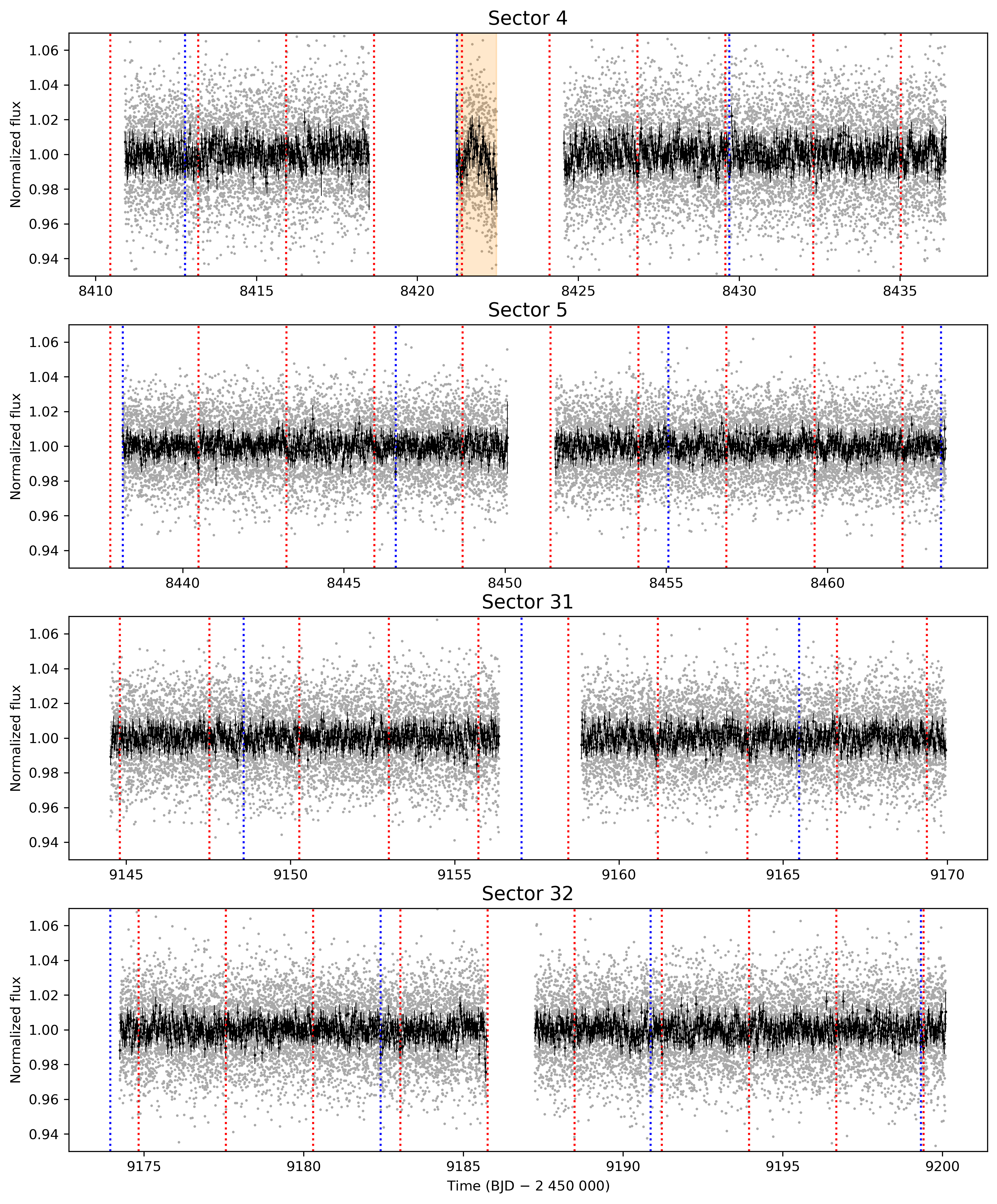}
    \caption{TESS photometry of LP~890-9. For each of the four sectors, the 2-minute data points (in grey) have been binned into 30-minute intervals to produce the black points, with error bars corresponding to the root-mean-square of the uncertainties of the points in the bins. The transits of LP~890-9\,b and c are indicated by red and blue dotted lines, respectively. The region marked in orange in sector 4 was impacted by thermal effects and thus excluded from our analysis.}
    \label{fig:TESS}
\end{figure*}

We retrieved the Presearch Data Conditioning Simple Aperture Photometry (PDCSAP, \citealt{2012PASP..124..985S,2012PASP..124.1000S,2014PASP..126..100S}) from the Mikulski Archive for Space Telescopes (MAST) and removed all data points for which the quality flag was not zero.  The resulting light curves are shown in Fig. \ref{fig:TESS}, together with the transit times. As noted in the data release notes\footnote{\url{https://archive.stsci.edu/missions/tess/doc/tess_drn/tess_sector_04_drn05_v04.pdf}} for sector 4, communications between the instrument and spacecraft were interrupted between BJD times \hbox{2 458 418.54} and 2 458 421.21, during which time no data or telemetry were collected, hence a gap in the light curve. The photometry obtained after this instrument anomaly was impacted by some thermal effects, as the camera temperature increased by $\sim$20 degrees before returning to nominal within three days. We thus decided to exclude this region of the light curve (marked in orange in \hbox{Fig. \ref{fig:TESS}}) from our analysis.

\begin{table*}[]
    \centering
    \begin{tabular}{l c c c c}
    \toprule
    \toprule
         \textbf{Date (UT)} & \textbf{Facility} & \textbf{Bandpass} & \textbf{Exp. time (s)} & \textbf{Notes} \\
         \hline
         10 Aug. 2021 & SSO/Europa & $I+z$ & 35 & b full \\
         10 Aug. 2021 & TRAPPIST-South & blue-blocking  & 120 & b full \\
         21 Aug. 2021 & SSO/Callisto & $i'$ & 92 & b full \\
         21 Aug. 2021 & SSO/Io & $I+z$ & 39 & b full \\
         1 Sept. 2021 & SSO/Europa & $i'$ & 92 & b full \\
         1 Sept. 2021 & SSO/Io & $z'$ & 49 & b full \\
         20 Sept. 2021 & SSO/Ganymede & $r'$ & 180 & b full \\
         20 Sept. 2021 & SSO/Io & $r'$ & 180 & b full \\
         20 Sept. 2021 & SSO/Europa & $r'$ & 180 & b full \\
         20 Sept. 2021 & ExTrA1 & 0.85$-$1.55 $\mu$m & 60 & b full \\
         20 Sept. 2021 & ExTrA2 & 0.85$-$1.55 $\mu$m & 60 & b full \\
         28 Sept. 2021 & MuSCAT3 & $g'$ & 120 & b full \\
         28 Sept. 2021 & MuSCAT3 & $r'$ & 120 & b full \\
         28 Sept. 2021 & MuSCAT3 & $i'$ & 110 & b full \\
         28 Sept. 2021 & MuSCAT3 & $z_\mathrm{s}$ & 100 & b full \\
         1 Oct. 2021 & SSO/Europa & $I+z$ & 39 & b full \\
         16 Oct. 2021 & SSO/Io & $I+z$ & 39 & c full \\
         23 Oct. 2021 & SSO/Europa & $I+z$ & 39 & b full \\
         28 Oct. 2021 & MuSCAT3 & $g'$ & 120 & b full \\
         28 Oct. 2021 & MuSCAT3 & $r'$ & 120 & b full \\
         28 Oct. 2021 & MuSCAT3 & $i'$ & 55 & b full \\
         28 Oct. 2021 & MuSCAT3 & $z_\mathrm{s}$ & 29 & b full \\
         2 Nov. 2021 & SSO/Europa & $I+z$ & 39 & c full \\
         8 Nov. 2021 & MuSCAT3 & $g'$ & 120 & b full \\
         8 Nov. 2021 & MuSCAT3 & $r'$ & 120 & b full \\
         8 Nov. 2021 & MuSCAT3 & $i'$ & 55 & b full \\
         8 Nov. 2021 & MuSCAT3 & $z_\mathrm{s}$ & 29 & b full \\
         11 Nov. 2021 & SSO/Europa & $z'$ & 49 & b full \\
         19 Nov. 2021 & SSO/Europa & $I+z$ & 39 & c full \\
         30 Nov. 2021 & SSO/Europa & $I+z$ & 39 & b full \\
         30 Nov. 2021 & SAINT-EX & $I+z$ & 128 & b partial \\
         3 Dec. 2021 & SSO/Europa & $I+z$ & 39 & b full \\
         2 Jan. 2022 & SSO/Europa & $I+z$ & 39 & b full \\
         17 Jan. 2022 & MuSCAT3 & $r'$ & 120 & c full \\
         17 Jan. 2022 & MuSCAT3 & $i'$ & 55 & c full \\
         17 Jan. 2022 & MuSCAT3 & $z_\mathrm{s}$ & 29 & c full \\
         \bottomrule
         \bottomrule
    \end{tabular}
    \caption{Ground-based transit light curves of LP~890-9 used in our global transit analysis. The last column shows which planet was observed and whether the transit was fully or only partially covered.}
    \label{tab:follow_up_obs}
\end{table*}

\subsection{Ground-based follow-up photometry}

We obtained follow-up photometry of multiple transit events of the candidate TOI-4306.01 with several ground-based facilities and different bandpasses, as part of TESS Follow-up Observing Program (TFOP) Sub-Group 1 (SG1) for Seeing-Limited Photometry. The goals of these observations were to confirm that the transit signal detected by TESS was on the expected star, assess its chromaticity, and obtain higher-precision transit light curves compared to the TESS data.
In addition to observations of transits associated with TOI-4306.01, we also monitored the target intensively with the SPECULOOS Southern Observatory (see Sect. \ref{sec:sso}), to search for possible additional transiting planets that may not have been detected by TESS and study the variability of the star. As mentioned previously, this monitoring revealed a second longer-period transiting planet candidate (identified by SPECULOOS as SPECULOOS-2\,c), whose orbital period was later confirmed by MuSCAT3 follow-up observations (see Sect. \ref{obs:muscat3}). We describe all the ground-based photometric observations in the following sections and summarise the transit light curves of both planets in Table \ref{tab:follow_up_obs}. 

\subsubsection{SPECULOOS-South}
\label{sec:sso}

The SPECULOOS Southern Observatory \citep[SSO, ][]{2018haex.bookE.130B,2018SPIE10700E..1ID,2018NatAs...2..344G,2018Msngr.174....2J} consists of four 1-m telescopes (Io, Europa, Ganymede, and Callisto) located at ESO Paranal Observatory in Chile. Each telescope is equipped with a deep-depletion 2K$\times$2K CCD detector optimised for the near infrared, providing a 12$\arcmin$ $\times$ 12$\arcmin$ field of view with a pixel scale of 0.35$\arcsec$ per pixel.

Fourteen transit light curves of TOI-4306.01 were obtained with SSO in several filters (see Table \ref{tab:follow_up_obs}): three in $r'$, two in $i'$, two in $z'$, and seven in the special `$I+z$' filter. As mentioned above, we also performed an intense photometric monitoring of the system outside of these transits, gathering in total 614.45 hours of observations spread over 119 nights between 9 August 2021 and 20 January 2022. For this monitoring, we used the $I+z$ filter with an exposure time of 39 seconds to optimise the S/N. These data revealed a second transit signal, with a transit depth of about 0.6\%. Three transits of this second object (identified by SPECULOOS as SPECULOOS-2\,c) were observed with SSO (see Table \ref{tab:follow_up_obs}), all in the $I+z$ filter. Each pair of consecutive transits was separated by 16.914 days. The only fractional period alias that remained possible based on the SSO data was 8.457 days (16.914/2 days). Shorter period aliases were excluded, as they would have produced several other detectable transits in the SSO data. The period of 8.457 days, which was $\sim$1.7 times more likely than the 16.914-days period based on geometric transit probability, was later confirmed thanks to photometric follow-up observations from Hawaii using MuSCAT3 (see Sect. \ref{obs:muscat3}).

The data were processed with the automatic SSO pipeline, which is described in detail in \cite{2020MNRAS.495.2446M}. This pipeline first performs standard image reduction steps, applying bias, dark, and flat-field corrections. Star detection, astrometric solving, and aperture photometry are then conducted using the \texttt{casutools} package \citep{2004SPIE.5493..411I}. For the observations presented here, the selected aperture radii are between 8 and 16 pixels (2.8--5.6$\arcsec$), depending on the filter and night. The target's light curve is then generated by an automated differential photometry algorithm, which uses a weighted ensemble of comparison stars to correct for most atmospheric and instrumental effects. Finally, this light curve is corrected for the effects of time-varying telluric water vapour (see also Pedersen et al., in prep.), which can significantly affect near-infrared differential photometry of very red objects such as LP~890-9.

\subsubsection{TRAPPIST-South}

We observed a transit of TOI-4306.01 on 10 August 2021 with the 0.6-m TRAPPIST-South telescope \citep{2011EPJWC..1106002G,2011Msngr.145....2J}, located at ESO La Silla Observatory in Chile. The wide `blue-blocking' filter (transmittance $>$90\% from 520 to $\sim$1100 nm) was used for these observations, with an exposure time of 120 seconds. The data were processed using the \texttt{prose} open-source\footnote{\url{https://github.com/lgrcia/prose}} Python framework, which is described in detail in \cite{2022MNRAS.509.4817G}. We used a photometric aperture of 8.7 pixels (5.6 $\arcsec$).

\subsubsection{ExTrA}

ExTrA \citep{Bon2015} is a near-infrared (0.85 to 1.55 $\mu$m) multi-object spectrograph fed by three 60-cm telescopes located at La Silla Observatory in Chile. One full transit of TOI-4306.01 was observed on 20 September 2021 using two ExTrA telescopes. We used 8$\arcsec$ aperture fibres and the low-resolution mode ($R {\sim} 20$) of the spectrograph with an exposure time of 60 seconds. Five fibre positioners are used at the focal plane of each telescope to select light from the target and four comparison stars. As comparison stars, we observed 2MASS J04172127-2815315, 2MASS J04191176-2812449, 2MASS J04181010-2753179, and 2MASS J04164312-2759006 with 2MASS $J$-magnitudes \citep{2006AJ....131.1163S} and \textit{Gaia} effective temperatures \citep{2018A&A...616A...1G} similar to the target. The resulting ExTrA data were analysed using a custom data reduction software, described in more detail in \cite{Coi2021}.

\subsubsection{MuSCAT3}
\label{obs:muscat3}

We observed three transits of TOI-4306.01 in the $g'$, $r'$, $i'$, and $z_\mathrm{s}$ bands simultaneously by using the multi-band imager MuSCAT3 on the Faulkes Telescope North of Las Cumbres Observatory on Haleakala, Hawaii \citep{2020SPIE11447E..5KN}. MuSCAT3 is equipped with four 2k$\times$ 2k CCD cameras for the four channels, each providing a pixel scale of 0.266$\arcsec$ pixel$^{-1}$ and a field of view of 9.1$\arcmin$ $\times$ 9.1$\arcmin$.
The first transit was observed on 28 September 2021 (UT) with exposure times of 120, 120, 110, and 100\,s in the $g'$, $r'$, $i'$ and $z_\mathrm{s}$ bands, respectively. The second transit was observed on 28 October 2021 (UT) with exposure times of 120, 120, 55, and 29\,s in the $g'$, $r'$, $i'$ and $z_\mathrm{s}$ bands, respectively. The third transit was taken on 8 November 2021 (UT) with the same exposure times as the second transit. All observations were conducted without defocussing. On the night of the first, second, and third transits, the lunar phase was 21, 22, and 3 days, and the angular distance between the target and the moon was 53, 70, and 125 degrees, respectively.

We also observed one transit of the second planet candidate detected by SSO (SPECULOOS-2\,c) with MuSCAT3 on 17 January 2022 (UT). The goal of these observations was to firmly confirm the orbital period of 8.457 days, since the less likely 16.914-days period would not have produced a transit during that night. We used exposure times of 120, 120, 55, and 29\,s for the $g'$, $r'$, $i’$ and $z_\mathrm{s}$ bands, respectively. The lunar phase and the moon distance were 14 days and 64 degrees, respectively. The sky background was bright because the almost-full moon was located nearby. Thus, the photometric precision in the $g'$ and $r'$ bands was not as good as for the previous observations of TOI-4306.01. Nonetheless, the transit was clearly detected in the $i'$ and $z'$ bands (see Fig.~\ref{fig:LCs_c}), thus confirming the orbital period of 8.457 days. We note that we did not use the $g'$-band light curve in our global transit analysis (see Sect.~\ref{sec:analysis}), as its scatter ($\sim$2.9\%) was too large for it to be useful.

The obtained images were calibrated by the {\tt BANZAI} pipeline \citep{2018SPIE10707E..0KM}. We performed aperture photometry on the calibrated images using a custom pipeline \citep{2011PASJ...63..287F}. We optimised the aperture radius and combination of comparison stars for each night and each band so that the dispersion of the light curve was minimised. The adopted aperture radii were 10--14 pixels, depending on the band and night.

\subsubsection{SAINT-EX}

A partial transit of TOI-4306.01 was observed on 30 November 2021 with the 1-m SAINT-EX telescope, located at the Observatorio Astron\'omico Nacional in the Sierra de San Pedro M\'artir in Mexico \citep{2018csss.confE..59S,2020A&A...642A..49D}. SAINT-EX is a twin of the SSO telescopes but the camera is operated using a different readout mode and higher gain so that it can observe brighter stars. For these observations, we used the $I+z$ filter with an exposure time of 128 seconds. The data were reduced using the custom PRINCE pipeline, which is detailed in \cite{2020A&A...642A..49D}. A weighted principal component analysis (PCA) approach \citep{2012PASP..124.1015B} was used to correct the light curve for systematics that are common between the target and comparison stars.

\subsection{Spectroscopy}

In this section, we describe the spectroscopic observations of LP~890-9 that we obtained to characterise the star (low-resolution near-infrared and optical spectra, Sect. \ref{spex} and \ref{kast}) and to measure radial velocities (high-resolution near-infrared spectra, Sect. \ref{ird}).

\subsubsection{IRTF/SpeX} 
\label{spex}

We obtained a low-resolution near-infrared spectrum of \hbox{LP~890-9} with the SpeX spectrograph \citep{spex} on the 3.2-m NASA Infrared Telescope Facility (IRTF) on two nights, 30\,August\,2021 and 23\,November\,2021 (UT).
The conditions on both nights were clear with seeing of 0.7-0.8$\arcsec$.
We used the short-wavelength cross-dispersed (SXD) mode with the 0.3$\arcsec$ $\times$ 15$\arcsec$ slit aligned to the parallactic angle, giving a 0.8--2.5\,$\mu$m spectrum with a resolving power of $R {\sim} 2000$.
In August, we collected three ABBA nod sequences (12 exposures) with integration times of 179.8\,seconds per exposure, giving a total integration time of 36\,minutes.
In November, we collected a single ABBA nod sequence with integration times of 599.6\,seconds per exposure, giving a total integration time of 40\,minutes.
We collected a set of standard SXD flat-field and arc-lamp exposures immediately after the science exposures, followed by a set of six exposures of the A0\,V star HD\,27016 ($V = 7.77$) in August and the A0\,V star HD\,22687 ($V = 8.55$) in November for flux and telluric calibration.
We reduced the data using Spextool v4.1 \citep{spextool}, following the standard instructions in the Spextool User's Manual\footnote{Available at \url{http://irtfweb.ifa.hawaii.edu/~spex/observer/}.}.
The final spectra have a median S/N of $\sim$100, with peaks in the $J$, $H$, and $K$ bands of 126, 136, and 126, respectively.

\subsubsection{Lick/Kast} 
\label{kast}

We obtained a low-resolution optical spectrum of LP~890-9 with the Kast Double Spectrograph on the 3m Shane Telescope at Lick Observatory on 14\,November\,2021 (UT) in clear conditions with 2$\arcsec$ seeing.
This instrument provides simultaneous blue and red optical spectra with parallel spectrograph cameras. Our analysis focussed on red optical data obtained with the 600/7500 grating and 1.5$\arcsec$-wide slit, providing 6000-9000~{\AA} spectra at an average resolution of $\lambda/\Delta\lambda$ $\approx$ 2500. 
The source was observed with 6 exposures of 600~s each within an hour of transit and at an average airmass of 2.5.
The flux calibrator Feige~110 \citep{1992PASP..104..533H,1994PASP..106..566H} was observed on the same night, and dome illuminated flat field and arc lamp exposures were obtained at the start of the night for pixel response and wavelength calibration.
Data were reduced using the KastRedux package\footnote{\url{https://github.com/aburgasser/kastredux}.} 
(Burgasser et al.\ in prep.), and included image reduction (flat-fielding, bad pixel masking, and linearity correction), optimal extraction, wavelength calibration, and flux density calibration.
No attempt was made to correct for telluric absorption. 
The data have a median S/N of $\sim$50 at 7400~{\AA}.

\subsubsection{Subaru/IRD} 
\label{ird}

To measure radial velocities (RVs) for LP~890-9, we obtained high-resolution near-infrared spectra with the InfraRed Doppler (IRD) spectrograph mounted on the Subaru 8.2-m telescope \citep{2012SPIE.8446E..1TT, 2018SPIE10702E..11K}. IRD is a fibre-fed, temperature-stabilised spectrograph, having a wavelength coverage of 970--1730 nm ($Y$, $J$, and $H$-bands) with a spectral resolution of $\approx 70,000$. For the simultaneous wavelength reference, laser-frequency comb (LFC) light was routinely injected into IRD using a secondary fibre. 
A total of 14 frames were secured for LP~890-9 between 10 September 2021 and 10 January 2022 (UT). Due to instrumental trouble in one of the two IRD detectors, only $H$-band spectra ($>1450$ nm) were obtained for the three frames taken in January 2022.
Given the faintness of the target ($J=12.258$), the exposure times were set to 1800 seconds for all frames. 

We reduced the raw IRD data as per the reduction procedure in \citet{2020PASJ...72...93H}, and extracted both stellar and LFC spectra, whose wavelengths were calibrated using the Thorium-Argon hollow cathode lamp. The reduced stellar spectra had a typical S/N of $\approx 25$ per pixel at 1000 nm. Using IRD's analysis pipeline for RV measurements \citep{2020PASJ...72...93H}, we computed relative RVs for the individual stellar spectra. In doing so, both $YJ$ and $H$-band spectra were analysed for all frames except the ones (three frames) in January 2022, for which we were only able to measure RVs for $H$-band spectra. The resulting RV measurements are given in Table \ref{tab:RVs}. The typical measurement errors are $10-15$ m/s and $7-10$ m/s for the $YJ$-band and the $H$-band spectra, respectively.

\subsection{High-angular-resolution imaging}

Exoplanet host stars can have spatially close companions which are bound or line-of-sight objects. These companions can create a false-positive transit signal if, for example, they are an eclipsing binary or other variable star. But mainly, close companions cause `third-light' flux contamination that leads to an underestimated planetary radius if not accounted for in the transit model \citep{2015ApJ...805...16C}. Companion stars can also cause non-detections of small planets residing with the same exoplanetary system \citep{2021AJ....162...75L}. Thus, to search for close-in (bound) companions unresolved in TESS or other ground-based follow-up observations, we obtained high-resolution imaging speckle observations of LP~890-9.

\begin{figure}
    \centering
    \includegraphics[width=\linewidth, height=\textheight, keepaspectratio]{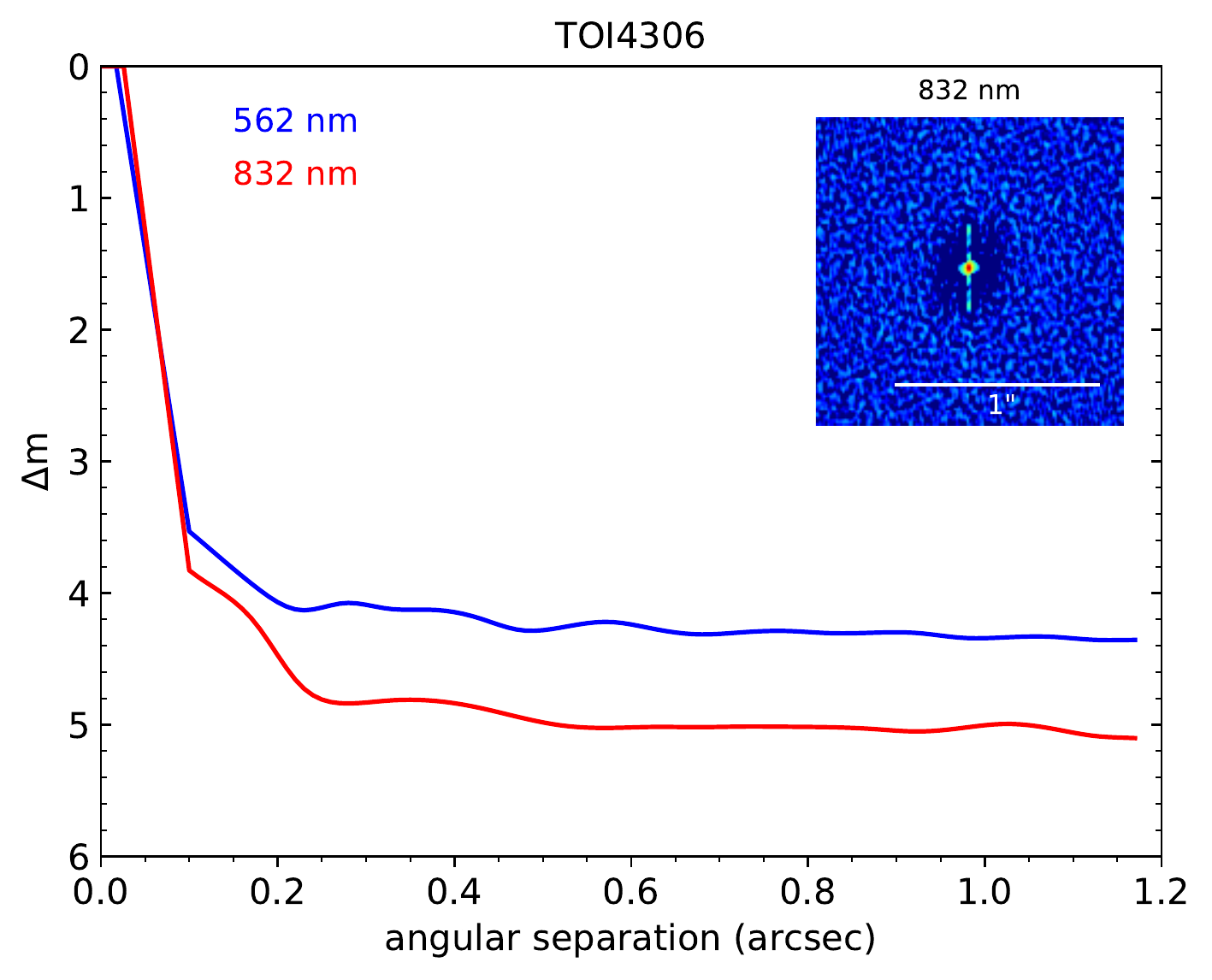}
    \caption{Zorro speckle imaging 5$\sigma$ contrast curves at 562 (blue) and \hbox{832 nm} (red), along with the 832 nm reconstructed image.}
    \label{fig:zorro}
\end{figure}

LP~890-9 was observed on 20 March 2022 (UT) using the Zorro speckle instrument on the Gemini South 8-m telescope \citep{2021FrASS...8..138S}. Zorro provides simultaneous speckle imaging in two bands (562\,nm and 832\,nm) with output data products including a reconstructed image with robust contrast limits on companion detections (e.g. \citealt{2016ApJ...829L...2H}). Twenty-five sets of 1000 $\times$ 0.06-sec exposures were collected for this faint star and subjected to Fourier analysis in our standard reduction pipeline \citep{2011AJ....142...19H}. Fig.~\ref{fig:zorro} shows our final 5$\sigma$ contrast curves and the 832 nm reconstructed speckle image. We find that LP~890-9 is a single star with no companion brighter than 4-5 magnitudes below that of the target star from $\sim$0.2$\arcsec$ out to 1.2$\arcsec$. At the distance of \hbox{LP~890-9} (32.3 pc), these angular limits correspond to spatial limits of 6.5 to 39 au.

\section{Stellar properties} \label{sec:star}

LP~890-9 is a relatively low-activity late-type M dwarf located at 32 pc. In this section, we describe the methodology used to determine its properties, which are given in Table \ref{table:star} together with its catalogued photometric and astrometric parameters.

\begin{table*}[hbt!]
\caption{\normalsize Properties of the host star.}
\begin{center}
\begin{tabular}{l c c}
\toprule
\toprule
\textbf{Property} & \textbf{Value} & \textbf{Source} \\
\midrule
\multicolumn{2}{l}{\textit{Designations}} & \\
LP & 890-9 & \cite{1979nlcs.book.....L} \\
TIC & 44898913 & \cite{2018AJ....156..102S} \\
TOI & 4306 & \\
SPECULOOS & 2 & \\
2MASS & J04163114-2818526 & \cite{2006AJ....131.1163S} \\
\textit{Gaia} EDR3 & 4886243456388510720 & \cite{gaia_edr3} \\
\midrule
\multicolumn{2}{l}{\textit{Astrometric properties}} & \\
RA (J2000) & 04:16:31.16 & \textit{Gaia} EDR3 \citep{gaia_edr3} \\
Dec (J2000) & $-$28:18:52.95 & \textit{Gaia} EDR3 \citep{gaia_edr3} \\
$\mathrm{\mu_{RA}}$ (mas $\mathrm{yr}^{-1}$) & $218.569 \pm 0.036$ & \textit{Gaia} EDR3 \citep{gaia_edr3} \\
$\mathrm{\mu_{Dec}}$ (mas $\mathrm{yr}^{-1}$) & $-251.145 \pm 0.047$ & \textit{Gaia} EDR3 \citep{gaia_edr3} \\
Parallax (mas) & $30.933 \pm 0.042$ & \textit{Gaia} EDR3 \citep{gaia_edr3} \\
Distance (pc) & $32.33 \pm 0.05$ & \textit{Gaia} EDR3 \citep{gaia_edr3} \\
$V_{\mathrm{tan}}$ (km/s) & $51.02 \pm 0.07$ & \textit{Gaia} EDR3 \citep{gaia_edr3} \\
RV (km/s) & $27.94 \pm 0.60$ & Sect.~\ref{sec:age} \\
$U$ (km/s) & $17.66 \pm 0.27$ & Sect.~\ref{sec:age} \\
$V$ (km/s) & $-57.77 \pm 0.30$ & Sect.~\ref{sec:age} \\
$W$ (km/s) & $5.48 \pm 0.39$ & Sect.~\ref{sec:age} \\
\midrule
\multicolumn{2}{l}{\textit{Photometric magnitudes}} & \\
TESS (mag) & 14.2683 $\pm $ 0.0076 & \cite{2018AJ....156..102S} \\
$V$ (mag) & 18.0 $\pm$ 0.2 & \cite{2018AJ....155..180M} \\
$g$ (mag) & 18.4422 $\pm$ 0.0072 & Pan-STARRS1 \citep{2016arXiv161205560C} \\
$r$ (mag) & 17.1365 $\pm$ 0.0050 & Pan-STARRS1 \citep{2016arXiv161205560C} \\
$i$ (mag) & 15.0975 $\pm$ 0.0026 & Pan-STARRS1 \citep{2016arXiv161205560C} \\
$z$ (mag) & 14.1593 $\pm$ 0.0013 & Pan-STARRS1 \citep{2016arXiv161205560C} \\
$y$ (mag) & 13.6454 $\pm$ 0.0044 & Pan-STARRS1 \citep{2016arXiv161205560C} \\
\textit{Gaia} (mag) & 15.7913 $\pm$ 0.0028 & \textit{Gaia} EDR3 \citep{gaia_edr3} \\
$J$ (mag) & 12.258 $\pm$ 0.023 & 2MASS \citep{2006AJ....131.1163S} \\
$H$ (mag) & 11.692 $\pm$ 0.025 & 2MASS \citep{2006AJ....131.1163S} \\
$K$ (mag) & 11.344 $\pm$ 0.023 & 2MASS \citep{2006AJ....131.1163S} \\
$W$1 (mag) & 11.129 $\pm$ 0.061 & WISE \citep{2014yCat.2328....0C} \\
$W$2 (mag) & 10.920 $\pm$ 0.020 & WISE \citep{2014yCat.2328....0C} \\
$W$3 (mag) & 10.715 $\pm$ 0.083 & WISE \citep{2014yCat.2328....0C} \\
\midrule
\multicolumn{2}{l}{\textit{Spectroscopic and derived properties}} & \\
Optical SpT & M6.0 $\pm$ 0.5 & Kast spectrum (Sect.~\ref{sec:spectro_analysis}) \\
Near-infrared SpT & M6.0 $\pm$ 0.5 & SpeX spectrum (Sect.~\ref{sec:spectro_analysis}) \\ 
$\mathrm{[Fe/H]}$ (dex) & $-0.028 \pm 0.089$ & SpeX spectrum (Sect.~\ref{sec:spectro_analysis}) \\
$T_{\mathrm{eff}}$ (K) & $2850 \pm 75$ & SED fit (Sect.~\ref{sec:rsms}) \\ 
$F_{\mathrm{bol}}$ ($10^{-11}$ $\mathrm{erg\,s^{-1}\,cm^{-2}}$) & $4.41 \pm 0.15$ & SED fit (Sect.~\ref{sec:rsms}) \\
$L_\star$ ($10^{-3}$ $L_\odot$) & $1.441 \pm 0.038$ & $F_{\mathrm{bol}}$ + parallax (Sect.~\ref{sec:rsms}) \\
$R_\star$ ($R_\odot$) & $0.1556 \pm 0.0086$ & $F_{\mathrm{bol}}$ + $T_{\mathrm{eff}}$ + parallax (Sect.~\ref{sec:rsms}) \\ 
$M_\star$ ($M_\odot$) & $0.118 \pm 0.002$ & Evol. modelling + $L_\star$ + $\mathrm{[Fe/H]}$ (Sect.~\ref{sec:rsms}) \\ 
log $g_\star$ (cgs) & $5.126_{-0.047}^{+0.050}$ & $M_\star$ + $R_\star$ \\ 
$\rho_\star$ ($\mathrm{g\,cm^{-3}}$) & $44.2_{-6.7}^{+8.3}$ & $M_\star$ + $R_\star$ \\ 
Age (Gyr) & 7.2$^{+2.2}_{-3.1}$ & $UVW$ + $\mathrm{[Fe/H]}$ (Sect.~\ref{sec:age}) \\
\bottomrule
\bottomrule
\end{tabular}
\end{center}
\label{table:star}
\end{table*}

\subsection{Spectroscopic analysis}
\label{sec:spectro_analysis}

\begin{figure}
    \centering
    \includegraphics[width=1.1\linewidth, height=\textheight, keepaspectratio]{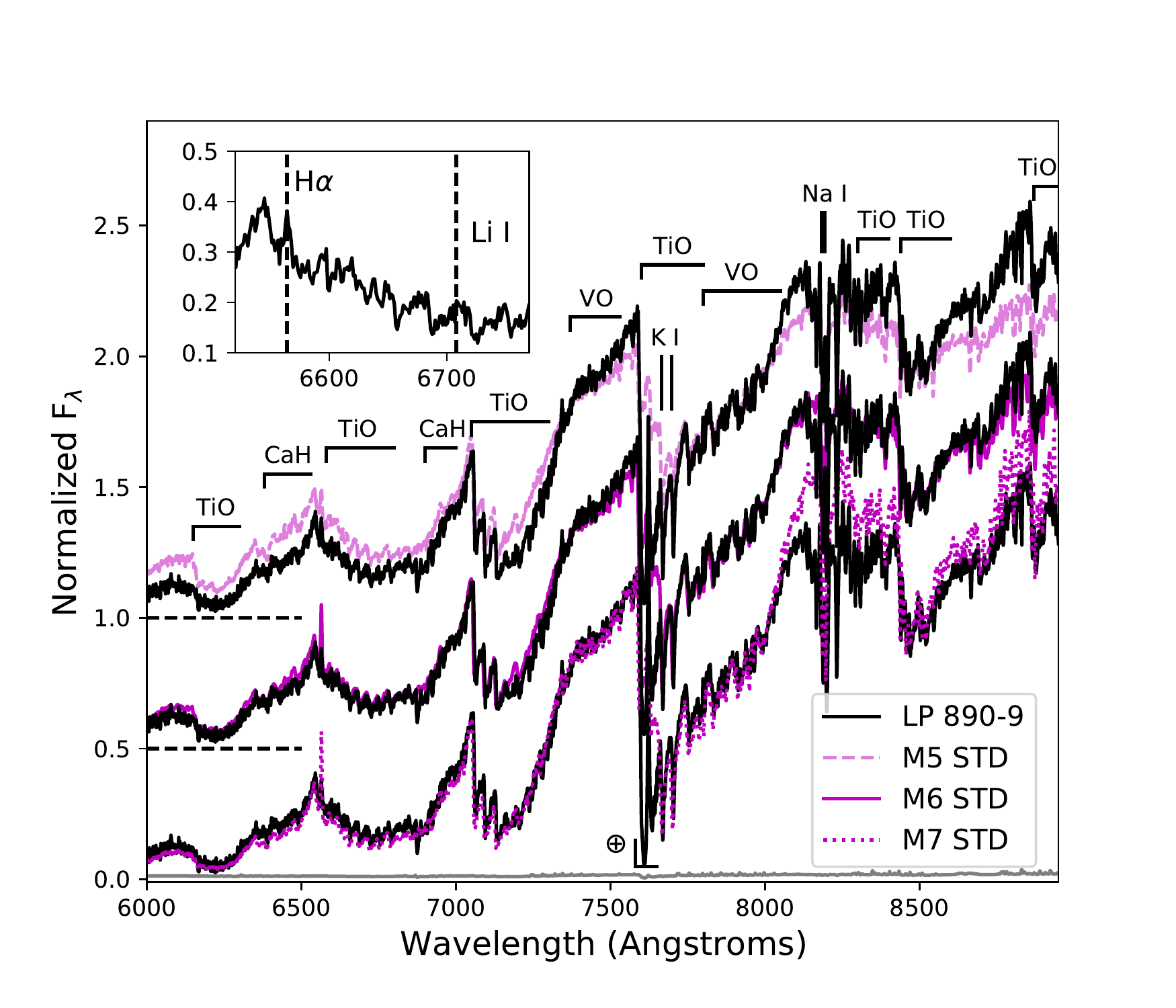}
    \caption{
        Lick/Kast red optical spectrum of LP~890-9 (solid black lines) compared to M5 (dashed magenta line), M6 (solid magenta line), and M7 (dotted magenta line) optical spectral templates from \citet{2007AJ....133..531B}. The three spectral templates are offset by a constant to facilitate comparison. Key atomic and molecular spectral features are labelled, and the inset box shows a close-up of the region around the 6563~{\AA} H$\alpha$ emission line and 6708~{\AA} Li~I absorption line (not detected). 
    }
    \label{fig:kast}
\end{figure}

Fig.~\ref{fig:kast} displays the reduced Kast spectrum compared to late-M dwarf SDSS optical spectral templates from \citet{2007AJ....133..531B}. The best overall match is M6.
We note that this is somewhat later than the index-based classification reported in \citet{2002AJ....123.2828C}, and index-based classifications derived from relations defined in \citet{2003AJ....125.1598L} and \citet{2007MNRAS.381.1067R}, all of which yield a classification of M5. Our template comparison rules this classification out. The spectrum shows a weak H$\alpha$ emission line with an equivalent width of $-1.5\pm0.3$~{\AA} corresponding to $\log_{10}(L_{\rm{H\alpha}}/L_{\rm{bol}})$ = $-4.58\pm0.10$ based on the $\chi$ factor relation of \citet{2014ApJ...795..161D}, making LP~890-9 a relatively low-activity M dwarf (cf. \citealt{2017ApJ...834...85N}).
Metallicity indices from \citet{2007ApJ...669.1235L,2013AJ....145..102L,2012AJ....143...67D}; and \citet{2019ApJS..240...31Z} all indicate a dwarf metallicity classification, with a modest subsolar metallicity ([Fe/H] $\approx$ $-$0.17 to $-$0.12 based on \citealt{2013AJ....145...52M}).

\begin{figure}
    \centering
    \includegraphics[width=\linewidth, height=\textheight, keepaspectratio]{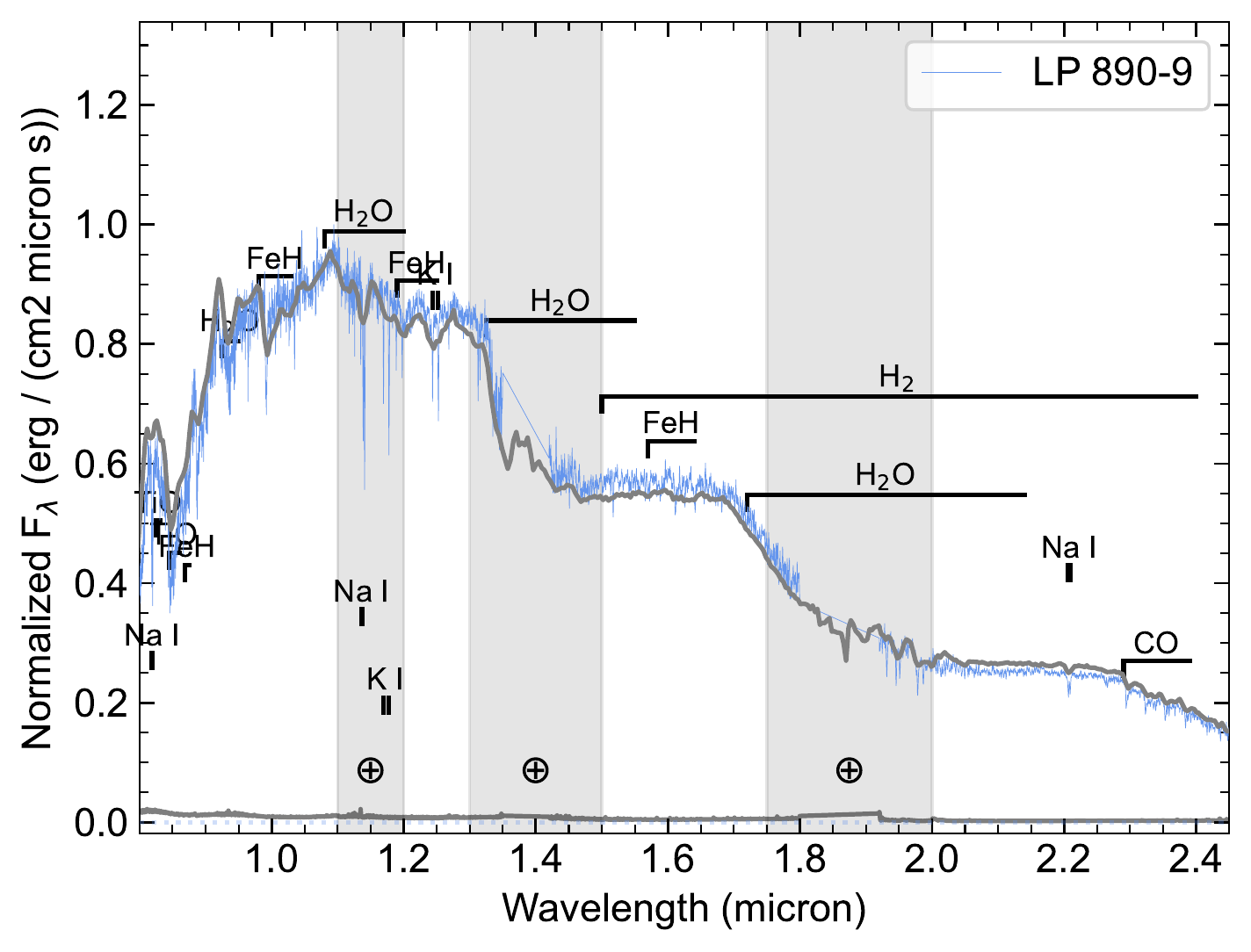}
    \caption{
        SpeX spectrum of LP~890-9 (blue) obtained in August 2021.
        The SpeX Prism spectrum of the M6.0 standard LHS\,1375 \citep{Kirkpatrick2010} is shown in grey for comparison.
        Prominent spectral features are highlighted, and regions of strong telluric absorption are shaded.
        The bottom line gives the measurement uncertainty.
    }
    \label{fig:spex}
\end{figure}

We performed a parallel set of analyses on the infrared SpeX spectra of LP~890-9 using the SpeX Prism Library Analysis Toolkit \citep[SPLAT, ][]{splat}. Results from both observations were identical. We obtained a spectral classification of M6.0 $\pm$ 0.5 by comparison to spectral standards \citep{Kirkpatrick2010}, as shown in Fig.~\ref{fig:spex}. With the SpeX spectrum, we can also obtain a metallicity estimate. We used SPLAT to measure the equivalent widths of the K-band Na\,\textsc{i} and Ca\,\textsc{i} doublets and the H2O--K2 index \citep{Rojas-Ayala2012}.
We then estimated the stellar metallicity using the \citet{Mann2014} relation between these observables and [Fe/H]. We used a Monte Carlo approach to calculate the uncertainty in our estimate, drawing $10^6$ samples from normal distributions given by the means and standard deviations of the measurements. We calculated the mean and standard deviation of the resulting values and, adding in quadrature the systematic uncertainty of the relation (0.07), we arrived at our final metallicity estimate of $\textrm{[Fe/H]} = -0.028 \pm 0.089$ based on the August data, consistent at $1\sigma$ with our estimate from the optical spectrum. The November data gave identical results within the uncertainties.

\subsection{SED fitting and evolutionary modelling}
\label{sec:rsms}

\begin{figure}
    \centering
    \includegraphics[width=0.47\textwidth]{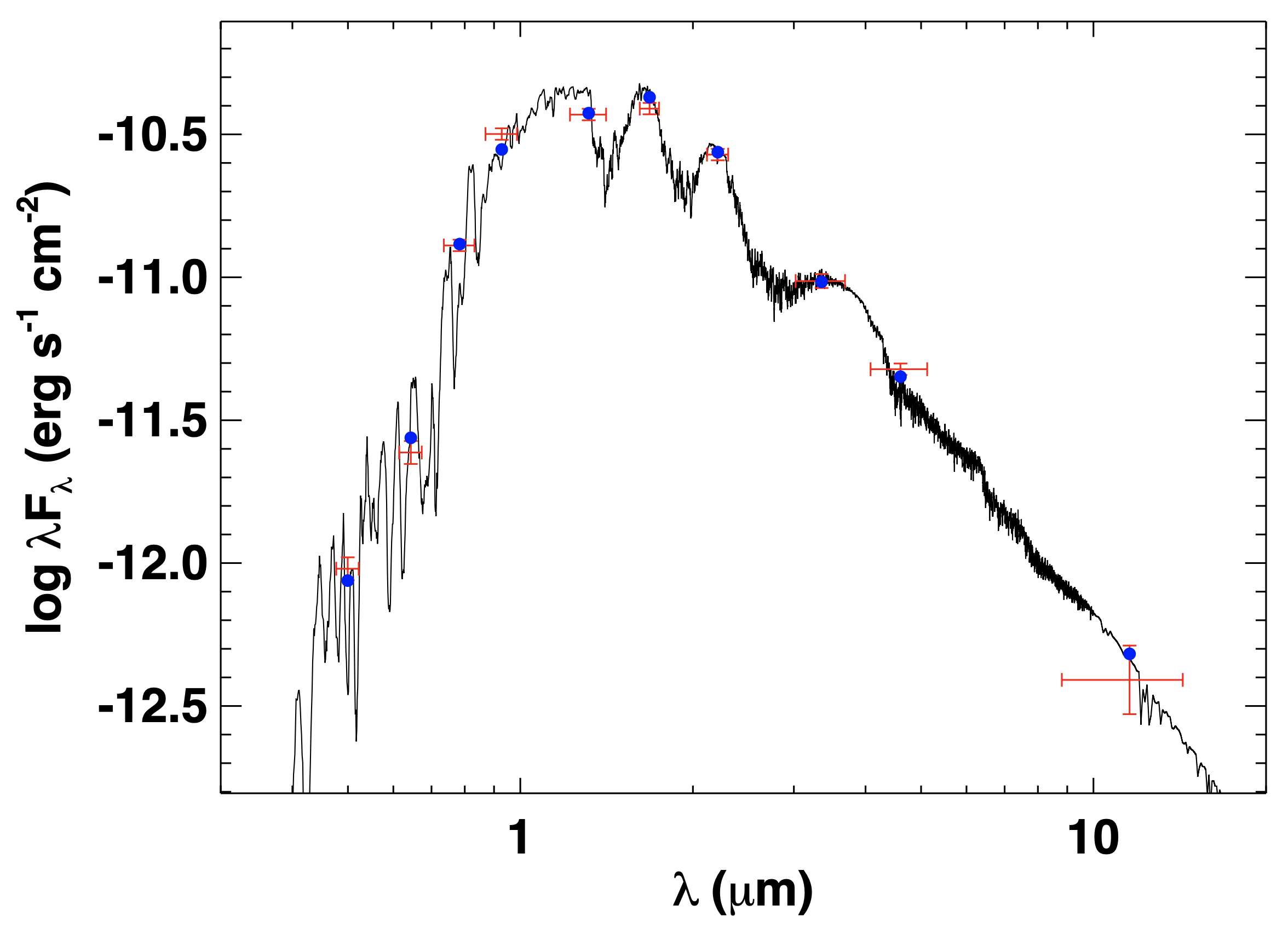}
    \caption{Spectral energy distribution (SED) of LP~890-9. The red symbols represent the observed photometric measurements (the horizontal bars represent the effective width of the bandpass). The blue symbols are the model fluxes from the best-fit NextGen atmosphere model (black curve).}
    \label{fig:sed}
\end{figure}

We performed an analysis of the broadband spectral energy distribution (SED) of the star together with the {\it Gaia\/} EDR3 parallax \citep[with no systematic offset applied; see, e.g.,][]{StassunTorres:2021}, in order to determine an empirical measurement of the stellar radius, following the procedures described in \citet{Stassun:2016} and \citet{Stassun:2017,Stassun:2018}. We pulled the $JHK_S$ magnitudes from {\it 2MASS} \citep{2006AJ....131.1163S}, the $W1W2W3$ magnitudes from {\it WISE} \citep{2014yCat.2328....0C}, and the $griz$ magnitudes from {\it Pan-STARRS} \citep{2016arXiv161205560C}. Together, the available photometry spans the full stellar SED over the wavelength range 0.4--10~$\mu$m (see Fig.~\ref{fig:sed}).  

We performed a fit using NextGen stellar atmosphere models, with the free parameters being the effective temperature ($T_{\rm eff}$) and metallicity ([Fe/H]). The remaining free parameter is the extinction $A_V$, which we fixed at zero due to the star's proximity. The resulting fit (Fig.~\ref{fig:sed}) has a reduced $\chi^2$ of 1.7, with best fit $T_{\rm eff} = 2850 \pm 75$~K and [Fe/H] = $0.0 \pm 0.5$. Integrating the model SED gives the bolometric flux at Earth, $F_{\rm bol} = 4.41 \pm 0.15 \times 10^{-11}$ erg~s$^{-1}$~cm$^{-2}$. Taking the $F_{\rm bol}$ and $T_{\rm eff}$, together with the {\it Gaia\/} EDR3 parallax, gives the stellar radius, $R_\star = 0.1556 \pm 0.0086\:R_\odot$. 

For the mass, we applied stellar evolutionary modelling, using the models for very low-mass stars presented in \citet{2019ApJ...879...94F}. We used as constraints the luminosity derived from $F_{\rm bol}$ and the {\it Gaia\/} EDR3 parallax ($L_\star = 1.4406 \pm 0.0375 \times 10^{-3} L_{\odot}$), the metallicity inferred in Sect.~\ref{sec:spectro_analysis}, and assuming an age $\gtrsim$ 2 Gyr (see Sect.~\ref{sec:age})\footnote{The luminosity of very low-mass stars evolves very slowly with time once the star has turned on core H-burning and has reached the main sequence. Hence, assuming any age $\gtrsim 2$ Gyr will provide the same resulting mass.}, we obtained a stellar mass of $0.118 \pm 0.002\:M_{\odot}$. This uncertainty reflects the error propagation on the stellar luminosity and metallicity, but also the uncertainty associated with the input physics of the stellar models \citep{2018ApJ...853...30V}. Considering the stellar radius estimate from SED fitting, this gives a mean stellar density of $\rho_\star = 44.2_{-6.7}^{+8.3}$~g~cm$^{-3}$ ($\rho_\star = 31.3_{-4.7}^{+5.9}$~$\rho_{\odot}$).

\subsection{Estimated stellar age}
\label{sec:age}

\begin{figure}
    \centering
    \vspace{-0.5cm}
    \includegraphics[width=0.47\textwidth]{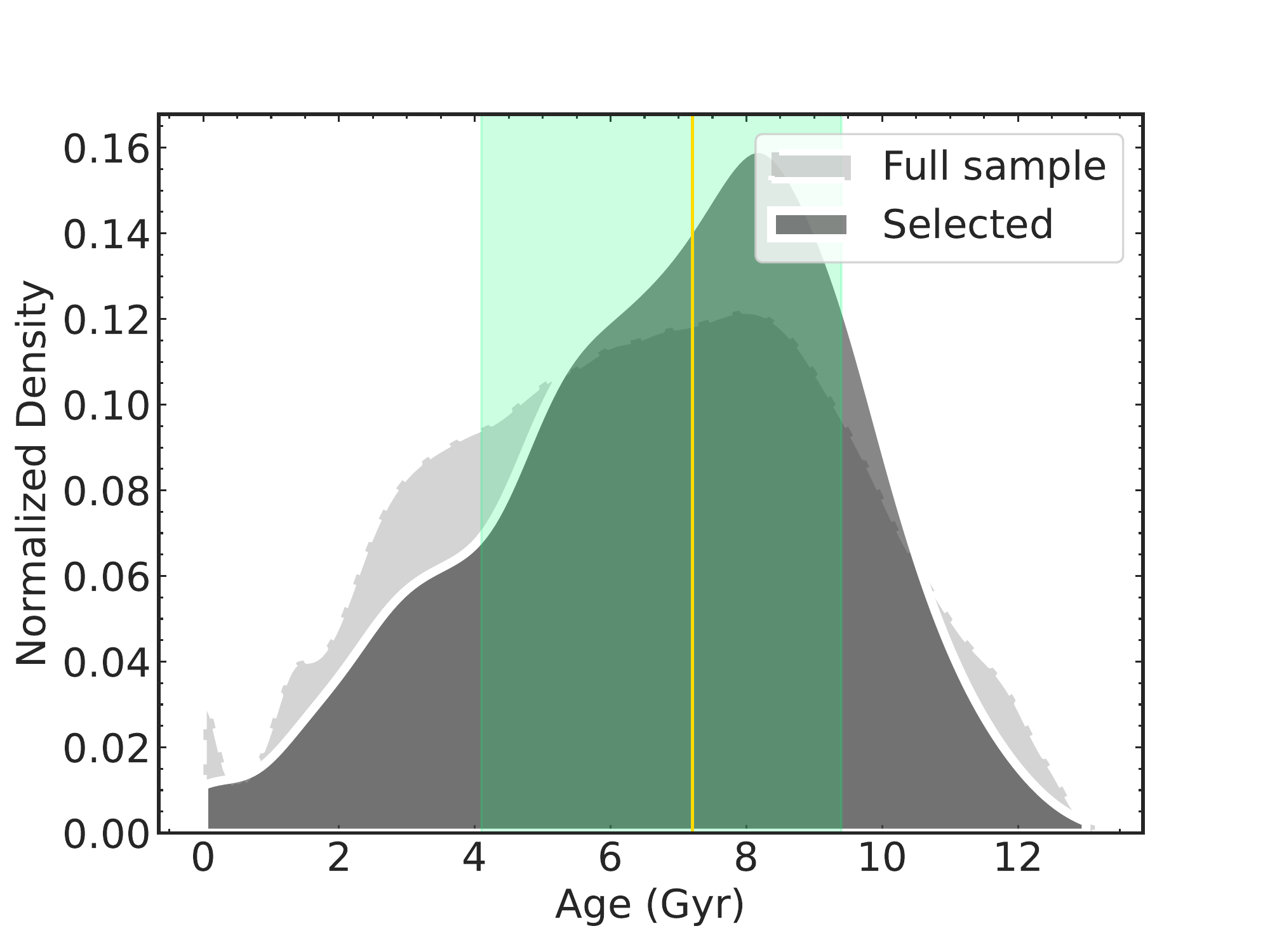}
    \caption{Distribution of ages in the full GALAH DR3 sample (light grey histogram) and sources that match the $UVW$ and metallicity of \hbox{LP~890-9} to within 10~km/s and 1$\sigma$, respectively (dark grey histogram). The latter has a broad peak at 7.2$^{+2.2}_{-3.1}$~Gyr (yellow line with uncertainties indicated by the green shaded region), which we take as an estimate for the age of LP~890-9.}
    \label{fig:age}
\end{figure}

The age of LP~890-9 was estimated by comparing its $UVW$ kinematics and metallicity to local stars, an adaptation of the method used to age-date TRAPPIST-1 \citep{2017ApJ...845..110B}. The comparison sample was drawn from the GALAH Data Release 3 catalogue \citep{2021MNRAS.506..150B}, for which ages have been estimated using the Bayesian Stellar Parameter Estimation (BSTEP) code by \citet{2018MNRAS.473.2004S}. The match set assumes individual $UVW$ velocities within 10~km/s of LP~890-9 and within 1$\sigma$ of the metallicity. Figure~\ref{fig:age} compares the distribution of ages in the full GALAH sample and the match set, which shows a broad peak at 7.2$^{+2.2}_{-3.1}$~Gyr (i.e., 4--9~Gyr).
This age is broadly consistent with the kinematics of the source, which are consistent with the thin disk Galactic population (8\% probability thick disk) based on \citet{2003AA...410..527B}.

\begin{figure*}[hbt!]
    \centering
    \includegraphics[width=0.90\textwidth]{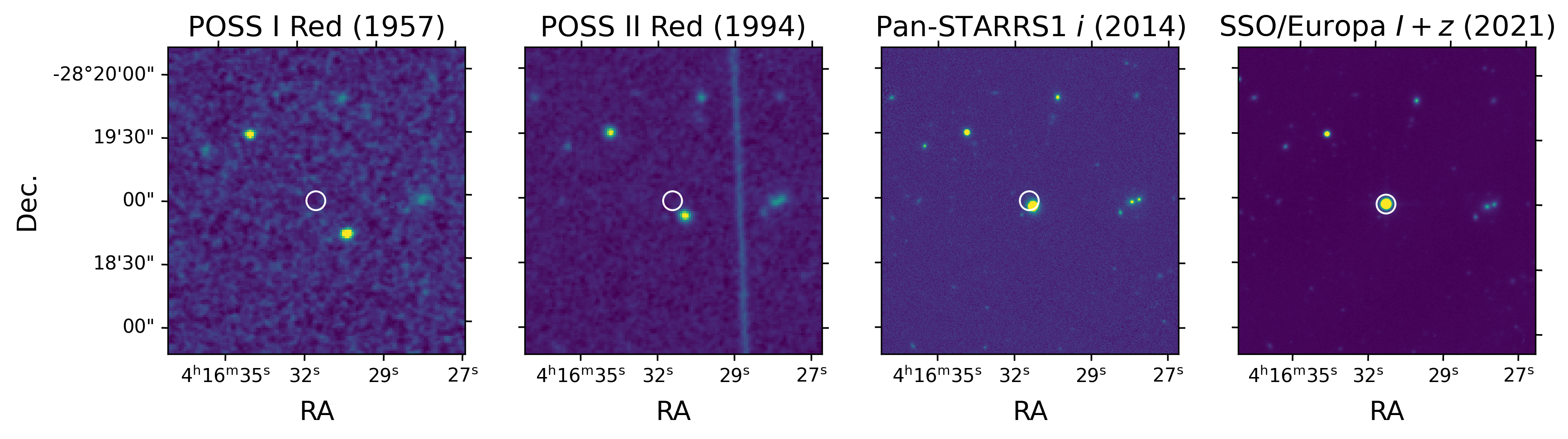}
    \caption{Imaging of the present-day position of LP~890-9 over 64 years, to assess for a current blend with an unresolved background object. Archival images from POSS I (1957), POSS II (1994), and Pan-STARRS (2014) are presented, along with a stack image from one night of observation with SSO/Europa (2021). For each image, the position of LP~890-9 during the SSO/Europa observations is shown as a white circle.}
    \label{fig:archival_im}
\end{figure*}

\section{Planet validation} \label{sec:vetting}

\subsection{TESS data validation report}

As an initial step for false-positive vetting and before obtaining the ground-based follow-up observations described in \hbox{Sect. \ref{sec:obs}}, we first closely examined the TESS Data Validation Report \citep{2018PASP..130f4502T,2019PASP..131b4506L} combining all four sectors (4, 5, 31, and 32) provided by the SPOC pipeline. \hbox{TOI-4306.01} successfully passed the eclipsing binary discrimination tests: the odd and even transit depths agreed to within 0.06$\sigma$, and no significant secondary eclipse was detected. TOI-4306.01 also passed the difference image centroid offset test, which did not reveal any significant offset between the transit source and the target (3.02$\pm$3.41 arcseconds), that would otherwise suggest a blend of multiple stars (e.g., background star or stellar system) and possible source confusion. In addition, the statistical bootstrap test estimated at only 2.6 $\times 10^{-17}$ the probability that the signal is a false alarm due to noise fluctuations in the light curve (e.g., stellar variability or residual instrumental systematics). 

We note that TOI-4306.01 failed the ghost diagnostic test, which involves correlating flux time series derived from photometric core and halo aperture pixels against the transit model. A higher correlation for the halo compared to the core may indicate that the transit signal does not originate from the target, but is caused instead by scattered light or a background object, such as a background eclipsing binary. For TOI-4306.01, the test returned a slightly higher correlation statistic\footnote{The correlation statistic is defined as $\frac{\tilde{b} \boldsymbol{\cdot} \tilde{s}}{\sqrt{\tilde{s} \boldsymbol{\cdot} \tilde{s}}}$ where $\tilde{b}$ is the whitened core or halo aperture flux time series and $\tilde{s}$ is the whitened transit model light curve for the given planet candidate \citep{2018PASP..130f4502T}.} for the halo (5.17) than the core (4.72). However, the recovery of the transit signal with ground-based telescopes (see Sect. \ref{sec:val_followup}) excludes the possibility that it was caused by scattered light or other instrumental artefacts in the TESS data. Furthermore, archival imaging (see Sect. \ref{sec:val_arch_imaging}) allows us to exclude any background object as the source of TOI-4306.01's transits.

\subsection{Follow-up photometry}
\label{sec:val_followup}

Owing to TESS's large pixel scale of 21$\arcsec$ per pixel, it is not uncommon that the photometric apertures used by SPOC to extract the light curves also include some flux from contaminating stars. In this context, ground-based follow-up photometry at a higher spatial resolution is important to explore the possible contamination from nearby sources and confirm that the target star is the source of the transits. Fig. \ref{fig:tpf} shows that LP~890-9 is relatively isolated. SPOC reports only a small amount of contamination from outside sources, ranging from 2.2 to 5.3\% of the total flux depending on the sector and aperture used (the PDCSAP photometry is corrected for this contamination). The closest known \textit{Gaia} DR2 source \citep{2018A&A...616A...1G}, labelled `2' in Fig. \ref{fig:tpf}, is 43.24$\arcsec$ away and about 3 magnitudes fainter (TESS-mag=17.02, \citealt{2018AJ....156..102S}). \hbox{LP~890-9} is clearly resolved in the SPECULOOS-South images, which have a pixel scale of 0.35$\arcsec$ per pixel, thus allowing the extraction of light curves with photometric apertures of only a few arcseconds (see Sect. \ref{sec:sso}). The smallest aperture that was tested was 1.4$\arcsec$ (4 pixels). No transit event was observed on any nearby star, while transits with a depth matching that of \hbox{TOI-4306.01} were clearly detected on the target star at the predicted times. 

Our follow-up transit photometry was obtained in seven different bandpasses $-$ $g'$, $r'$, $i'$, blue-blocking, $I+z$, $z'$, and 0.85-1.55 $\mu$m $-$ covering a wavelength range from $\sim$0.4 to 1.6 $\mu$m. Together with the TESS transit photometry, this allowed us to check for a wavelength dependence of the transit depth, that could indicate a blend with a stellar eclipsing binary either in the background/foreground or bound to the target star. We found that the transit depths in the individual bandpasses are in very good agreement and do not show any chromatic dependence (besides the effect of stellar limb-darkening) within the accuracy of the measurements (see Sect. \ref{sec:ddFs} and Table \ref{tab:ddFs}).

\subsection{Archival imaging}
\label{sec:val_arch_imaging}

The high proper motion of LP~890-9 (333 mas $\mathrm{yr}^{-1}$, \citealt{gaia_edr3}) makes it possible to investigate archival images for a possible background object that could be blended with the target at its current position \citep[see, e.g.,][]{2006ApJ...641L..57B}. We inspected POSS I/DSS \citep{1963bad..book..481M}, POSS II/DSS2 \citep{1991PASP..103..661R}, and Pan-STARRS1 \citep{2016arXiv161205560C} images spanning 64 years with our recent observations. None of these archival images show any source at the present-day target location (see \hbox{Fig. \ref{fig:archival_im}}). Given the detection limits of these images, we can thus rule out background sources brighter than $\sim$5 magnitudes below that of LP~890-9.


\begin{table*}[!htbp] 
   \centering
   \caption{Properties of the LP~890-9 planetary system based on our global transit analysis (see Sect. \ref{sec:glob_an}).}
	\begin{tabular}{lccc}
    	\toprule
    	\toprule
    	\textbf{Parameters} & \multicolumn{2}{c}{\textbf{Values}} & \textbf{Priors} \\
    	\midrule
    	\midrule
    	\vspace{0.1cm}
    	\bf{Star} & \multicolumn{2}{c}{\textbf{LP~890-9}} & \\
    	\vspace{0.1cm}
    	Luminosity, $L_{\star}$ ($10^{-3}\:L_\odot$) & \multicolumn{2}{c}{$1.438_{-0.037}^{+0.037}$} & $\mathcal{N}$(1.441, 0.038$^2$) \\
    	\vspace{0.1cm}
    	Effective temperature, $T_{\mathrm{eff}}$ (K) & \multicolumn{2}{c}{$2871_{-45}^{+32}$} & $-$ \\
    	\vspace{0.1cm}
    	Mass, $M_{\star}$ ($M_{\odot}$) & \multicolumn{2}{c}{$0.118 \pm 0.002$} & $\mathcal{N}$(0.118, 0.002$^2$) \\
    	\vspace{0.1cm}
    	Radius, $R_{\star}$ ($R_{\odot}$) & \multicolumn{2}{c}{$0.1532_{-0.0024}^{+0.0048}$} & $-$ \\
    	\vspace{0.1cm}
    	Density, $\rho_{\star}$ ($\rho_{\odot}$) & \multicolumn{2}{c}{$32.9_{-3.0}^{+1.5}$} & $-$ \\
    	\vspace{0.1cm}
    	Log surface gravity, log $g_{\star}$ (cgs) & \multicolumn{2}{c}{$5.139_{-0.028}^{+0.013}$} & $-$ \\
    	\vspace{0.1cm}
    	Quadratic limb-darkening coefficient $u_{1,\:g'}$ & \multicolumn{2}{c}{$0.813 \pm 0.049$} & $\mathcal{N}$(0.830, 0.053$^2$) \\
    	\vspace{0.1cm}
    	Quadratic limb-darkening coefficient $u_{2,\:g'}$ & \multicolumn{2}{c}{$0.091_{-0.074}^{+0.062}$} & $\mathcal{N}$(0.140, 0.091$^2$) \\
    	\vspace{0.1cm}
    	Quadratic limb-darkening coefficient $u_{1,\:r'}$ & \multicolumn{2}{c}{$0.841 \pm 0.047$} & $\mathcal{N}$(0.847, 0.051$^2$) \\
    	\vspace{0.1cm}
    	Quadratic limb-darkening coefficient $u_{2,\:r'}$ & \multicolumn{2}{c}{$0.062_{-0.070}^{+0.064}$} & $\mathcal{N}$(0.086, 0.091$^2$) \\
    	\vspace{0.1cm}
    	Quadratic limb-darkening coefficient $u_{1,\:i'}$ & \multicolumn{2}{c}{$0.531 \pm 0.037$} & $\mathcal{N}$(0.532, 0.039$^2$) \\
    	\vspace{0.1cm}
    	Quadratic limb-darkening coefficient $u_{2,\:i'}$ & \multicolumn{2}{c}{$0.298_{-0.090}^{+0.084}$} & $\mathcal{N}$(0.311, 0.107$^2$) \\
    	\vspace{0.1cm}
    	Quadratic limb-darkening coefficient $u_{1,\:z'}$ & \multicolumn{2}{c}{$0.404_{-0.034}^{+0.033}$} & $\mathcal{N}$(0.417, 0.035$^2$) \\
    	\vspace{0.1cm}
    	Quadratic limb-darkening coefficient $u_{2,\:z'}$ & \multicolumn{2}{c}{$0.292_{-0.089}^{+0.086}$} & $\mathcal{N}$(0.331, 0.104$^2$) \\
    	\vspace{0.1cm}
    	Quadratic limb-darkening coefficient $u_{1,\:I+z}$ & \multicolumn{2}{c}{$0.375 \pm 0.027$} & $\mathcal{N}$(0.384, 0.028$^2$) \\
    	\vspace{0.1cm}
    	Quadratic limb-darkening coefficient $u_{2,\:I+z}$ & \multicolumn{2}{c}{$0.243_{-0.076}^{+0.078}$} & $\mathcal{N}$(0.303, 0.089$^2$) \\
    	\vspace{0.1cm}
    	Quadratic limb-darkening coefficient $u_{1,\:\mathrm{blue-blocking}}$ & \multicolumn{2}{c}{$0.416 \pm 0.023$} & $\mathcal{N}$(0.417, 0.023$^2$) \\
    	\vspace{0.1cm}
    	Quadratic limb-darkening coefficient $u_{2,\:\mathrm{blue-blocking}}$ & \multicolumn{2}{c}{$0.277_{-0.070}^{+0.069}$} & $\mathcal{N}$(0.286, 0.073$^2$) \\
    	\vspace{0.1cm}
    	Quadratic limb-darkening coefficient $u_{1,\:\mathrm{0.85-1.55\, \mu m}}$ & \multicolumn{2}{c}{$0.238 \pm 0.015$} & $\mathcal{N}$(0.238, 0.015$^2$) \\
    	\vspace{0.1cm}
    	Quadratic limb-darkening coefficient $u_{2,\:\mathrm{0.85-1.55\, \mu m}}$ & \multicolumn{2}{c}{$0.203_{-0.061}^{+0.059}$} & $\mathcal{N}$(0.212, 0.060$^2$) \\
        \vspace{0.1cm}
    	Quadratic limb-darkening coefficient $u_{1,\:\mathrm{TESS}}$ & \multicolumn{2}{c}{$0.410 \pm 0.026$} & $\mathcal{N}$(0.409, 0.026$^2$) \\
    	\vspace{0.1cm}
    	Quadratic limb-darkening coefficient $u_{2,\:\mathrm{TESS}}$ & \multicolumn{2}{c}{$0.300_{-0.085}^{+0.081}$} & $\mathcal{N}$(0.292, 0.084$^2$) \\
    	\hline
        \vspace{0.1cm}
        \bf{Planets} & \bf{b} & \bf{c} \\
        \vspace{0.1cm}
        Transit depth, $\mathrm{d}F$ (ppm) & $6238_{-153}^{+168}$ & $6659_{-273}^{+282}$ & $-$ \\
        \vspace{0.1cm}
        Transit impact parameter, $b$ ($R_{\star}$) & $0.15_{-0.10}^{+0.14}$ & $0.697_{-0.019}^{+0.024}$ & $-$ \\
        \vspace{0.1cm}
        Orbital period, $P$ (days) & $2.7299025_{-0.0000040}^{+0.0000034}$ & $8.457463 \pm 0.000024$ & $-$ \\
        \vspace{0.1cm}
        Mid-transit time, $T_{0}$ ($\mathrm{BJD_{TDB}} - 2\:450\:000$) & 9447.82637 $\pm$ 0.00014 & $9520.70598 \pm 0.00031$ & $-$ \\
        \vspace{0.1cm}
        Transit duration, $W$ (min) & $50.74_{-0.47}^{+0.51}$ & $57.56_{-0.91}^{+0.99}$ & $-$ \\
        \vspace{0.1cm}
        Orbital inclination, $i_{\mathrm{p}}$ (deg) & $89.67_{-0.33}^{+0.22}$ & $89.287_{-0.047}^{+0.026}$ & $-$ \\
        \vspace{0.1cm}
        Orbital semi-major axis, $a_{\mathrm{p}}$ (au) &  $0.01875 \pm 0.00010$ & $0.03984 \pm 0.00022$ & $-$ \\
        \vspace{0.1cm}
        Scale parameter, $a_{\mathrm{p}}/R_{\star}$ & $26.32_{-0.82}^{+0.39}$ & $55.94_{-1.74}^{+0.82}$ & $-$ \\
        \vspace{0.1cm}
        Radius, $R_{\mathrm{p}}$ ($R_{\oplus}$) & $1.320_{-0.027}^{+0.053}$ & $1.367_{-0.039}^{+0.055}$ & $-$ \\
        \vspace{0.1cm}
        Stellar irradiation, $S_{\mathrm{p}}$ ($S_{\oplus}$) & $4.09 \pm 0.12$ & $0.906 \pm 0.026$ & $-$ \\
        \vspace{0.1cm}
        Equilibrium temperature,$^{a}$ $T_{\mathrm{eq}}$ (K) & $396 \pm 3$ & $272 \pm 2$ & $-$ \\
        \vspace{0.1cm}
        RV semi-amplitude,$^{b}$ $K$ (m/s) & < 25.1 & < 33.0 & $-$ \\
        \vspace{0.1cm}
        Mass,$^{b}$ $M_{\mathrm{p}}$ ($M_{\oplus}$) & < 13.2 & < 25.3 & $-$ \\
        \bottomrule
        \bottomrule
        \end{tabular}
\justifying
\small
\textbf{Notes.} For each parameter, we indicate the median of the posterior distribution function, along with the 1-$\sigma$ credible intervals. For the priors given in the last column, $\mathcal{N}(\mu, \sigma^2)$ represents a normal distribution of mean $\mu$ and variance $\sigma^2$. $^{a}$ The equilibrium temperature corresponds to a case with null Bond albedo and an efficient heat recirculation from the dayside to the nightside hemispheres of the planet. $^{b}$ 2$\sigma$ upper limits from our 2-planet RV fit (see Sect. \ref{sec:rv_analysis}).
\label{tab:param}
\end{table*}

\subsection{Statistical validation}

To fully vet the two planet candidates, we used the open-source software package {\fontfamily{pcr}\selectfont TRICERATOPS}\footnote{\url{https://github.com/stevengiacalone/triceratops}} \citep{2020ascl.soft02004G,2021AJ....161...24G}, which uses a Bayesian framework that incorporates prior knowledge of the target star, planet occurrence rates, and stellar multiplicity to calculate the probability that the transit signal is due to a planet transit or another astrophysical source. To consider a planet as statistically validated, we require its False Positive Probability (FPP), which is the sum of probabilities for all false positive scenarios, to be less than 0.01 (1\%). This condition was defined by \cite{2021AJ....161...24G} based on results obtained with {\fontfamily{pcr}\selectfont TRICERATOPS} for 68 TOIs (TESS Objects of Interest) that were previously identified as either confirmed planets or astrophysical false positives by TFOP using follow-up observations.

Using the phase-folded SSO $I+z$ transit light curve (which provides the tightest photometric constraints) and the Zorro speckle imaging 5$\sigma$ contrast curve at 832 nm as inputs to {\fontfamily{pcr}\selectfont TRICERATOPS}, we found FPPs of 6.5 $\times$ $10^{-6}$ and 3.2 $\times$ $10^{-7}$ for the inner and outer candidates, respectively. These values were obtained for each planet individually, not accounting for the fact that they are both transiting in the same system. Applying a `multiplicity boost' to account for the a priori higher likelihood of a candidate in a multi-transiting system to be a real planet would allow to further reduce these FPPs, by a factor of $\sim$20--50 \citep{2012ApJ...750..112L,2021ApJS..254...39G}. In any case, these FPPs are well below 1\%, so we consider that the two transit signals correspond to validated planets. Henceforth, we denote the inner one (TOI-4306.01) as LP~890-9\,b and the second outer transiting planet (SPECULOOS-2\,c) as LP~890-9\,c.

\begin{figure*}[hbt!]
    \centering
    \includegraphics[width=0.85\textwidth]{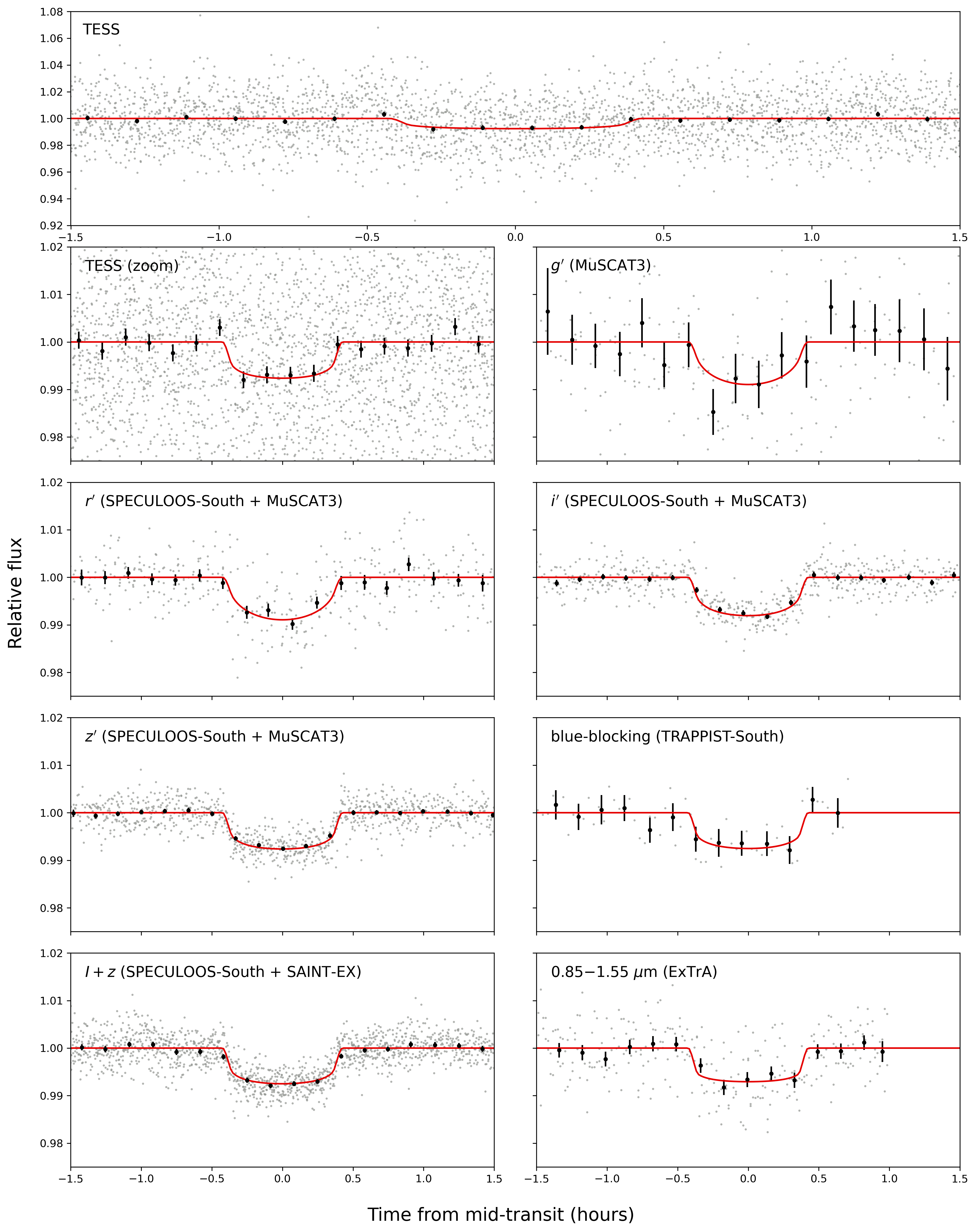}
    \caption{Phase-folded detrended transit photometry of LP~890-9\,b in each observed bandpass. The unbinned data points are shown in grey, while the black circles with error bars correspond to 10-min bins. The best-fit transit model is shown in red, with a different limb darkening for each bandpass.}
    \label{fig:LCs_b}
\end{figure*}

\begin{figure*}[hbt!]
    \centering
    \includegraphics[width=0.90\textwidth]{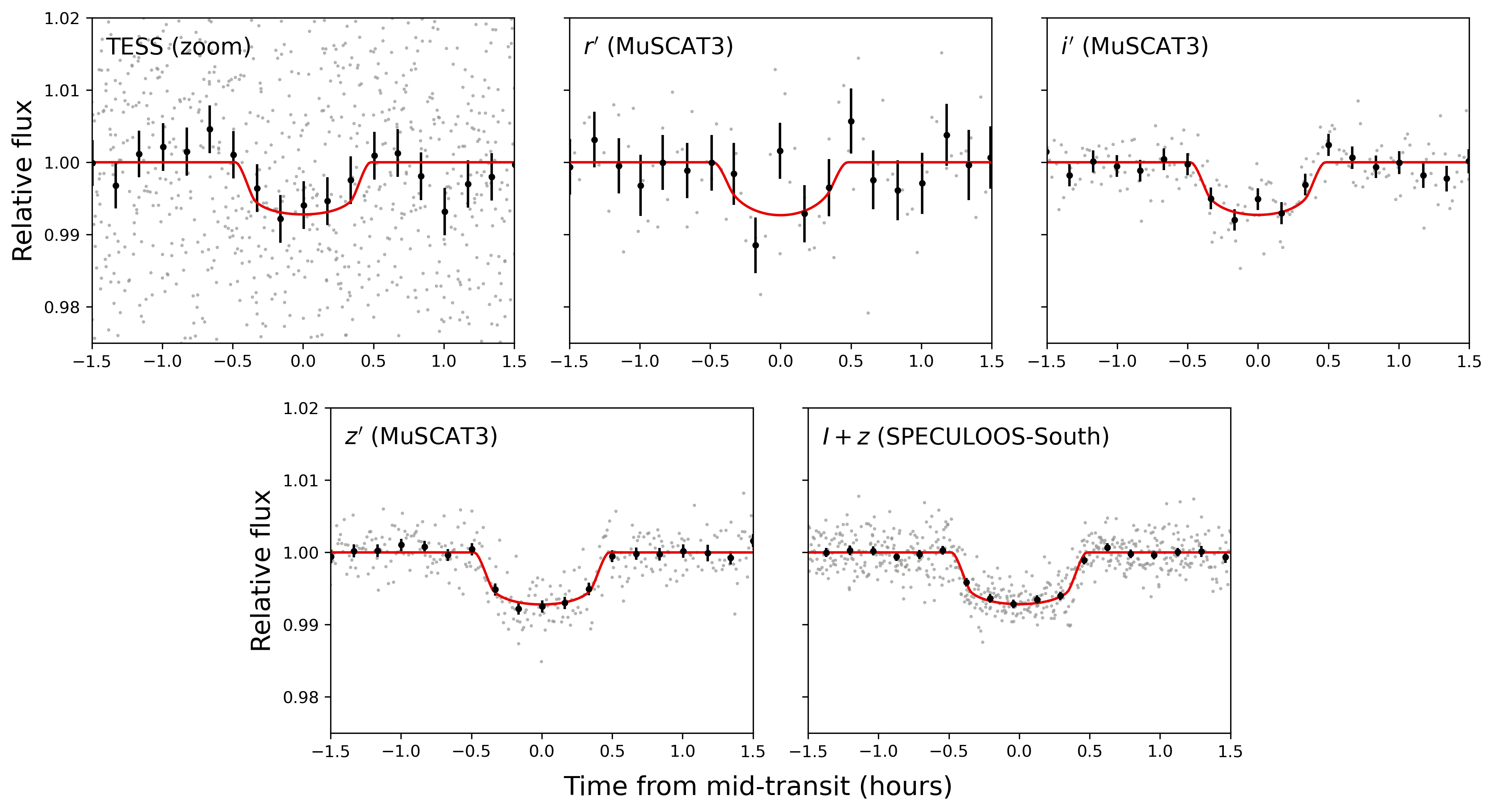}
    \caption{Phase-folded detrended transit photometry of LP~890-9\,c in each observed bandpass. The unbinned data points are shown in grey, while the black circles with error bars correspond to 10-min bins. The best-fit transit model is shown in red, with a different limb darkening for each bandpass. We note that the MuSCAT3 $r'$- and $i'$- light curves show a possible hump around mid-transit which could be related to spot crossing.}
    \label{fig:LCs_c}
\end{figure*}

\section{Photometric analysis} \label{sec:phot}

\subsection{Transit analysis} \label{sec:analysis}

\subsubsection{Derivation of the system parameters}
\label{sec:glob_an}

We performed a joint fit of all the transit light curves described in Sect. \ref{sec:obs} using the most recent version of the adaptive MCMC code presented in \citeauthor{2012A&A...542A...4G} (\citeyear{2012A&A...542A...4G}, see also \citealt{2014A&A...563A..21G}). This implementation makes uses of the Metropolis-Hastings algorithm \citep{1953JChPh..21.1087M,1970Bimka..57...97H} and the Gibbs sampler \citep{4767596}. The data were modelled using the quadratic limb-darkening transit model of \cite{2002ApJ...580L.171M} multiplied by a photometric baseline model, different for each light curve, aimed at representing the photometric variations caused by other astrophysical, instrumental, or environmental effects. Fitting the transit signals simultaneously with the possibly underlying correlated noise (instead of pre-detrending the data) ensures a better propagation of the uncertainties to the derived system parameters of interest. For each light curve, we explored a large range of baseline models consisting of low-order polynomials with respect to, e.g., time, airmass, full width at half maximum of the point spread function, target's location on the detector, background, or any combination of these parameters. The minimal baseline model is a simple constant to account for any out-of-transit flux offset. We show in Table \ref{tab:baselines} the baseline model selected for each light curve based on the Bayesian Information Criterion (BIC, \citealt{1978AnSta...6..461S}). This table also gives for each light curve the two scaling factors, $\beta_{w}$ and $\beta_{r}$, that were applied to the photometric error bars to account respectively for over- or under-estimated white noise and the presence of residual correlated (red) noise (see \citealt{2012A&A...542A...4G} for details).

The transit model parameters sampled by the MCMC for each of the two planets were: the log of the orbital period \hbox{(log $P$)}, the mid-transit time ($T_0$), the transit depth ($\mathrm{d}F=R_{\mathrm{p}}^2/R_\star^2$ where $R_{\rm{p}}$ is the radius of the planet and $R_\star$ the stellar radius), the cosine of the orbital inclination (\hbox{cos $i_{\rm{p}}$}), and the two parameters $\sqrt{e_{\rm{p}}}$ cos $\omega_{\rm{p}}$ and $\sqrt{e_{\rm{p}}}$ sin $\omega_{\rm{p}}$ (with $e_{\rm{p}}$ the eccentricity and $\omega_{\rm{p}}$ the argument of periastron). We also fitted the log of the stellar density (log $\rho_\star$), the log of the stellar mass (log $M_\star$), and the effective temperature ($T_{\rm{eff}}$). Finally, we fitted for each bandpass the combinations $q_1=(u_1+u_2)^2$ and $q_2=0.5\,u_1 (u_1+u_2)^{-1}$ of the quadratic limb-darkening coefficients ($u_1$ and $u_2$), following the triangular sampling scheme advocated by \cite{2013MNRAS.435.2152K}. The coefficients of the baseline models were not sampled by the MCMC, but determined from the residuals at each step of the procedure using a singular value decomposition method \citep{1992nrfa.book.....P}.

The priors used in our analysis are listed in the last column of Table \ref{tab:param}. We assumed normal prior distributions for $M_\star$ and the luminosity $L_\star$ based on the values derived in Sect. \ref{sec:rsms} (Table \ref{table:star}). Normal priors for the quadratic limb-darkening coefficients ($u_1$ and $u_2$) were computed for each bandpass using the \texttt{PyLDTk} code \citep{Parviainen2015} and the library of PHOENIX high-resolution synthetic spectra of \cite{Husser2013}. We increased the widths of these computed priors by a conservative factor of ten to account for model-dependent uncertainties.

We performed two different analyses, assuming either eccentric or circular orbits for the two planets (setting in this case their respective $\sqrt{e_{\rm{p}}}$ cos $\omega_{\rm{p}}$ and $\sqrt{e_{\rm{p}}}$ sin $\omega_{\rm{p}}$ values to zero). For each analysis, we ran two chains of 250 000 steps (including 20\% burn-in) and checked their convergence by using the statistical test of \cite{1992StaSc...7..457G}, ensuring that the test values for all sampled parameters were <1.01. The physical parameters of the system were deduced from the above transit model parameters at each step of the MCMC. For each planet, the semi-major axis in stellar radii ($a_{\rm{p}}/R_\star$) was computed from $\rho_\star$ and $P$ using Kepler's third law \citep{2003ApJ...585.1038S}. $R_\star$ was deduced from $\rho_\star$ and $M_\star$. $L_\star$ was computed from $R_\star$ and $T_{\rm{eff}}$. For each planet, $R_{\rm{p}}$, $i_{\rm{p}}$, $a_{\rm{p}}$, $e_{\rm{p}}$, $\omega_{\rm{p}}$, the stellar irradiation ($S_{\rm{p}}$), and equilibrium temperature ($T_{\rm{eq}}$) were then easily obtained from the above parameters. We did not find evidence for eccentric orbits for any of the two planets based on the transit photometry, so we adopted the results of the circular analysis as our nominal solution (this is also consistent with simulations of the system's tidal evolution, see Sect. \ref{sec:tides}).

The posterior distribution functions for the main fitted transit parameters are displayed in Fig. \ref{fig:posteriors}. Table \ref{tab:param} gives the medians and 1-$\sigma$ credible intervals of the posterior distribution functions obtained for the system parameters. The corresponding best-fit transit models are shown in Fig. \ref{fig:LCs_b} (planet b) and \ref{fig:LCs_c} (planet c), together with the detrended phase-folded photometry. Most notably, we find that the two planets have similar radii of $1.320_{-0.027}^{+0.053}$ and $1.367_{-0.039}^{+0.055}$ $R_\oplus$, respectively, and that they orbit rather close to the second-order 3:1 mean motion resonance, but not exactly within ($P_\mathrm{c}/P_\mathrm{b}=3.098$). We also note that the stellar density $\rho_\star = 32.9_{-3.0}^{+1.5}\:\rho_\odot$ derived from our global transit fit is in very good agreement with the value of $31.3_{-4.7}^{+5.9}\:\rho_\odot$ expected based on the stellar radius from our SED fit and the stellar mass from our evolutionary modelling (see Table \ref{table:star}).

\subsubsection{Check for transit chromaticity}
\label{sec:ddFs}

To assess the transit chromaticity, we repeated the same global analysis as in Sect. \ref{sec:glob_an}, but allowing this time some transit depth variations between the different bandpasses. The transit depths found for the two planets in each observed bandpass are given in Table \ref{tab:ddFs}. For each planet, the transit depths are all in agreement (at the $\sim$1.1$\sigma$ level) and do not show any chromatic dependence.

\begin{table}[hbt!]
\centering
\caption{Transit depths returned by our global data analysis for which we allowed transit depth variations between the different bandpasses (see Sect. \ref{sec:ddFs}).}
\begin{tabular}{lcc}
\toprule
\toprule
Bandpass & Transit depth & Transit depth \\
 & planet b (ppm) & planet c (ppm) \\
\midrule
\vspace{0.1cm}
$g'$ & $6634_{-2170}^{+2190}$ & $-$ \\
\vspace{0.1cm}
$r'$ & $6100_{-604}^{+628}$ & $4657_{-2180}^{+2400}$ \\
\vspace{0.1cm}
$i'$ & $5971_{-288}^{+296}$ & $6227_{-887}^{+870}$ \\
\vspace{0.1cm}
$z'$ & $6337_{-265}^{+266}$ & $7203_{-500}^{+509}$ \\
\vspace{0.1cm}
$I+z$ & $6353_{-221}^{+227}$ & $6545_{-311}^{+323}$ \\
\vspace{0.1cm}
blue-blocking & $6267_{-1620}^{+1650}$ & $-$ \\
\vspace{0.1cm}
0.85--1.55 $\mu$m & $5771 \pm 1250$ & $-$ \\
\vspace{0.1cm}
TESS & $5937_{-857}^{+850}$ & $6221_{-1920}^{+1880}$ \\
\bottomrule
\bottomrule
\end{tabular}
\label{tab:ddFs}
\end{table}

\begin{table}[hbt!]
\centering
\caption{Transit timings returned by our global data analysis for which we allowed TTVs from the linear transit ephemerides defined by the orbital period $P$ and mid-transit time $T_0$ given in Table \ref{tab:param} (see Sect. \ref{sec:TTVs}).}
\vspace{-0.1cm}
\begin{tabular}{lcc}
\toprule
\toprule
Epoch & Transit timing & TTV \\
 & ($\mathrm{BJD_{TDB}} - 2\,450\,000$) & (min) \\
\midrule
\multicolumn{3}{c}{\textit{Planet b}}\\
\vspace{0.1cm}
-4 & $9436.90651_{-0.00034}^{+0.00033}$ & $-0.37_{-0.50}^{+0.48}$ \\
\vspace{0.1cm}
0 & $9447.82619_{-0.00036}^{+0.00035}$ & $-0.26_{-0.51}^{+0.50}$ \\
\vspace{0.1cm}
4 & $9458.74582_{-0.00054}^{+0.00049}$ & $-0.23_{-0.77}^{+0.70}$ \\
\vspace{0.1cm}
11 & $9477.85672_{-0.00075}^{+0.00065}$ & $2.05_{-1.07}^{+0.93}$ \\
\vspace{0.1cm}
14 & $9486.04532_{-0.00036}^{+0.00038}$ & $0.46_{-0.52}^{+0.54}$ \\
\vspace{0.1cm}
15 & $9488.7757_{-0.0020}^{+0.0022}$ & $1.18_{-2.91}^{+3.19}$ \\
\vspace{0.1cm}
23 & $9510.61356_{-0.00100}^{+0.00082}$ & $-0.81_{-1.44}^{+1.17}$ \\
\vspace{0.1cm}
25 & $9516.07459_{-0.00056}^{+0.00054}$ & $0.96_{-0.81}^{+0.78}$ \\
\vspace{0.1cm}
29 & $9526.99350 \pm 0.00027$ & $-0.06 \pm 0.39$ \\
\vspace{0.1cm}
30 & $9529.72301_{-0.00066}^{+0.00072}$ & $-0.62_{-0.95}^{+1.03}$ \\
\vspace{0.1cm}
37 & $9548.83271_{-0.00089}^{+0.00078}$ & $-0.07_{-1.28}^{+1.12}$ \\
\vspace{0.1cm}
38 & $9551.56258_{-0.00046}^{+0.00047}$ & $-0.11_{-0.66}^{+0.67}$ \\
\vspace{0.1cm}
49 & $9581.5911_{-0.0020}^{+0.0015}$ & $-0.63_{-2.92}^{+2.24}$ \\
\midrule
\multicolumn{3}{c}{\textit{Planet c}}\\
\vspace{0.1cm}
-2 & $9503.79132_{-0.00063}^{+0.00067}$ & $0.40_{-0.91}^{+0.96}$ \\
\vspace{0.1cm}
0 & $9520.70594_{-0.00059}^{+0.00057}$ & $-0.04_{-0.85}^{+0.82}$ \\
\vspace{0.1cm}
2 & $9537.6207 \pm 0.0011$ & $-0.34_{-1.53}^{+1.52}$ \\
\vspace{0.1cm}
9 & $9596.82308_{-0.00052}^{+0.00053}$ & $-0.07_{-0.75}^{+0.76}$ \\
\bottomrule
\bottomrule
\end{tabular}
\vspace{-0.4cm}
\label{tab:TTVs}
\end{table}

\begin{figure}
    \centering
    \includegraphics[width=0.96\linewidth, height=0.97\textheight, keepaspectratio]{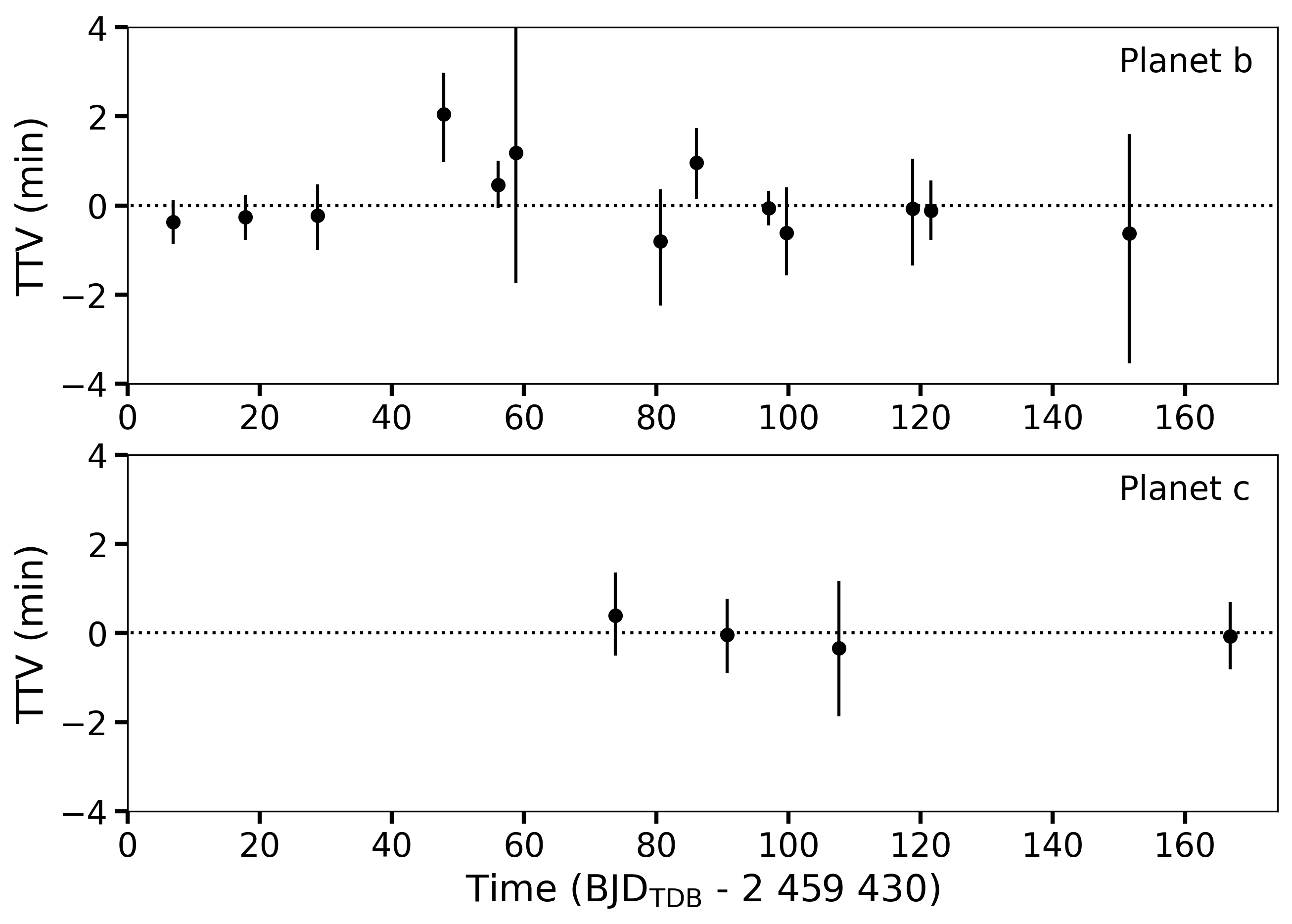}
    \caption{TTVs of the two planets measured from the ground-based transit photometry (see Sect. \ref{sec:TTVs}).}
    \label{fig:TTVs}
\end{figure}

\subsubsection{Search for transit timing variations}
\label{sec:TTVs}

As mentioned in Sect. \ref{sec:glob_an}, the system is within 10$\%$ of the second-order 3:1 mean motion resonance. To explore our photometric dataset for possible transit timing variations (TTVs), we repeated the same global analysis as in Sect. \ref{sec:glob_an}, but allowing this time for each transit of the two planets a timing offset with respect to the reference transit ephemerides defined by the $P$ and $T_0$ values reported in \hbox{Table \ref{tab:param}}. We limited our search for TTVs to the ground-based transit photometry, as TESS individual transits do not have a high enough S/N to allow precise measurements of their timings. Table \ref{tab:TTVs} presents the timings and corresponding TTVs that we obtained for the 13 transits of planet b and 4 transits of \hbox{planet c} that were observed from the ground. \hbox{Fig. \ref{fig:TTVs}} shows the TTVs as a function of time. We do not detect any significant TTVs in the current dataset, nor any trend that could suggest the existence of additional planets in the system.

\subsection{Searches for additional transiting planets and detection limits} 
\label{sec:search}

\subsubsection{TESS photometry}
\label{sec:tess_search}

As mentioned in Sect. \ref{obs:tess}, LP~890-9 was observed by TESS in four sectors. While the TESS Science Office issued an alert for TOI-4306.01 (LP~890-9\,b), it did not report any other transit signal that could be related to \hbox{LP~890-9\,c} or other planets in the system. However, due to the detection threshold (MES=7.1$\sigma$) to issue an alert by the automatic SPOC and Quick Look (QLP) pipelines, some lower S/N transit signals may have gone unnoticed. We thus performed our own search for transit signals in the TESS data using our custom {\fontfamily{pcr}\selectfont  SHERLOCK}\footnote{{The \fontfamily{pcr}\selectfont  SHERLOCK} (\textbf{S}earching for \textbf{H}ints of \textbf{E}xoplanets f\textbf{R}om \textbf{L}ightcurves \textbf{O}f spa\textbf{C}e-based see\textbf{K}ers) code is fully available on GitHub: \url{https://github.com/franpoz/SHERLOCK}} pipeline \citep{pozuelos2020,2020A&A...642A..49D}. 

{\fontfamily{pcr}\selectfont  SHERLOCK} is an end-to-end pipeline that combines six different modules that allow exploring the TESS data fast and robustly. These six modules consist of (1) downloading and preparing the light curves from their online repositories, (2) searching for planetary candidates, (3) performing a semi-automatic vetting of the interesting signals, (4) computing a statistical validation, (5) modelling the signals to refine their ephemerides, and (6) computing observational windows from ground-based observatories to trigger a follow-up campaign. 
By default, {\fontfamily{pcr}\selectfont  SHERLOCK} works with the PDCSAP light curve and applies a multi-detrend approach employing the bi-weight algorithm provided in the \texttt{w{\={o}}tan} package \citep{wotan2019} to optimise the transit search. This strategy allows the user to maximise the S/N and the signal detection efficiency (SDE) during the transit search, which is performed over the nominal PDCSAP light curve, jointly with the new detrended light curves, employing the \texttt{transit least squares} (TLS) package \citep{hippke_TLS_2019}. TLS is optimised for detecting shallow periodic transits using an analytical transit model based on stellar parameters. The transit search is carried out in a loop; once a signal is found, it is stored and masked, and then the search keeps running until no more signals above a user-defined S/N threshold are found in the dataset. Each of these search-find-mask actions is called a `run'. Our experience points out that results found beyond five runs are less reliable due to the accumulated gaps in a given light curve after many mask-and-run iterations. Thus, in our search, we allowed a maximum of five runs. In addition, {\fontfamily{pcr}\selectfont  SHERLOCK} allows one to apply a preliminary Savitzky-Golay filter \citep{savgol}, which enhances the detection of shallow transits with the associated risk of obtaining more false-positive detections that the user needs to vet carefully. This filter is very useful when dealing with threshold-crossing events with low S/N like TOI-4306.01, which was found with a MES value of 8.3 by SPOC. 

We first tried to recover this candidate by running two suites of searches, one with a pure TLS search and the other with the Savitzky-Golay filter applied previously. In both cases, we successfully recovered the candidate issued by TESS in the first run, with signal-to-noise ratios ranging from 3.2 to 3.9 and 6.1 to 7.1, respectively. The very low signal-to-noise ratios reported by the pure TLS search motivated us to keep using only the Savitzky-Golay filter strategy. Then, we performed two independent transit searches for extra planets by considering the four sectors simultaneously. (1) We focussed our search on orbital periods ranging from 0.5 to 40 days where a minimum of two transits was required to claim a detection. (2) We then focussed on longer orbital periods, ranging from 40 to 80 days, where single events could be recovered. None of these strategies yielded positive results, with all the signals found being attributable to systematics, noise, or variability.  

Following \cite{wells2021}, several scenarios may explain the lack of extra signals in TESS data: (1) no other planets exist in the system; (2) if they do exist, they do not transit; or (3) they do exist, but the photometric precision of the data is not good enough to detect them, or they have periods longer than the ones explored in this study. If scenario (2) or (3) is true, extra planets might be detected by means of RV follow-up as discussed in Sect.~\ref{sec:RV_prospects}. On the other hand, we know from the ground-based photometric follow-up that there is at least a second transiting planet in the system, LP~890-9\,c. The fact that it was undetected by both the SPOC and {\fontfamily{pcr}\selectfont  SHERLOCK} transit searches suggests that we are limited in this case by its detectability in the TESS data (scenario 3). 

To explore the detection limits of the TESS data, we used our {\fontfamily{pcr}\selectfont  MATRIX ToolKit} \footnote{{The \fontfamily{pcr}\selectfont  MATRIX ToolKit} (\textbf{M}ulti-ph\textbf{A}se \textbf{T}ransits \textbf{R}ecovery from \textbf{I}njected e\textbf{X}oplanets \textbf{T}ool\textbf{K}it) code is available on GitHub: \url{https://github.com/PlanetHunters/tkmatrix}} \citep{matrix2022}, which was used successfully in a number of previous studies \citep[see, e.g.,][]{wells2021,nicole2022}.
{\fontfamily{pcr}\selectfont MATRIX} performs transit injection-and-recovery experiments by inserting synthetic planets over the PDCSAP light curve, combining all the sectors available. The user defines the ranges in the $R_{\mathrm{p}}$--$P$ parameter space to be examined. In our case, we explored the ranges of 0.5 to 3.0\,$R_{\oplus}$ with steps of 0.28\,$R_{\oplus}$, and 0.5--30.0 days with steps of 0.33 days. In addition, each combination of $R_{\mathrm{p}}$--$P$ was explored using four different phases, that is, different values of $T_{0}$. Hence, we explored a total of 3600 scenarios. For simplicity, the synthetic planets were injected assuming their impact parameters and eccentricities equal zero. Moreover, we detrended the light curves using a bi-weight filter with a window size of 1 day, which was found to be the optimal value during the {\fontfamily{pcr}\selectfont SHERLOCK} executions, and masked the transits corresponding to LP~890-9\,b. We considered a synthetic planet to be recovered when its epoch was detected with 1~hour accuracy and the recovered period was within 5\,\% of the injected period ($P_{\mathrm{injected}}$). In addition, for each period found that did not match the injected period, {\fontfamily{pcr}\selectfont MATRIX} checked if it could correspond to the first harmonic (2$\times P_{\mathrm{injected}}$) or sub-harmonic ($P_{\mathrm{injected}}/2$). If this was the case, then the signal was also considered as recovered. It is worth noting that since we injected the synthetic signals in the PDCSAP light curve, then these signals were not affected by the PDCSAP systematic corrections. In this respect, the detection limits that we find here should be considered as an optimistic scenario \citep[see, e.g.,][]{pozuelos2020,eisner2020}. As demonstrated previously during the {\fontfamily{pcr}\selectfont SHERLOCK} executions, using a preliminary Savitzky-Golay filter before the transit search yields higher S/N detections. We thus used the same strategy during the injection-and-recovery experiments. Finally, for each scenario, we performed up to three runs (as defined above for {\fontfamily{pcr}\selectfont SHERLOCK}) to try to recover the injected signal. Under these considerations, our transit search performed above via {\fontfamily{pcr}\selectfont SHERLOCK}, with a larger number of runs (five) and the multi-detrend approach, can be considered more efficient than the injection-and-recovery map obtained by {\fontfamily{pcr}\selectfont MATRIX}, which still offers a reasonable estimation of the detection limits.   

\begin{figure}
    \centering
    \includegraphics[width=0.48\textwidth]{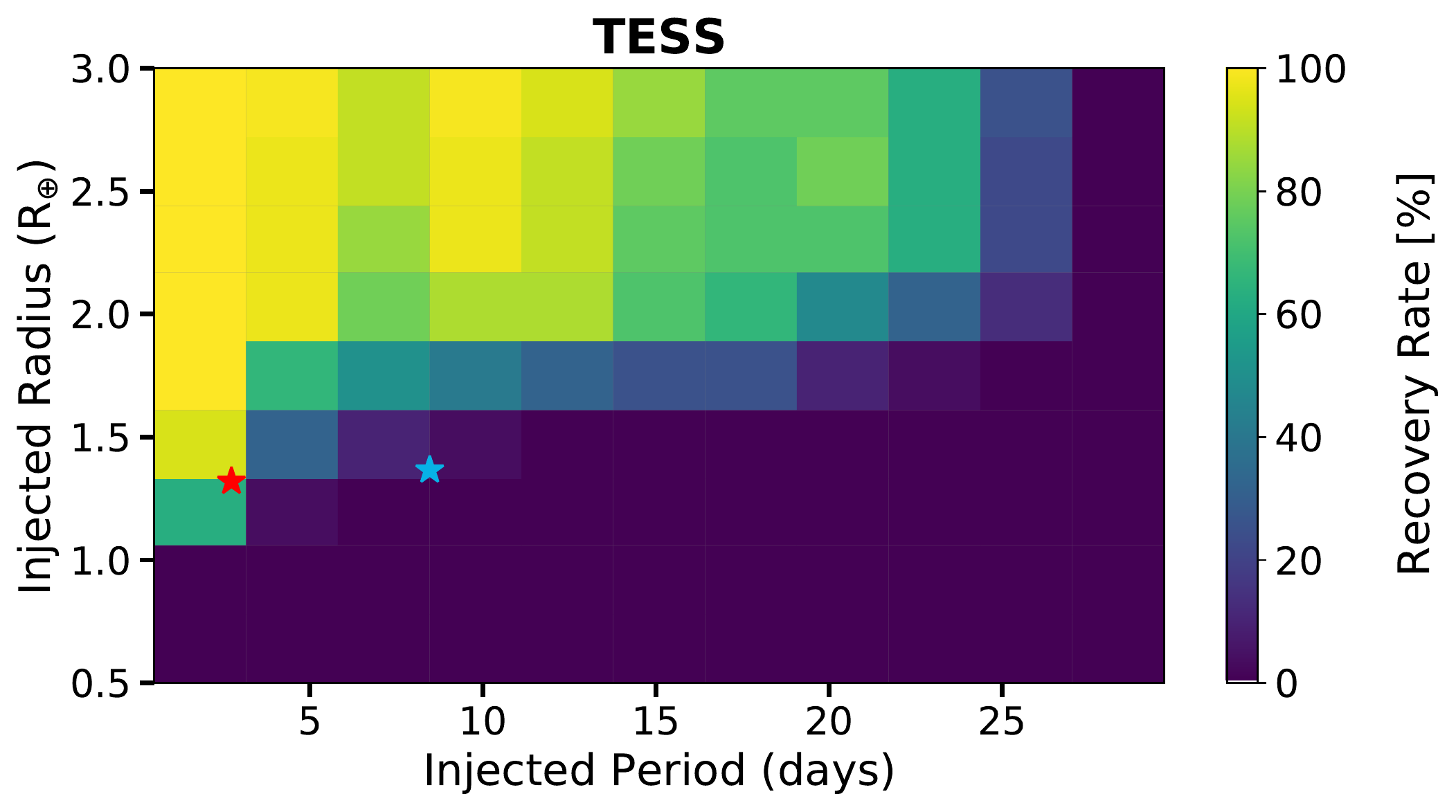}
    \includegraphics[width=0.48\textwidth]{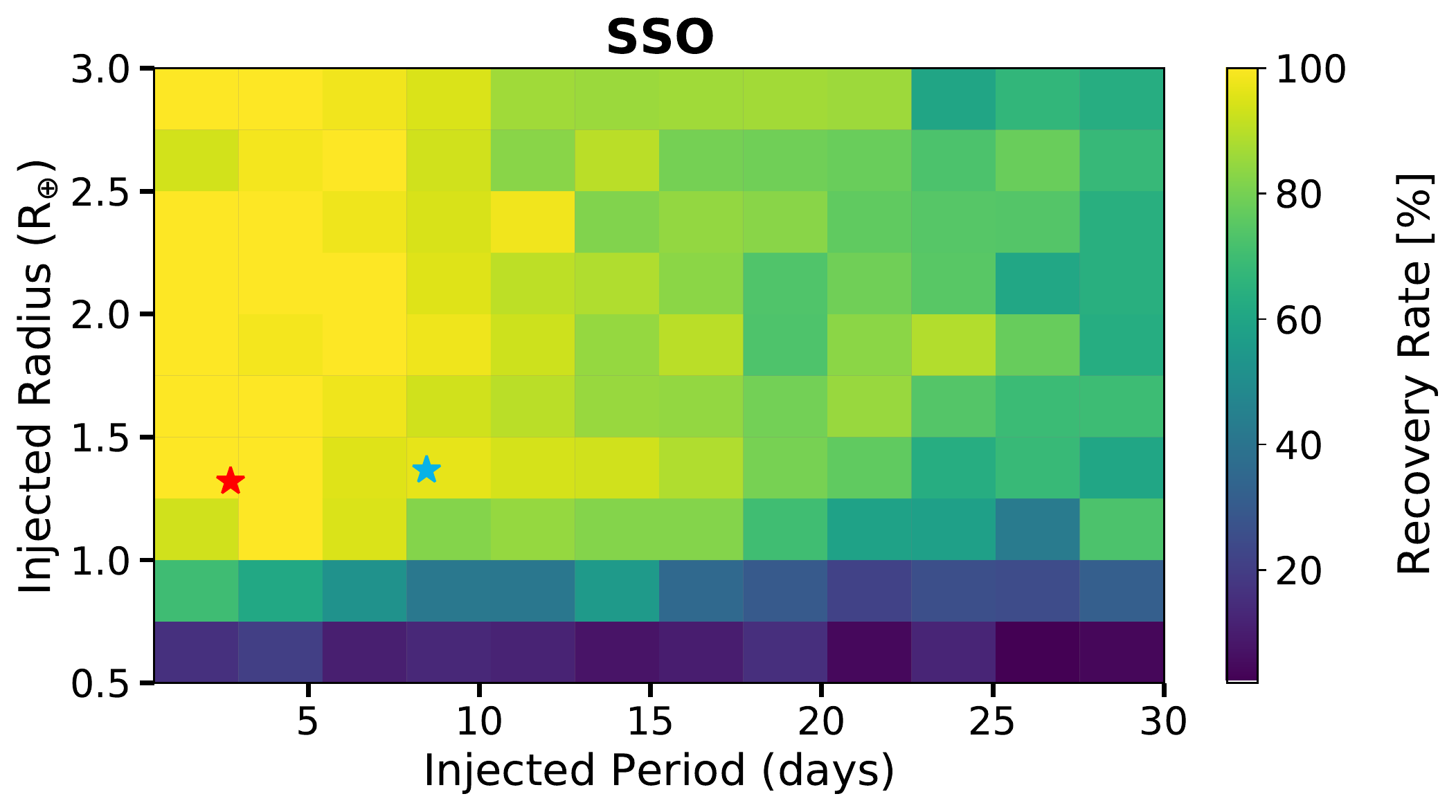}
    \caption{Results from the transit injection-and-recovery tests performed on the TESS (upper panel) and SSO (lower panel) data to assess the detectability of transiting planets in the LP~890-9 system. \textit{Upper panel:} We explored a total of 3600 different scenarios. Each pixel evaluated 32 scenarios, that is, 32 light curves with injected planets having different $P$, $R_{\mathrm{p}}$, and $T_{0}$. Larger recovery rates are presented in yellow and green, while lower recovery rates are shown in blue and darker hues. LP~890-9\,b (recovery rate $\sim$60\,$\%$) and c (recovery rate $\sim$6\,$\%$) are displayed as red and blue stars, respectively (see Sect. \ref{sec:tess_search} for details). \textit{Lower panel:} We injected 6000 artificial planets into the combined $I+z$-filter SSO light curve. On average, each box in this plot contains 50 injection scenarios with differing $P$, $R_{\rm{p}}$, $i_{\rm{p}}$, and $\phi$. Similarly to the upper panel, we include LP~890-9\,b (recovery rate of 100\,$\%$) and c (recovery rate of 96\,$\%$) as red and blue stars, respectively (see Sect. \ref{sec:specphotom} for details).}
    \label{fig:recovery}
\end{figure}

The results, shown in Fig.~\ref{fig:recovery} (upper panel), allow us to rule out planets whose sizes are $>$1.8\,$R_{\oplus}$ and orbital periods $<$14~days, with recovery rates ranging from 100 to $\sim$80\,$\%$. The same planetary sizes but with orbital periods between 14 and 24~days have recovery rates ranging from $\sim$80 to 40\,$\%$. Beyond an orbital period of 24 days, the sensitivity decreases to $\sim$20--0\,$\%$. Smaller planets with sizes between 1.0 and 1.8\,$R_{\oplus}$ have good recovery rates, between 100 and $\sim$60\,$\%$, for short orbital periods $<$3.2~days. LP~890-9\,b lies in this region, with a recovery rate of $\sim$60\,$\%$. On the other hand, when exploring longer orbital periods $>$3.2~days, the detectability of such small planets with sizes between 1.0 and 1.8\,$R_{\oplus}$ decreases to $\sim$50--0\,$\%$. In this region lies LP~890-9\,c, with a recovery rate of only $\sim$6\,$\%$. This result is consistent with the non-detection of LP~890-9\,c in the TESS data during our transit search with {\fontfamily{pcr}\selectfont SHERLOCK}. For the whole range of periods explored, planets with $R_{\mathrm{p}} \leq$1.0\,$R_{\oplus}$ can not be found, which hints at a clear limitation of the TESS data to discover transiting (sub-)Earth-sized planets in this system.

\subsubsection{SPECULOOS-South photometry}
\label{sec:specphotom}

\begin{figure}
    \centering
    \includegraphics[width=\columnwidth]{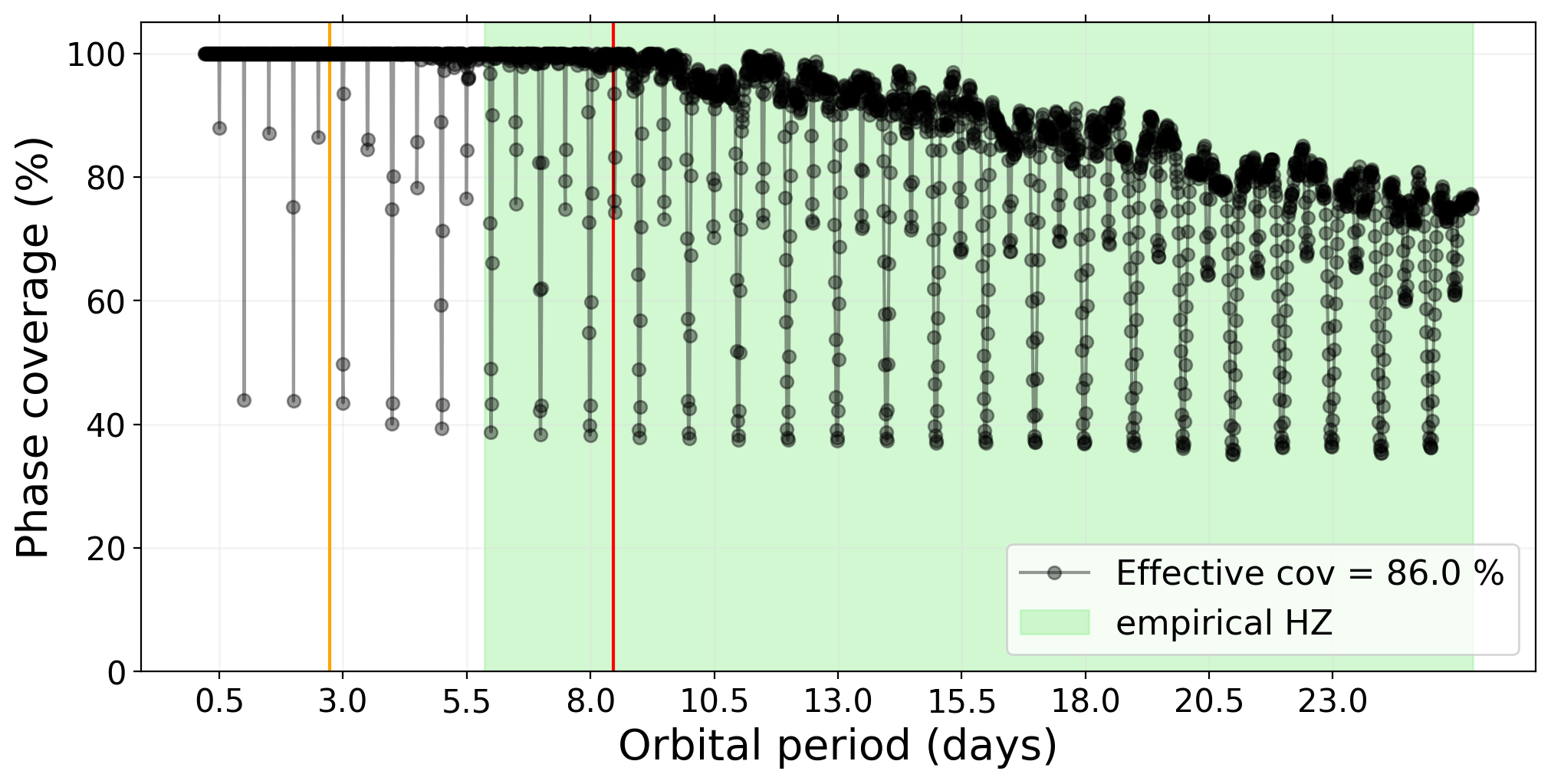}
    \caption{Phase coverage as a function of the orbital period using all photometric data obtained with SSO for LP~890-9. The effective coverage of 86\% is the integral of the black curve. The orbital periods of planets b and c are shown as orange and red lines, respectively. The figure also shows the empirical (optimistic) HZ in green, i.e. the region between the `Recent Venus' and `Early Mars' limits (\citealt{2013ApJ...765..131K,Kopparapu2014}, and Sect. \ref{sec:HZ}).}
    \label{fig:phase_coverage_TOI-4306}
\end{figure}

After the announcement of the TOI-4306.01 planet candidate by TESS in July 2021, we initiated a blind search for other transiting planets in the system with SSO and detected the first transit of planet c after about 2 months of daily monitoring. We stopped our follow-up in mid-January 2022, only when we were confident that the empirical (optimistic) habitable zone (HZ) of the star (\citealt{2013ApJ...765..131K,Kopparapu2014}, see also Sect. \ref{sec:HZ}) had been reasonably well explored. To quantify this, we used a metric that we introduced in \cite{2021A&A...645A.100S}, the effective phase coverage. This metric gives an estimation of how the phase of a hypothetical planet would be covered for a range of periods. In that regard, we computed the percentage of phase covered for each orbital period from $P=0.1$ day to a given $P=P_{\mathrm{max}}$ and took the effective phase coverage for periods $\leq$ $P_{\mathrm{max}}$ to be the integral over the period range. We stopped our photometric monitoring of LP~890-9 when we reached an effective coverage of at least 80\% for $P_{\mathrm{max}}=25.823$ days, which is the orbital period associated to the outer limit of the empirical HZ (\citealt{2013ApJ...765..131K,Kopparapu2014}, see also Sect. \ref{sec:HZ}). Fig. \ref{fig:phase_coverage_TOI-4306} shows the evolution of the phase coverage as a function of the orbital period of a putative transiting planet based on the SSO photometry. The periodic drops in phase coverage that are visible for all integer periods are expected for ground-based observations from a unique site. The effective phase coverage for the outer edge of the empirical HZ is 86\%. As a comparison, we detected the first transit of planet c when its effective phase coverage was only 57\%, and observed its second transit when the effective coverage was 78\%.

\begin{figure*}
    \centering
    \includegraphics[width=\textwidth]{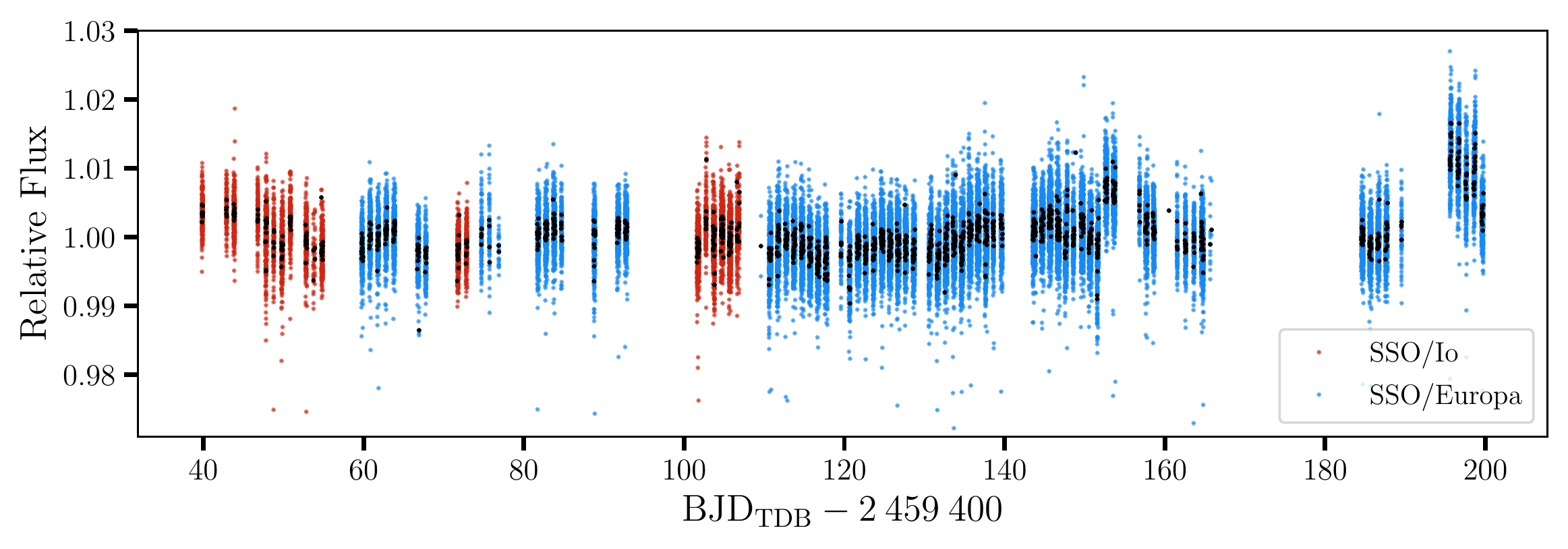}
    \caption{Global $I+z$-filter light curves for the SSO/Io (red) and SSO/Europa (blue) observations. The black points correspond to 20-min bins.}
    \label{fig:SSO_global}
\end{figure*}

We ran an automatic transit search through the SSO $I+z$ photometry to check for potential additional transit signals that may have been missed by our visual inspection of the data. Transit-finding tools often require a rather `flat' light curve to work optimally. Therefore, we first stitched together each night's normalised $I+z$ differential light curve, as these light curves are optimised for the atmospheric conditions of the night, combining light curves from all SSO telescopes. This `combined' light curve was then cleaned for bad weather, as described in \citet{2020MNRAS.495.2446M}, and frames were removed where the atmospheric conditions would significantly affect the photometry. We also removed all observations between 28 December 2021 and 3 January 2022 inclusive, as our autoguiding software, \textsc{donuts} \citep{2013PASP..125..548M}, did not run on these nights and the corresponding light curves were significantly more noisy. Then, to model remaining photometric variations (caused by intrinsic stellar variability and nightly atmospheric variations) without modelling transit features, we employed a Gaussian Process (GP) approach \citep{2006gpml.book.....R}. We used a squared exponential kernel \citep{2006gpml.book.....R} implemented by the \textsc{george} package \citep{2014ApJ...795...64F,2015ITPAM..38..252A}:
\begin{equation}
    k(t_{i},t_{j}) = A \exp\left( -\frac{(t_{j}-t_{i})^2}{l^2} \right)
\end{equation}
where $i$ and $j$ are two different data points in the light curve (taken at times $t_{i}$ and $t_{j}$), $A$ is the amplitude of correlation, and $l$ is the length-scale over which correlations decay. To apply the GP regression, the light curve was binned every 10 minutes and sigma-clipped twice to lower the significance of any infrequent, short features. In essence, binning and sigma-clipping have the same effect; they remove short-duration structures to prevent over-fitting with the GP. The first sigma-clip was a basic 3$\sigma$ clip of the entire light curve, to remove the most extreme outliers. We also included a second, `running' sigma-clip to account for any photometric variability. Here, we binned the light curve into hours, longer than a typical transit duration, to produce a running median (a series of points representing the median of each hour bin) and a running standard deviation (defined similarly). We then computed the median running standard deviation, $\sigma$, and clipped the data points more than 3$\sigma$ above or below the running median. Even though this target does not demonstrate significant photometric variability, this method can also be used to sigma-clip rapidly-varying stars.

For a transit-finding algorithm, we used the \textsc{astropy} \citep{2013A&A...558A..33A,2018AJ....156..123A} implementation of Box Least Squares, or BLS \citep{2002A&A...391..369K}. We considered using Transit Least Squares, or TLS \citep{hippke_TLS_2019}; however, for very red stars, such as those in SPECULOOS's target list, the detection efficiencies of BLS and TLS converge while TLS is more demanding in terms of computing time and power \citep{hippke_TLS_2019}. We limited BLS to search for 10,000 periods between 0.5 and 30 days, with a transit duration between 0.02--0.09 days. The highest significance period is flagged, and the corresponding transits masked before re-running the BLS transit search on the light curve. This automatic, iterative process is run until the S/N of the transits is less than 5. For LP~890-9, we first successfully detected planet b with a period of 2.730 days. Next, planet c was detected with a period of 8.457 days. All $I+z$ transits from planets b and c in Table \ref{tab:follow_up_obs} were recovered. There were no additional, convincing detections with S/N$\geq$5.

To assess the detection efficiency of our transit-search pipeline and the detection limits of the data themselves, we performed injection and recovery tests on the SSO $I+z$ light curve. We generated transits for 6000 artificial planets using \textsc{PyTransit} \citep{2015MNRAS.450.3233P}, a transit light curve modelling package that implements the \citet{2002ApJ...580L.171M} model. We used the limb-darkening coefficients $u_{1}=0.384\pm0.028$ and $u_{2}=0.303\pm0.089$ for the $I+z$ filter from Sect. \ref{sec:analysis}. For each of the planets injected, their parameters (radius $R_{\rm{p}}$, orbital period $P$, and inclination $i_{\rm{p}}$) were drawn from the following distributions:
\begin{equation}
    \begin{split}
        \frac{R_{\rm{p}}}{R_{\oplus}} &\sim \mathcal{U}(0.5,3.0) \\
        \cos{i_{\rm{p}}} &\sim \mathcal{U}(\cos{i_{\rm{min}}},\cos{i_{\rm{max}}}) \\
        \frac{P}{\rm{days}} &\sim \mathcal{U}(0.5,30.0) 
    \end{split}
\end{equation}
where $\mathcal{U}(a,b)$ represents a uniform distribution between $a$ and $b$. $i_{\rm{min}}$ and $i_{\rm{max}}$ are the minimum and maximum inclinations for a transiting planet. The host mass and radius were assumed constant, using the values in Table \ref{table:star}. However, the inclination limits depend on the orbital period; therefore, when drawing the planetary parameters, we drew individual inclinations from a range set by each period. We only considered circular orbits (with $e$=0) as we do not know the underlying eccentricity distribution. Close-in planets are also expected to experience significant circularisation of their orbits \citep{2017NatAs...1E.129L}. The time at which the first transit was injected was also randomly drawn from a uniform distribution, $\phi \sim \mathcal{U}$(0,1), where $\phi$ is the phase of the period, such that the first transit was injected at $\phi P$ from the start of observations.

We injected each of the 6000 artificial planets into the target's differential light curve and ran the transit-search pipeline. The planets were injected before the GP-detrending to allow for the possibility that the GP could over-model and remove or warp the transit signals. If we were to inject planets after GP-detrending, this could artificially inflate the recovery results, as it is easier to detect a transit in a `flat' light curve than in a light curve exhibiting time-dependent structure. From BLS, we only considered the most likely transit and did not look at other peaks in the periodogram. A planet was successfully recovered if at least one transit was detected by BLS above a S/N threshold of 5, with an epoch detected within an hour of the injected transit epoch. We did not include any requirements on the recovered period due to large gaps in the data from the day-night cycle and bad weather conditions. These gaps mean that we often recover period aliases. Additionally, when testing periods up to 30 days, we have a high chance of only detecting a single transit with no meaningful period information. As we only require one `real' transit for recovery, it is possible that we can recover a planet even if we miss some of its transits, or incorrectly detect noise as transits. The lack of a period condition on planet recovery may thus result in an overestimation of our recovery fractions. Therefore, this injection-recovery framework is most useful for detecting individual transit signals (including single transits) which would then be followed up with manual vetting.

For all synthetic planets with radii $R_{\rm{p}}$ = 0.5--3.0\,$R_{\oplus}$ and periods $P$ = 0.5--30\,days, we recovered $\sim$72\%. The lower panel of Fig.~\ref{fig:recovery} shows how the recovery rates vary in the $R_{\rm{p}}$ -- $P$ parameter space. Comparing these results with the ones obtained in Sect. \ref{sec:tess_search} for the TESS data (upper panel of Fig.~\ref{fig:recovery}), we see that the SSO data are sensitive to smaller planets and/or longer orbital periods. 
SSO has a high detection potential for short-period ($P\,\leq\,10$ days) planets in the Earth to super-Earth regime ($R_{\rm{p}}\,\geq\,1\,R_{\oplus}$) with recovery rates ranging from 86 to 100\%, and an average of $\sim$97\%. This is consistent with the detection of both the b and c planets in the SSO data, with respective recovery rates of 100 and 96 \%. The sensitivity is still reasonable for smaller planets with 0.75\,$R_{\oplus}\leq R_{\rm{p}}< 1\,R_{\oplus}$ in the same period range ($P$\,$<$\,10\,days), with an average of $\sim$57\%. However, it drops to $\sim$15\% for even smaller planets with 0.5\,$R_{\oplus} \leq\,R_{\rm{p}} <\,0.75\,R_{\oplus}$. 
Due to the long observation span of LP~890-9 with the SSO and our decision not to include a period criterion in our recovery, we obtain good recovery rates (above $\sim$60\%) for Earth-sized planets and super-Earths ($R_{\rm{p}}\,\geq\,1\,R_{\oplus}$) with periods longer than 10 days, up to 30 days. For the same range of periods, the sensitivity drops to $\sim$25--50\% for planets with 0.75\,$R_{\oplus}\leq R_{\rm{p}}< 1\,R_{\oplus}$ and below 15\% for the smallest planets with 0.5\,$R_{\oplus} \leq\,R_{\rm{p}} <\,0.75\,R_{\oplus}$. Based on these results, we conclude that it is unlikely there are any additional transiting short-period (super-)Earth-sized planets hidden in this system. However, it is still possible that there could be more transiting bodies, either too small or too far away from their host to be detected in the SSO data.


\begin{figure}
    \centering
    \includegraphics[width=\columnwidth]{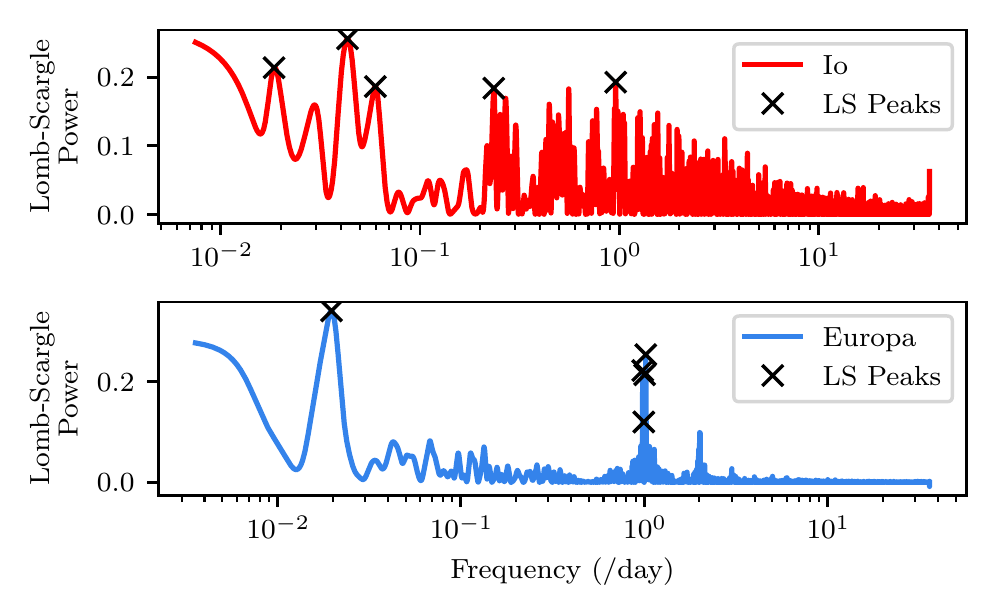}
    \caption{Lomb-Scargle periodograms for SSO/Io (top, red) and SSO/Europa (bottom, blue). The five highest peaks are marked by black crosses in both cases. For SSO/Europa, the most significant peak is around 50--55\,days and the next highest four are at $\sim$1 day and were determined to be aliases. The results for SSO/Io are less clear due to large gaps and much less data. The most significant peaks are found at 23.0, 54.0, and 4.3 days. The peak around 54 days could correspond to the one detected for SSO/Europa, but the other two are not present in SSO/Europa's periodogram.}
    \label{fig:SSO_LS_results}
\end{figure}

\subsection{Stellar variability} \label{sec:activity}

We searched for flares and rotational variability in the SSO $I+z$-filter photometry (which has a higher precision than the TESS photometry for this late-type target) using the techniques described in \cite{2022MNRAS.513.2615M}. The light curves used in this section were cleaned for bad weather and poor photometric conditions, as described in Sect. \ref{sec:specphotom}. However, unlike in Sect. \ref{sec:specphotom}, we did not use the nightly normalised light curves. Instead, we used the raw flux values from aperture photometry across all nights observed with the same telescope as one time series, and performed our differential photometry process on the entire time series, to obtain what we call a `global' light curve (see also \citealt{2020MNRAS.495.2446M}). Doing so allows to preserve the flux relationships between nights and study long-term photometric trends, but limits our ability to optimise the photometry night-by-night (e.g., aperture size, comparison stars). The global light curves for SSO/Io and SSO/Europa are presented in Fig. \ref{fig:SSO_global}.

The search for flares was done in two parts: a simple, automatic flare-detection algorithm was run first to extract all the flare candidates (based on the method described in \citealt{2020MNRAS.497.3790L}), followed by a manual vetting process to confirm them. We detected no conclusive flares (flare structures that could not be attributed to atmospheric effects) in any of the light curves using this method.

To search for a rotation period, we applied the Lomb-Scargle periodogram analysis \citep{1976Ap&SS..39..447L, 1982ApJ...263..835S} to the global light curves, binned every 20 minutes. The SSO light curves are not uniformly sampled as there are gaps, not only from the day/night cycle, but also from bad weather and changes to SSO's observation strategy. Therefore, careful treatment of the resulting periodograms is essential to remove aliases. We searched for periods between the Nyquist limit (twice the bin size) and the entire observation window. We removed all peaks in the periodogram with a false alarm probability below 3$\sigma$ (0.0027). We visually inspected the periodograms of LP~890-9, shown in Fig. \ref{fig:SSO_LS_results}, and the corresponding phase-folded light curves for all possible rotation periods (i.e., all significant peaks in this periodogram). In addition, by comparing with periodograms of the time stamps, airmass, and FWHM, we eliminated signals which arose from the non-uniform sampling and from ground-based systematics. The most likely rotation period detected for the SSO/Europa light curve (individually and combined with SSO/Io) was found to be around 50--55\,days. However, as this value is approximately a third of the entire observation span of SSO/Europa, more observations would be required to confirm. The Lomb-Scargle results for SSO/Io were less clear, with the most significant peaks at 23.0, 54.0, and 4.3 days. The signal around 54 days could correspond to the one detected for SSO/Europa, but the other two are not present in SSO/Europa's periodogram. These results were consistent regardless of whether the transits of planets b and c were masked. However, when applying the Lomb-Scargle analysis to the light curves with a longer binning of 60 minutes, we found the same result for SSO/Europa, but no significant peaks in the periodogram for SSO/Io. Additionally, there were unexplained jumps in the light curve on the nights of 3 December 2021 and 15 January 2022 that did not correlate with FWHM, sky background level, target's position on the CCD, airmass, nor with the photometric aperture or any instrumental parameter (CCD temperature, exposure time, ...). If these jumps are due to unknown systematics, then they will affect the reliability of our rotation analysis.

\begin{figure}
    \centering
    \includegraphics[width=\linewidth, height=\textheight, keepaspectratio]{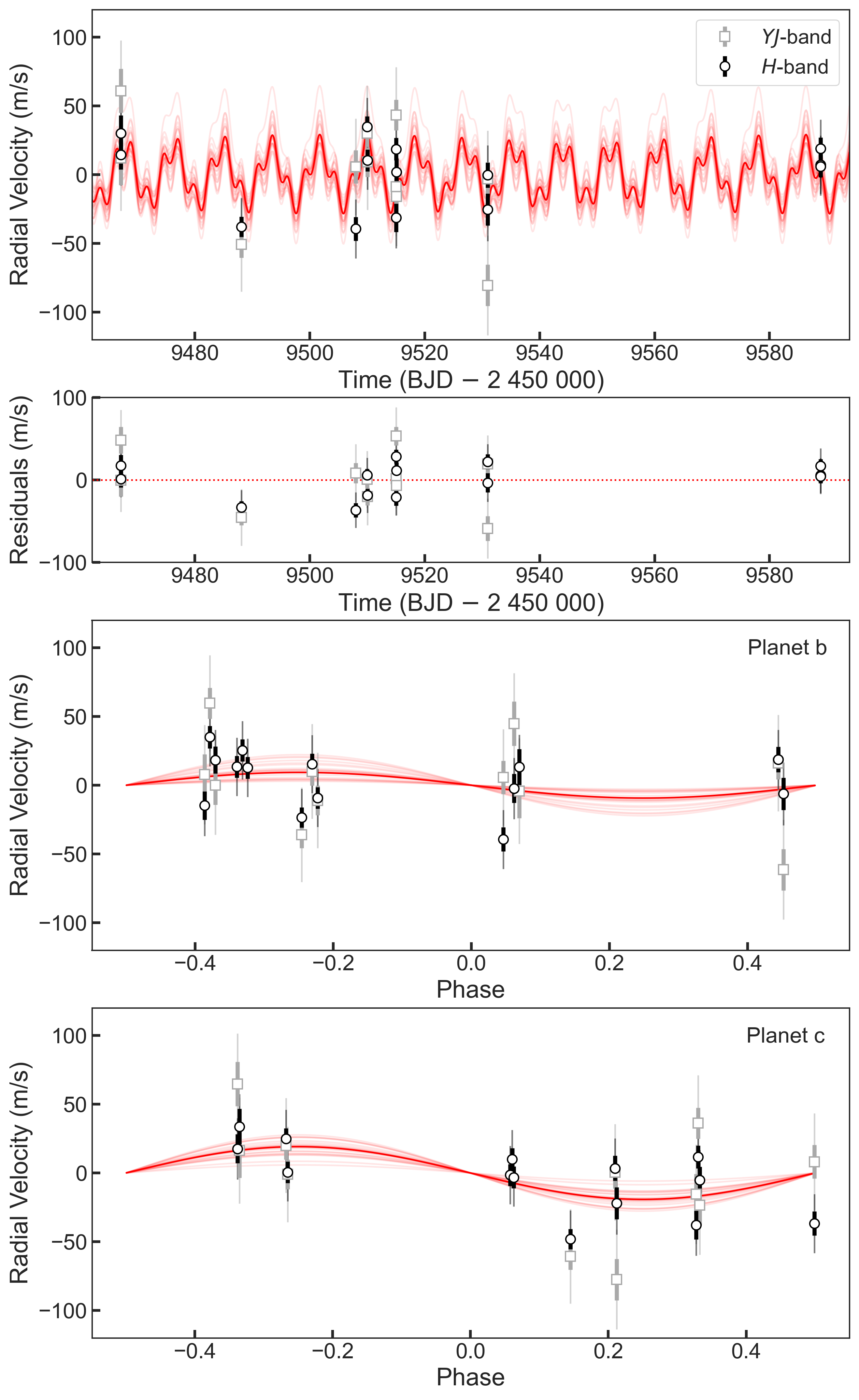}
    \caption{Subaru/IRD radial velocities obtained in the $YJ$- (grey squares) and $H$- (black circles) bands. The data are plotted with both their original error bars (thick solid lines) and the error bars enlarged by the best-fit jitter terms (thinner transparent lines). The top panel displays the data as a function of time. The solid red line shows the 2-planet model generated from the posterior median, while semi-transparent red lines show 25 models randomly drawn from the posteriors. The residuals around the median solution are shown in the second panel. The two lower panels show the data folded on the orbital periods of planets b and c, respectively, after removing the RV component from the other planet.}
    \label{fig:rvs}
\end{figure}

\section{Radial velocity analysis} 
\label{sec:rv_analysis}

We analysed the Subaru/IRD data using the \texttt{juliet} package \citep{2019MNRAS.490.2262E}, which is built over \texttt{radvel} \citep{2018PASP..130d4504F} for the modelling of radial velocities and the \texttt{dynesty} \citep{2020MNRAS.493.3132S} dynamic nested sampling algorithm for estimating Bayesian posteriors and evidences. We modelled the RVs using a sum of two Keplerians (one for each planet) and assumed circular orbits. We imposed normal priors on the orbital period $P$ and mid-transit time $T_0$ of each planet based on the results of our global transit analysis reported in Table \ref{tab:param}. We assumed wide uniform priors between 0 and 500 m/s for the RV semi-amplitudes $K$. We treated the $YJ$-band and $H$-band RVs as independent datasets, thus fitting for each of them a different zero point (systemic RV) and extra jitter term. For each dataset, this jitter is added quadratically to the measurement errors of the data points, to account for any underestimation of the uncertainties or any excess noise not captured by the model. Fig. \ref{fig:rvs} shows the results of our RV fit. The data are plotted with both their original error bars (thick solid lines) and the error bars enlarged by the best-fit jitter terms (thinner transparent lines). These jitters are large \hbox{($33.0_{-7.4}^{+10.5}$ m/s} for the $YJ$-band and $19.6_{-4.3}^{+5.7}$ m/s for the $H$-band) and the datasets are rather limited, so that the RV semi-amplitudes of the two planets are poorly constrained and we were only able to derive upper limits: $K<25.1$ m/s for \hbox{planet b} and $K<33.0$ m/s for planet c (2$\sigma$ upper limits). The derived upper limits on the planetary masses (using the orbital inclinations from our transit analysis, Table \ref{tab:param}) are $M_\mathrm{p} < 13.2\: M_\oplus$ for \hbox{planet b} and $M_\mathrm{p} < 25.3\: M_\oplus$ for planet c at 2$\sigma$. Both are well within the planetary regime.

The large extra jitters returned by our 2-planet fit suggest that there is some variability in the data that is not captured by the model. One possible explanation is that there may be an additional planet in the system which produces a detectable RV signal but was not detected in our photometry (either because it is not transiting or because it is beyond the detection limits of our photometric dataset). We thus tried adding a third planet as a third Keplerian in the model, using a wide uniform prior between 1 and 125 days (the time baseline of the RV data) for its orbital period. Comparing the Bayesian evidences $Z$ of the two models, we found that the 3-planet model is only weakly disfavoured compared to the 2-planet one ($\Delta$ ln $Z$ = $-$0.65). The orbital period found for the third Keplerian is $64_{-9}^{+11}$ days, with a 2$\sigma$ upper limit on its semi-amplitude of 49 m/s. Instead of adding a third Keplerian in the model, we also tried including a linear trend that could be created by a third planet with an orbital period longer than the time baseline of the RV data. We found this model to be marginally disfavoured ($\Delta$ ln $Z$ = $-$3.72) compared to the 2-planet model. Since the 2-planet model is the simplest of the three models we tested and it has the highest Bayesian evidence, it appears to be the best model given the data at hand. However, additional planets in the system might be revealed with more RV data. 
The variability seen in the current RV dataset may also be (at least partly) related to stellar activity. In this context, it is worth noting that the period of $64_{-9}^{+11}$ days found for the third Keplerian in the above 3-planet fit is within the range of the possible rotation period of 50--55\,days suggested by the SSO photometry (Sect.~\ref{sec:activity}). Finally, it is also possible that the RV measurements are impacted by systematic effects. Indeed, large and currently unexplained RV variabilities have been found in Subaru/IRD data obtained for other faint and apparently quiet M dwarfs (see, e.g., \citealt{2022arXiv220113274M,2022AJ....163..298M}). 

Using the probabilistic mass--radius relationship of \cite{2017ApJ...834...17C}, we find predicted masses of $2.3_{-0.7}^{+1.7}$ and \hbox{$2.5_{-0.8}^{+1.8}$ $M_\oplus$} for LP~890-9\,b and c, respectively. Combined with the orbital parameters of the two planets, these masses would correspond to respective radial-velocity semi-amplitudes of $4.4_{-1.4}^{+3.2}$ and $3.3_{-1.1}^{+2.4}$ m/s, significantly smaller than the measurement errors and variability of the current Subaru/IRD data. Additional RV measurements, possibly obtained with other available or next-generation instruments, are required to improve the preliminary mass constraints derived here and investigate the possible presence of other planets in the system. We discuss some prospects for RV follow-up in Sect. \ref{sec:RV_prospects}.


\section{Dynamical analysis} \label{sec:dynamics}

\subsection{Expected TTVs} 
\label{sec:exp_ttvs}

As described in Sect. \ref{sec:TTVs}, we do not detect any significant TTVs in the current photometric dataset, although the two planets are close to the second-order 3:1 mean motion resonance (but not exactly within: $P_\mathrm{c}/P_\mathrm{b}=3.098$). In general, the lack of observed TTVs might be caused by either their low amplitude or the poor baseline coverage provided by the observational data. In our case, the ground-based observations that we used to explore TTVs were obtained over $\sim$150 days, which is much longer than the super-period of 2.64 days \citep{lithwick2012} for LP~890-9\,b and c. Hence, our ground-based data should show any TTVs if they were large enough. To test the expected TTV amplitudes in the system, we used the \texttt{TTVFast2Furious} package \citep{hadden2019} following the strategy presented by \cite{cloutier2020}. We ran 10$^{3}$ realisations with orbital periods and times of mid-transit sampled from the posterior distributions given in Table~\ref{tab:param}. The planetary masses were drawn from their predictions from the probabilistic mass--radius relationship of \cite{2017ApJ...834...17C} (see Sect. \ref{sec:rv_analysis}) and the argument of periastron following a random distribution from 0 to 2$\pi$. We found maximum TTV values for LP~890-9\,b and c of only 1.5 and 1.0\,s, respectively. As a comparison, the typical timing uncertainties returned by our global transit analysis are on the order of 1 minute (see Table~\ref{tab:TTVs}).
Based on the low expected TTV amplitudes, we conclude that the LP~890-9 system is not suitable for a determination of the planetary masses (and dynamical studies in general) using the TTV method. This is in contrast with the TRAPPIST-1 system, where the outer five planets are close to first-order\footnote{The lower the order of the mean motion resonance, the larger the induced TTV amplitude.} resonances with adjacent planets, resulting in large-amplitude TTVs (from $\sim$5 to 100 minutes) -- an important feature which made it possible to derive precise planetary masses and eccentricities \citep{2021PSJ.....2....1A}.

\subsection{Stability} 
\label{sec:stability}

From the current dataset, we concluded that the LP~890-9 system is composed of at least two planets orbiting in nearly circular orbits and with similar planetary sizes. 
To assess the global stability of the system, we used the Mean Exponential Growth factor of Nearby Orbits \citep[MEGNO;][]{cincottasimo1999,cincottasimo2000,Cincotta2003}, $Y(t)$ parameter. This chaos index is widely used to explore the stability of extrasolar planetary systems 
\cite[e.g.,][]{Hinse2015,Jenkins2019,delrez2021}. In short, its time-averaged mean value, $\langle Y(t) \rangle$, amplifies any stochastic behaviour, which can be used to distinguish between quasi-periodic orbits, if $\langle Y(t\rightarrow \infty)\rangle \rightarrow$ 2, or chaotic trajectories, if $\langle Y(t\rightarrow \infty)\rangle \rightarrow \infty$. We made use of the MEGNO implementation with the N-body integrator {\scshape rebound} \citep{rein2012}, which employs the Wisdom-Holman WHfast code \citep{rein2015}. 
In this context, we studied a set of scenarios, where the planetary properties of LP~890-9\,b and c were drawn from the posterior distributions and derived parameters given in Table~\ref{tab:param}. In particular, semi-major axes, orbital inclinations, and planetary masses were sampled from uniform distributions using their nominal values up to 3$\sigma$ of their uncertainties. In addition, the mean anomalies and arguments of periastron were sampled randomly from 0 to 2$\pi$. We ran two suites of simulations: first, we considered circular orbits, and secondly, slightly eccentric orbits using a uniform distribution of 0.10$\pm$0.05 for both planets. In each experiment, we generated 1,000 scenarios, and they were integrated over 10$^{6}$ orbits of the outer planet, with a time-step of 5$\%$ of the orbital period of the inner planet. We found that the LP~890-9 system is highly stable in both experiments with $\langle Y(t) \rangle$ clustering between 1.999--2.001 in 99.3$\%$ and 93.8$\%$ of cases for circular and eccentric orbits, respectively.

Having shown the stability of the LP~890-9 system on its nominal configuration and within 3$\sigma$ of the planetary properties, we considered possible stability limitations for the planetary masses and the orbital eccentricities, the two main parameters that have a major impact on the orbital dynamics and that are poorly constrained by the current data. To this end, we extended our analyses by building stability maps following \cite{delrez2021}, that is, by exploring the $M_{\mathrm{b}}$--$M_{\mathrm{c}}$ and $e_{\mathrm{b}}$--$e_{\mathrm{c}}$ parameter spaces. We explored planetary masses ranging from 1 to 20 $M_{\oplus}$ and eccentricities ranging from 0.0 to 0.6 for both planets. In these sets of simulations, we took 20 values from each range, so each stability map has $20{\times}20$ pixels. Moreover, we set 10 different initial conditions for each realisation, randomly sampling the argument of periastron and the mean anomaly from 0 to 2$\pi$. Hence, each stability map consisted of 4000 scenarios. We then averaged the results of the 10 different initial conditions to obtain the averaged value of $\langle Y\rangle$ for each pixel of the stability map. The integration time and time-step were the same as previously used. We found that the system is fully stable over the whole range of masses explored with $\Delta \langle Y(t) \rangle$= 2.0 $-$ $\langle Y(t) \rangle <10^{-5}$. On the other hand, we found that the eccentricities have to satisfy the condition $e_{\mathrm{b}}\leq 0.39-1.11\times e_{\mathrm{c}}$. This implies that the maximum eccentricity for planet b (when planet c has a circular orbit) is 0.39, and the maximum eccentricity for planet c (when planet b has a circular orbit) is 0.35. For eccentricity values that do not meet this condition, the system may become unstable. By combining the results obtained along this subsection and assuming that there are no more planets in the system, we conclude that the dynamical architecture of the LP~890-9 system resides in a stable set of parameters, allowing for long-term evolution.

\subsection{Tidal evolution} 
\label{sec:tides}

The two planets of the LP~890-9 system reside in close-in orbits and, consequently, due to the short distance from the host star, are affected by tides. Indeed, tidal interactions between the star and planets in closely-packed systems have an impact on planetary orbits, altering the obliquities ($\epsilon$) and rotational periods ($P_{\mathrm{rot}}$) on short time scales, driving the planets 
into a particular configuration called tidal locking, where the tidal torques fix the rotation rate to a specific frequency, and the planetary and stellar spins are aligned.
On longer time scales, the tidal interactions can also induce the circularisation of the orbits and reduce their semi-major axes \citep[see, e.g.,][]{bolmont2014,barnes2017}. This scenario is interesting since a tidally-locked planet in a circular orbit rotates synchronously (i.e., its rotational and orbital periods are equal), meaning that the same hemisphere of the planet is always facing the star, a configuration that can have important implications for the planet's global climate.

To explore the evolution of the LP~890-9 system under tidal interactions, we employed the constant time-lag model (CTL), where a planet is modelled as a weakly viscous fluid deformed by the gravitational interactions \cite[see, e.g.,][]{Mignard1979TheI,Hut1981TidalSystems.,Eggleton1998TheFriction,Leconte2010IsEccentricity}. We used the implementation of CTL in the {\sc{posidonius}} N-body code \citep{blancocuaresma2017}. The main factor in the CTL model that controls the tidal dissipation of a given planet is the product of two free parameters: the degree-2 potential Love number, $k_{2}$, and the constant time-lag, $\Delta \tau$. The $k_{2}$ accounts for the elastic properties and the self-gravity of the planets, with values ranging from 0 to 1.5. On the other hand, $\Delta \tau$ refers to the lag between the line connecting the two centers of mass (star-planet) and the direction of the tidal bulges, which can span orders of magnitudes \citep{barnes2017}. Although the masses (and thus bulk densities) of the LP~890-9 planets are currently poorly constrained (Sect.~\ref{sec:rv_analysis}), their small sizes imply that they are most likely terrestrial. Following \cite{turbet2018} and \cite{bolmont2020}, who studied the tidal effects for the TRAPPIST-1 planets, we thus used Earth's value $k_{2,\oplus} \Delta \tau_{\oplus} = 213$~s \citep{neron1997} as a reference and explored dissipation factors ranging from 0.1$\times k_{2,\oplus} \Delta \tau_{\oplus}$ (for dry rocky planets such as Mars) to 10$\times k_{2,\oplus} \Delta \tau_{\oplus}$ (for planets dominated by fluid-solid friction), to evaluate a range of possible tidal behaviours. 

In a first step, we assessed the time needed by the LP~890-9 planets to evolve towards the tidal locking configuration. To this end, we conducted a suite of simulations considering different scenarios combining three different initial planetary rotational periods $P_{\mathrm{0,rot}}$ (10, 100, and 1000~h), three different initial obliquities $\epsilon_{0}$ (15$\degr$, 45$\degr$, and 75$\degr$) and three dissipation factors (0.1$\times$, 1$\times$, and 10$\times k_{2,\oplus} \Delta \tau_{\oplus}$). We found that the slowest evolution towards tidal locking happened for the particular combination of initial conditions given by $\epsilon_{0}=75\degr$ and $P_{\mathrm{0,rot}}=10$~h for dry terrestrial planets (0.1$\times k_{2,\oplus}\Delta\tau_{\oplus}$). In such a case, we found time scales of ${\sim}1.5\times10^{4}$ and ${\sim}1.5\times10^{6}$~yr, for planets b and c, respectively. Hence, these values might be considered as upper limits on the time scale required for each planet to evolve into a tidally-locked configuration. Considering the estimated age of the system (Sect.~\ref{sec:age}), the two planets are likely tidally locked.

In a second step, we evaluated the time needed for the circularisation of the orbits due to tidal interactions. From the global transit analysis presented in Sect.~\ref{sec:glob_an}, we did not find any evidence for eccentric orbits. However, the eccentricities of the planets are poorly constrained by the transit photometry, and high-precision RV measurements are needed for a proper estimation. On the other hand, when studying the dynamical architecture of the LP~890-9 system in Sect.~\ref{sec:stability}, we found that it may tolerate a certain level of planetary eccentricities allowing for long-term dynamical evolution. Then, assuming initial slightly eccentric orbits of 0.01 for both planets, we performed three simulations considering dissipation factors of 0.1$\times$, 1$\times$, and 10$\times k_{2,\oplus} \Delta \tau_{\oplus}$. We found that the eccentricities of both planets decreased to reach mean equilibrium values of ${\sim}1.8\times10^{-4}$ and ${\sim}1.7\times10^{-5}$ for planets b and c, respectively. The time scales were of the order of ${\sim}10^{9}$, ${\sim}10^{8}$ and ${\sim}10^{7}$~yr, for the different dissipation factors considered, being longer for dry rocky planets. These small residual eccentricities, inferior to 10$^{-3}$, are the result of the competition between tidal damping and planet-planet excitations \citep{turbet2018}. Then, considering that the system is likely older than 2~Gyr (Sect.~\ref{sec:age}), planetary orbits may have had time to evolve towards these low eccentricities due to tidal interactions. Thus, should future observations reveal larger eccentricities than the equilibrium values found here, this could suggest that other unknown planets in the system might be exciting the orbits.

\begin{figure*}[!htbp] 
    \centering
    \includegraphics[width=0.80\textwidth, height=\textheight, keepaspectratio]{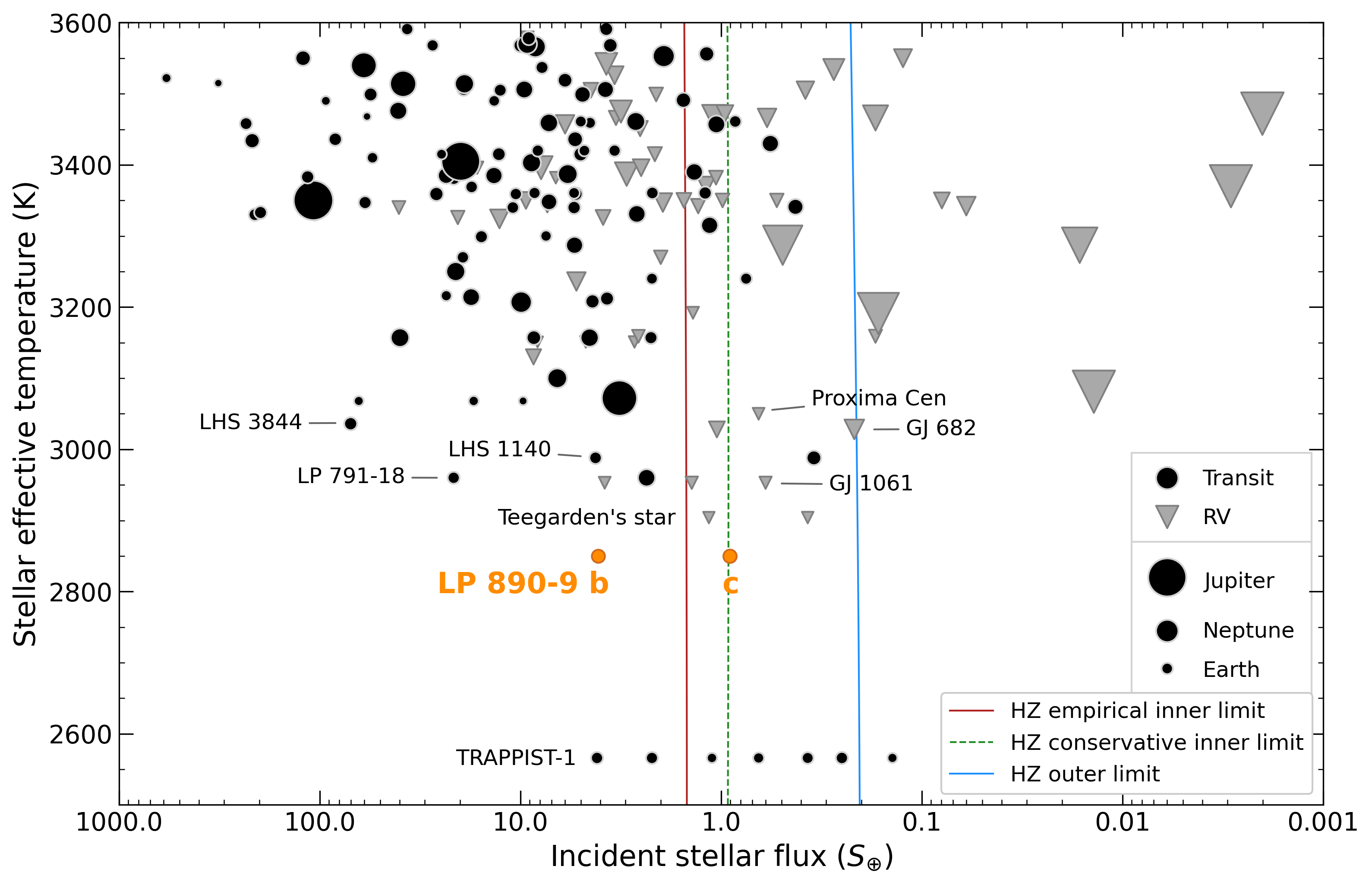}
    \caption{Exoplanets found around cool dwarfs with an effective temperature lower than 3600 K using the transit (black circles) or radial velocity (grey triangles) techniques. Data were taken from NASA Exoplanet Archive on 20 April 2022. The planets are shown as a function of their incident stellar flux and the effective temperature of their host star. The size of the symbols is proportional to the planetary radius, estimated from the mass using the probabilistic mass--radius relationship of \cite{2017ApJ...834...17C} when it is unknown (for non-transiting RV planets). LP~890-9\,b and c are shown as orange circles. The inner (Recent Venus) and outer (Early Mars) boundaries of the empirical (optimistic) HZ are shown as solid red and blue lines, respectively (\citealt{Kopparapu2014} and Sect. \ref{sec:HZ}). The dashed green line shows an alternative conservative inner edge limit from 3D models \citep{Leconte2013}. The outer edge of the HZ in 3D models agrees with incident stellar flux for Early Mars insulation. We note that the three limits shown do vary with the stellar effective temperature, but this is not obvious on this plot due to the log-scale of the $x$-axis.}
    \label{fig:Teff_vs_S}
\end{figure*}

\section{Discussion} \label{sec:discussion}

\subsection{LP~890-9 in the context of other planetary systems around cool dwarfs}

Fig. \ref{fig:Teff_vs_S} shows the planets that have been found so far around cool dwarfs using the transit and radial velocity techniques, as a function of the stellar flux they receive and the effective temperature of their host star. Aside from LP~890-9, only five planetary systems are known around hosts with \hbox{$T_{\mathrm{eff}}<3000$ K:} TRAPPIST-1 \citep{2016Natur.533..221G,2017Natur.542..456G}, Teegarden's star \citep{2019A&A...627A..49Z}, GJ 1061 \citep{2020MNRAS.493..536D}, LP 791-18 \citep{2019ApJ...883L..16C}, and LHS 1140 \citep{2017Natur.544..333D,2019AJ....157...32M}. With an effective temperature of $2850 \pm 75$ K (Sect. \ref{sec:star}), LP~890-9 is the second-coolest star found to host planets after TRAPPIST-1. Interestingly, all these systems -- including LP~890-9 -- are multi-planet systems, in line with the observation that compact multi-planetary systems should be a relatively common outcome of planet formation around low-mass stars (e.g., \citealt{2019A&A...631A...7C}, \citealt{2020MNRAS.491.1998M}, \citealt{Mulders2021}).

\subsection{Prospects for RV follow-up} 
\label{sec:RV_prospects}

As mentioned in Sect. \ref{sec:rv_analysis}, the expected RV semi-amplitudes of the two planets based on the mass--radius relationship of \cite{2017ApJ...834...17C} are $4.4_{-1.4}^{+3.2}$ and \hbox{$3.3_{-1.1}^{+2.4}$ m/s}, respectively. Given the faintness of the star, detecting such small signals will be challenging, even for state-of-the-art instrumentation, and will require a substantial investment of observing time. 

Besides its late spectral type, the star is too faint to recover the expected RV semi-amplitudes with the SPIROU infrared spectrograph at the CFHT \citep{donati20} or to be measured with the upcoming NIRPS infrared spectrograph at the 3.6\,m ESO telescope (limiting magnitude $H$=9, \citealt{wildi17}). The expected RV semi-amplitudes are also too small to be recovered with CRIRES+ at the VLT (RV precision\footnote{\url{https://www.eso.org/sci/facilities/paranal/instruments/crires/inst.html}}$\sim$20 m/s, \citealt{dorn14}). Simply scaling from the existing RV constraints derived from the current Subaru/IRD data ($K_{\mathrm{b}}<25.1$ m/s and $K_{\mathrm{c}}<33.0$ m/s) and the expected RV semi-amplitudes given above suggests that $\sim$31 and $\sim$100 times more data would be needed to constrain the masses of planets b and c, respectively, with that facility.

In the optical range, currently available high-resolution spectrographs, like ESPRESSO at the VLT \citep{pepe21} and MAROON-X at Gemini-North \citep{MAROON-X2021}, would also require substantial amounts of observing time to detect the RV signals of the planets. Even assuming an inactive and slow rotating star (stellar jitter $\sim$1.5 m/s which is probably optimistic, see Sect.~\ref{sec:rv_analysis}), we estimate that about 60 spectra with an average exposure time of 1800s would be required with MAROON-X to recover the RV signals of both planets with 4--5$\sigma$. With the future ANDES spectrograph at the ELT \citep{Marconi21}, a 7--9$\sigma$ detection should be feasible by investing only half of this observing time.

To summarise, while a 4--5$\sigma$ detection of the RV signals would require a large amount of observing time with current high-resolution spectrographs, measuring the masses of the two planets with at least 9$\sigma$ accuracy will require waiting for next-generation instruments to become available in the next decade. Furthermore, such an RV follow-up will also allow to investigate the origin of the large variability seen in the Subaru/IRD data (Sect. \ref{sec:rv_analysis}) and may reveal some additional planets in the system.

\begin{figure*}[!htbp] 
    \centering
    \includegraphics[width=0.80\textwidth]{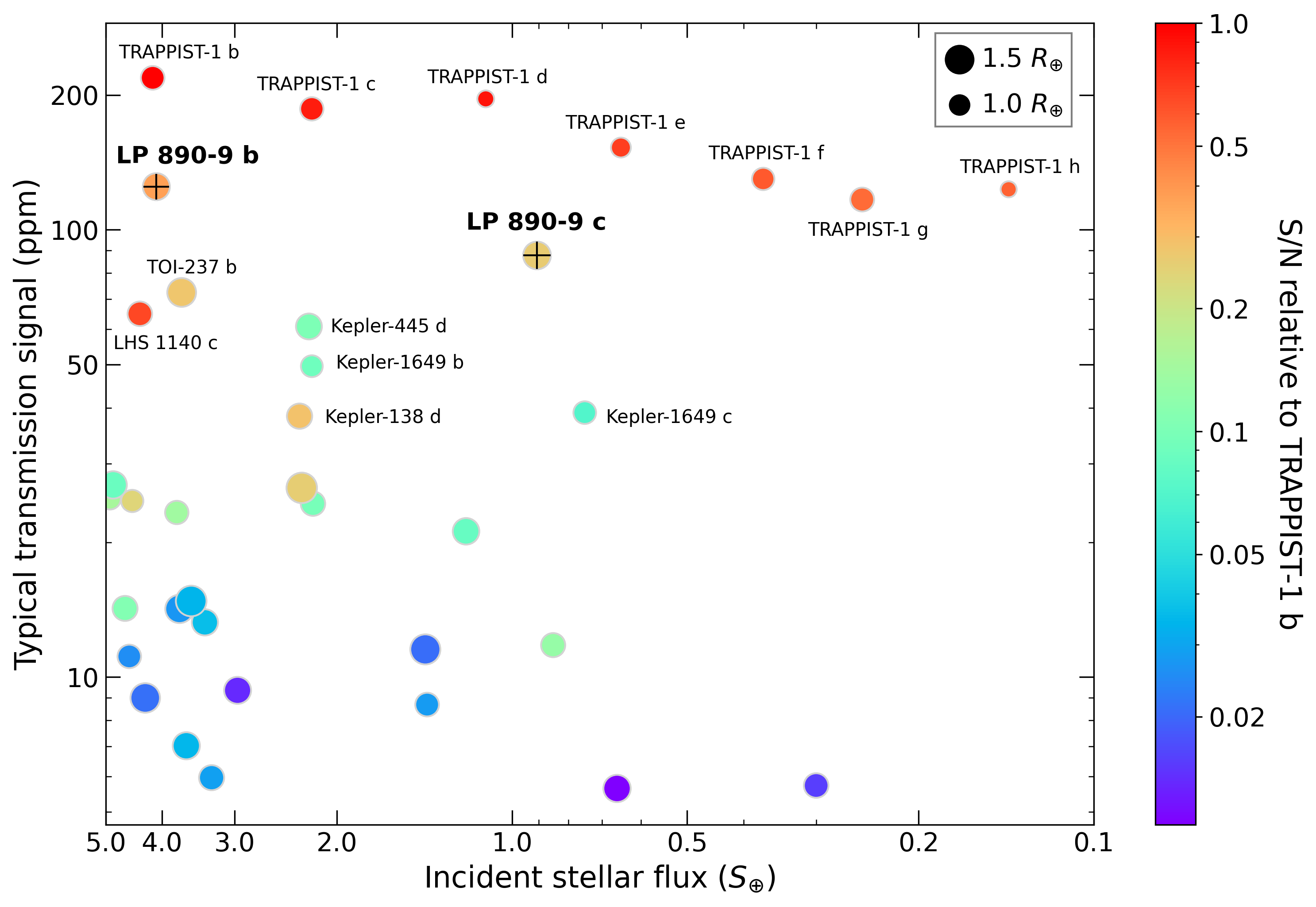}
    \caption{Expected transmission signals (see Sect. \ref{sec:atm} for details) of currently known transiting terrestrial exoplanets ($R_{\rm{p}}<1.6 \:R_\oplus$) as a function of their incident stellar flux. Data were taken from NASA Exoplanet Archive on 29 March 2022. We only show planets with mild incident irradiations, lower than five times that received by the Earth ($S_{\rm{p}}<5 \:S_\oplus$). The symbols are colour-coded as a function of their S/N relative to TRAPPIST-1\,b, obtained by scaling the signal amplitude with the hosts' brightness in the $J$-band and using TRAPPIST-1 b's S/N as a reference. The size of the symbols is proportional to the planetary radius. Transmission signals less than 5 ppm have been removed to enhance the readability of the figure.}
    \label{fig:atmSNR}
\end{figure*}

\subsection{Potential habitability}
\label{sec:HZ}

The habitable zone (HZ) is a concept that can be used to prioritise detected rocky exoplanets for follow-up observations. It is defined as the circular region around one or multiple stars, where liquid water is possible on the surface of a geologically-active rocky planet (e.g., \citealt{Kasting1993, 2013ApJ...765..131K, Kaltenegger2013, Haghighipour2013, Kane2013, Ramirez2016}). The width and distance of the HZ region depends in a first approximation on the incident stellar flux and the atmospheric composition of the planet. 

The limits of the conservative HZ is set by the greenhouse effect of CO$_{2}$ and H$_{2}$O vapour. At the outer edge, condensation and scattering by CO$_{2}$ outstrips its greenhouse capacity. At the inner edge, the mean surface temperatures exceed the critical point of water, triggering a runaway greenhouse environment and rapid water loss to space on short timescales (see, e.g., \citealt{Kasting1993, Kopparapu2014, Kaltenegger2017}). Modelling clouds is complex especially because no data exists that can provide comparison data sets for, e.g., fast rotating or slow rotating, very hot or very cold Earth-like planets. However, both 1D and 3D models generally find wider boundaries when considering cloud feedback than the original classical HZ but models differ in their specific results (see, e.g., the review in  \citealt{Kaltenegger2017}). The updated conservative HZ limits \citep{Kopparapu2014} have been scaled to the 3D models by \cite{Leconte2013} to account for some cloud feedback and are used here.

The concept of the empirical (or optimistic) HZ relies on observations in our own Solar System \citep{Kasting1993}. It is based on the incident flux on Venus and Mars when there were no more hints of liquid water on their surface. In our Solar System, that translates to 1.76 current Earth's incident irradiation ($S_{\oplus}$) for Venus around 1 billion years ago and 0.32 $S_{\oplus}$ for Mars about 3.8 billion years ago, providing empirical proxies for the incident flux limits of habitability. The inner edge of the empirical HZ is not well known because of the geological active history of Venus' surface. The outer edge of the HZ in 3D models agrees with incident stellar flux for Early Mars insulation. The empirical HZ provides an empirical dataset for the search for habitable planets, complementary to modelling efforts. 

Cool stars warm an Earth-like planet's surface more effectively than hotter stars (see, e.g., \citealt{Kasting1993}). Thus, for LP~890-9, the inner limits of the HZ are 1.49 $S_{\oplus}$ for the empirical and \hbox{0.92 $S_{\oplus}$} for the conservative HZ respectively \hbox{\citep{2013ApJ...765..131K}.} The outer limit of the HZ for LP~890-9 is found at a stellar flux of about 0.21 $S_{\oplus}$. While the HZ limits can change with atmospheric composition (e.g., \citealt{Pierrehumbert2011,Ramirez2017,Ramirez2018}), here we used the empirical HZ and the conservative HZ defined for a planet with an N$_2$-H$_2$O-CO$_2$ dominated atmosphere like Earth to identify the HZ limits for the LP~890-9 system. 

With an incident stellar flux of $4.09 \pm 0.12$ $S_{\oplus}$, LP~890-9\,b is too close to the star to be within the HZ. However, \hbox{LP~890-9\,c} receives $0.906 \pm 0.026$ $S_{\oplus}$, which places it within the limits of both the empirical and the conservative HZ. It orbits close to the inner edge of the conservative HZ, and in the inner third of the empirical HZ of LP~890-9 (see Fig. \ref{fig:Teff_vs_S}). Due to increasing stellar irradiation towards the inner edge of the HZ, LP~890-9\,c should develop a water-rich atmosphere because more surface water would evaporate than on Earth due to increased temperatures. Such an increase in water vapour in the atmosphere should be detectable with JWST and upcoming Extremely Large Telescopes (see Sect. \ref{sec:atm}).

While LP~890-9\,c orbits within the HZ of its host star, due to its position close to a cool M star, a potential slow rotation due to tidal forces could maintain habitability even for higher incident fluxes than the nominal limits of the HZ, due to an increased cloud coverage on the day-side of the planet. This makes planets that could be synchronously locked at the inner edge of the HZ, like LP~890-9\,c, very interesting test-cases for this hypothesis (see, e.g., \citealt{Yang2013, Wolf2014, Kopparapu2016}).

\subsection{Prospects for atmospheric characterisation}
\label{sec:atm}

Rocky exoplanets orbiting in the HZ of nearby late-type M dwarf stars provide unique opportunities for characterising their atmospheres as well as searching for biosignature gases. To assess the potential of the system for atmospheric characterisation in the context of other known transiting terrestrial planets, we followed the same approach as for K2-315\,b in \cite{2020AJ....160..172N}. We retrieved the currently known transiting exoplanets with a reported radius below 1.6 $R_\oplus$ from NASA Exoplanet Archive\footnote{\url{https://exoplanetarchive.ipac.caltech.edu/}}. We focussed here on temperate planets with mild incident irradiations, lower than five times that received by the Earth ($S_{\rm{p}}<5 \:S_\oplus$), and estimated their typical signal amplitude in transit transmission spectroscopy as:

\begin{align}
    \Delta \mathrm{d}F &= \frac{2R_{\rm{p}}h_{\rm{eff}}}{R_\star^2}
\end{align}

\noindent
where $h_{\rm{eff}}$ is the effective atmospheric height (that is, the extent of the atmospheric annulus probed in transmission). $h_{\rm{eff}}$ is directly proportional to the planet's atmospheric scale height $H$: $h_{\rm{eff}} = N_{\rm{H}}\:H = N_{\rm{H}}\:\frac{kT}{\mu g}$, where $k$ is Boltzmann's constant, $T$ is the atmospheric temperature, $\mu$ is the atmospheric mean molecular mass, and $g$ is the surface gravity. $N_{\rm{H}}$ is a number ranging typically between 6 and 10, depending on the presence of clouds, the spectral resolution and range covered \citep{2009ApJ...690.1056M,2013Sci...342.1473D}. We assumed here $N_{\rm{H}}=7$, $\mu=20$ amu (relevant for scenarios of secondary atmospheres), and $T$ to be equal to the equilibrium temperature, calculated for a sphere of null Bond albedo and an efficient heat redistribution. For planets with unknown masses, we estimated $g$ using the probabilistic mass--radius relationship of \cite{2017ApJ...834...17C}.

Fig. \ref{fig:atmSNR} shows the estimated transmission signals as a function of the planets' incident stellar flux. The points are colour-coded as a function of their S/N relative to TRAPPIST-1\,b, obtained by scaling the signal amplitude with the hosts' brightness in the $J$-band and using TRAPPIST-1 b's S/N as a reference. LP~890-9\,b compares well to the outer \hbox{TRAPPIST-1} planets in terms of potential for atmospheric characterisation, while LP~890-9\,c stands out as the second-most favourable habitable-zone terrestrial planet after the TRAPPIST-1 planets. The projected signals in transmission are $\sim$125 and $\sim$88 ppm for \hbox{LP~890-9\,b} and c, respectively, with corresponding signal-to-noise ratios relative to TRAPPIST-1\,b of 0.38 and 0.26. This means that assuming comparable secondary atmospheres, LP~890-9\,b and c would require respectively 7 and 15 times the amount of in-transit time to be characterised to the same extent as \hbox{TRAPPIST-1\,b}. Assessing the presence of a 10 bar CO$_2$ atmosphere with JWST/NIRSpec would thus require around 14 and 30 transits for planets b and c, respectively (scaled from \citealt{2019AJ....158...27L}). Repeating the same exercise to compare LP~890-9\,c with \hbox{TRAPPIST-1\,d} and e, which receive closer amounts of stellar flux, we find a S/N of 0.29 relative to TRAPPIST-1\,d and 0.38 relative to \hbox{TRAPPIST-1\,e}. Characterising LP~890-9\,c to the same extent as \hbox{TRAPPIST-1\,d} (resp. TRAPPIST-1\,e) would thus require 12 (resp. 7) times the amount of in-transit time. Based on these estimations, about 90 transits would be needed to detect a 1 bar H$_2$O atmosphere around LP~890-9\,c with JWST/NIRSpec (scaled again from \citealt{2019AJ....158...27L}). This rough comparison assumes a similar atmosphere for all planets. However, since \hbox{LP~890-9\,c} is located close to the inner limit (runaway greenhouse) of the conservative HZ (see Sect.~\ref{sec:HZ}), it may have a thicker H$_2$O atmosphere which would boost its atmospheric signals. 

\section{Conclusions} \label{sec:conclusion}

We have presented the discovery and initial characterisation of the LP~890-9 system, which hosts two temperate super-Earths transiting a nearby M6 dwarf. LP~890-9 is the second-coolest star found to host planets after TRAPPIST-1. The seven terrestrial planets orbiting TRAPPIST-1 have garnered the interest of a broad community across very diverse scientific disciplines: planet formation and evolution (e.g., \citealt{2017A&A...604A...1O,2019A&A...631A...7C,2022NatAs...6...80R}), star-planet interactions (e.g., \citealt{2017MNRAS.464.3728B,2017MNRAS.469L..26O,2018PNAS..115..260D}), multi-planet dynamics (e.g., \citealt{2018A&A...613A..68G,2021PSJ.....2....1A,2022A&A...658A.170T}), interior modelling (e.g., \citealt{2018ApJ...865...20D,2018A&A...613A..37B,2021AsBio..21.1325B}), atmospheric observations (e.g., \citealt{2016Natur.537...69D,2017AJ....154..121B,2018NatAs...2..214D}) and modelling (e.g., \citealt{2019AJ....158...27L,2020ApJ...898L..33P,2021MNRAS.505.3562L}), or climate predictions (e.g., \citealt{turbet2018,2021arXiv210911457T,2021arXiv210911459S}), among others. The discovery of the remarkable \hbox{LP~890-9} system presented in this work offers another rare opportunity to study temperate terrestrial planets around our smallest and coolest neighbours.

\begin{acknowledgements}

Funding for the TESS mission is provided by NASA's Science Mission Directorate. We acknowledge the use of public TESS data from pipelines at the TESS Science Office and at the TESS Science Processing Operations Center. This research has made use of the Exoplanet Follow-up Observation Program website, which is operated by the California Institute of Technology, under contract with the National Aeronautics and Space Administration under the Exoplanet Exploration Program. Resources supporting this work were provided by the NASA High-End Computing (HEC) Program through the NASA Advanced Supercomputing (NAS) Division at Ames Research Center for the production of the SPOC data products. This paper includes data collected by the TESS mission that are publicly available from the Mikulski Archive for Space Telescopes (MAST).
The research leading to these results has received funding from the European Research Council (ERC) under the FP/2007--2013 ERC grant agreement n$^{\circ}$ 336480, and under the European Union's Horizon 2020 research and innovation programme (grants agreements n$^{\circ}$ 679030 \& 803193/BEBOP); from an Action de Recherche Concert\'{e}e (ARC) grant, financed by the Wallonia--Brussels Federation, from the Balzan Prize Foundation, from the BELSPO/BRAIN2.0 research program (PORTAL project), from the Science and Technology Facilities Council (STFC; grants n$^\circ$ ST/S00193X/1, ST/00305/1, and ST/W000385/1), and from F.R.S-FNRS (Research Project ID T010920F). This work was also partially supported by a grant from the Simons Foundation (PI Queloz, grant number 327127), as well as by the MERAC foundation (PI Triaud).
TRAPPIST is funded by the Belgian Fund for Scientific Research (Fond National de la Recherche Scientifique, FNRS) under the grant PDR T.0120.21, with the participation of the Swiss National Science Fundation (SNF). 
This work is partly supported by MEXT/JSPS KAKENHI Grant Numbers JP15H02063, JP17H04574, JP18H05439, JP18H05442, JP19K14783, JP21H00035, JP21K13975, JP21K20376, JP22000005, Grant-in-Aid for JSPS Fellows Grant Number JP20J21872, JST CREST Grant Number JPMJCR1761, the Astrobiology Center of National Institutes of Natural Sciences (NINS) (Grant Numbers AB031010, AB031014), and
Social welfare juridical person SHIYUKAI (chairman MASAYUKI KAWASHIMA). This paper is based on data collected at the Subaru Telescope, which is located atop Maunakea and operated by the National Astronomical Observatory of Japan (NAOJ). We wish to recognise and acknowledge the very significant cultural role and reverence that the summit of Maunakea has always had within the indigenous Hawaiian community. This paper is based on observations made with the MuSCAT3 instrument, developed by the Astrobiology Center and under financial supports by JSPS KAKENHI (JP18H05439) and JST PRESTO (JPMJPR1775), at Faulkes
Telescope North on Maui, HI, operated by the Las Cumbres Observatory.
Some of the observations in the paper made use of the High-Resolution Imaging instrument Zorro obtained under Gemini LLP Proposal Number: GN/S-2021A-LP-105. Zorro was funded by the NASA Exoplanet Exploration Program and built at the NASA Ames Research Center by Steve B. Howell, Nic Scott, Elliott P. Horch, and Emmett Quigley. Zorro was mounted on the Gemini North (and/or South) telescope of the international Gemini Observatory, a program of NSF's OIR Lab, which is managed by the Association of Universities for Research in Astronomy (AURA) under a cooperative agreement with the National Science Foundation on behalf of the Gemini partnership: the National Science Foundation (United States), National Research Council (Canada), Agencia Nacional de Investigación y Desarrollo (Chile), Ministerio de Ciencia, Tecnología e Innovación (Argentina), Ministério da Ciência, Tecnologia, Inovações e Comunicações (Brazil), and Korea Astronomy and Space Science Institute (Republic of Korea).
We acknowledge funding from the European Research Council under the ERC Grant Agreement n. 337591-ExTrA.
This work has been carried out within the framework of the NCCR PlanetS supported by the Swiss National Science Foundation.
This work is based upon observations carried out at the Observatorio Astron\'omico Nacional at the Sierra de San Pedro M\'artir (OAN-SPM), Baja California, M\'exico. We warmly thank the entire
technical staff of the Observatorio Astron\'omico Nacional at San Pedro Mártir for their unfailing support to SAINT-EX operations.
Research at Lick Observatory is partially supported by a generous gift from Google.
L.D. is an F.R.S.-FNRS Postdoctoral Researcher. 
M.G. and E.J. are F.R.S.-FNRS Senior Research Associates. 
V.V.G. is an F.R.S.-FNRS Research Associate. 
B.V.R. thanks the Heising-Simons Foundation for support.
Y.G.M.C. acknowledges support from UNAM-PAPIIT IG-101321.
B.-O.D. acknowledges support from the Swiss National Science Foundation (PP00P2-163967 and PP00P2-190080).
M.N.G. acknowledges support from the European Space Agency (ESA) as an ESA Research Fellow.
A.H.M.J.T acknowledges funding from the European Research Council (ERC) under the European Union's Horizon 2020 research and innovation programme (grant agreement n$^\circ$ 803193/BEBOP), from the MERAC foundation, and from the Science and Technology Facilities Council (STFC; grants n$^\circ$ ST/S00193X/1, ST/00305/1, and ST/W000385/1).
E.D. acknowledges support from the innovation and research Horizon 2020 program in the context of the Marie Sklodowska-Curie subvention 945298.
V.K. acknowledges support from NSF award AST2009343.
This publication benefits from the support of the French Community of Belgium in the context of the FRIA Doctoral Grant awarded to M.T.
\end{acknowledgements}

\vspace{-0.4cm}


\bibliographystyle{aa} 
\bibliography{ref}

\begin{thebibliography}{205}
\expandafter\ifx\csname natexlab\endcsname\relax\def\natexlab#1{#1}\fi

\bibitem[{{Agol} {et~al.}(2021){Agol}, {Dorn}, {Grimm}, {Turbet}, {Ducrot},
  {Delrez}, {Gillon}, {Demory}, {Burdanov}, {Barkaoui}, {Benkhaldoun},
  {Bolmont}, {Burgasser}, {Carey}, {de Wit}, {Fabrycky}, {Foreman-Mackey},
  {Haldemann}, {Hernandez}, {Ingalls}, {Jehin}, {Langford}, {Leconte},
  {Lederer}, {Luger}, {Malhotra}, {Meadows}, {Morris}, {Pozuelos}, {Queloz},
  {Raymond}, {Selsis}, {Sestovic}, {Triaud}, \& {Van
  Grootel}}]{2021PSJ.....2....1A}
{Agol}, E., {Dorn}, C., {Grimm}, S.~L., {et~al.} 2021, The Planetary Science
  Journal, 2, 1

\bibitem[{{Aller} {et~al.}(2020){Aller}, {Lillo-Box}, {Jones}, {Miranda}, \&
  {Barcel{\'o} Forteza}}]{2020A&A...635A.128A}
{Aller}, A., {Lillo-Box}, J., {Jones}, D., {Miranda}, L.~F., \& {Barcel{\'o}
  Forteza}, S. 2020, \aap, 635, A128

\bibitem[{{Ambikasaran} {et~al.}(2015){Ambikasaran}, {Foreman-Mackey},
  {Greengard}, {Hogg}, \& {O'Neil}}]{2015ITPAM..38..252A}
{Ambikasaran}, S., {Foreman-Mackey}, D., {Greengard}, L., {Hogg}, D.~W., \&
  {O'Neil}, M. 2015, IEEE Transactions on Pattern Analysis and Machine
  Intelligence, 38, 252

\bibitem[{{Astropy Collaboration} {et~al.}(2018){Astropy Collaboration},
  {Price-Whelan}, {Sip{\H{o}}cz}, {G{\"u}nther}, {Lim}, {Crawford}, {Conseil},
  {Shupe}, {Craig}, {Dencheva}, {Ginsburg}, {VanderPlas}, {Bradley},
  {P{\'e}rez-Su{\'a}rez}, {de Val-Borro}, {Aldcroft}, {Cruz}, {Robitaille},
  {Tollerud}, {Ardelean}, {Babej}, {Bach}, {Bachetti}, {Bakanov}, {Bamford},
  {Barentsen}, {Barmby}, {Baumbach}, {Berry}, {Biscani}, {Boquien}, {Bostroem},
  {Bouma}, {Brammer}, {Bray}, {Breytenbach}, {Buddelmeijer}, {Burke},
  {Calderone}, {Cano Rodr{\'\i}guez}, {Cara}, {Cardoso}, {Cheedella}, {Copin},
  {Corrales}, {Crichton}, {D'Avella}, {Deil}, {Depagne}, {Dietrich}, {Donath},
  {Droettboom}, {Earl}, {Erben}, {Fabbro}, {Ferreira}, {Finethy}, {Fox},
  {Garrison}, {Gibbons}, {Goldstein}, {Gommers}, {Greco}, {Greenfield},
  {Groener}, {Grollier}, {Hagen}, {Hirst}, {Homeier}, {Horton}, {Hosseinzadeh},
  {Hu}, {Hunkeler}, {Ivezi{\'c}}, {Jain}, {Jenness}, {Kanarek}, {Kendrew},
  {Kern}, {Kerzendorf}, {Khvalko}, {King}, {Kirkby}, {Kulkarni}, {Kumar},
  {Lee}, {Lenz}, {Littlefair}, {Ma}, {Macleod}, {Mastropietro}, {McCully},
  {Montagnac}, {Morris}, {Mueller}, {Mumford}, {Muna}, {Murphy}, {Nelson},
  {Nguyen}, {Ninan}, {N{\"o}the}, {Ogaz}, {Oh}, {Parejko}, {Parley}, {Pascual},
  {Patil}, {Patil}, {Plunkett}, {Prochaska}, {Rastogi}, {Reddy Janga},
  {Sabater}, {Sakurikar}, {Seifert}, {Sherbert}, {Sherwood-Taylor}, {Shih},
  {Sick}, {Silbiger}, {Singanamalla}, {Singer}, {Sladen}, {Sooley},
  {Sornarajah}, {Streicher}, {Teuben}, {Thomas}, {Tremblay}, {Turner},
  {Terr{\'o}n}, {van Kerkwijk}, {de la Vega}, {Watkins}, {Weaver}, {Whitmore},
  {Woillez}, {Zabalza}, \& {Astropy Contributors}}]{2018AJ....156..123A}
{Astropy Collaboration}, {Price-Whelan}, A.~M., {Sip{\H{o}}cz}, B.~M., {et~al.}
  2018, \aj, 156, 123

\bibitem[{{Astropy Collaboration} {et~al.}(2013){Astropy Collaboration},
  {Robitaille}, {Tollerud}, {Greenfield}, {Droettboom}, {Bray}, {Aldcroft},
  {Davis}, {Ginsburg}, {Price-Whelan}, {Kerzendorf}, {Conley}, {Crighton},
  {Barbary}, {Muna}, {Ferguson}, {Grollier}, {Parikh}, {Nair}, {Unther},
  {Deil}, {Woillez}, {Conseil}, {Kramer}, {Turner}, {Singer}, {Fox}, {Weaver},
  {Zabalza}, {Edwards}, {Azalee Bostroem}, {Burke}, {Casey}, {Crawford},
  {Dencheva}, {Ely}, {Jenness}, {Labrie}, {Lim}, {Pierfederici}, {Pontzen},
  {Ptak}, {Refsdal}, {Servillat}, \& {Streicher}}]{2013A&A...558A..33A}
{Astropy Collaboration}, {Robitaille}, T.~P., {Tollerud}, E.~J., {et~al.} 2013,
  \aap, 558, A33

\bibitem[{{Bailey}(2012)}]{2012PASP..124.1015B}
{Bailey}, S. 2012, \pasp, 124, 1015

\bibitem[{{Bakos} {et~al.}(2006){Bakos}, {P{\'a}l}, {Latham}, {Noyes}, \&
  {Stefanik}}]{2006ApJ...641L..57B}
{Bakos}, G.~{\'A}., {P{\'a}l}, A., {Latham}, D.~W., {Noyes}, R.~W., \&
  {Stefanik}, R.~P. 2006, \apjl, 641, L57

\bibitem[{{Barclay} {et~al.}(2018){Barclay}, {Pepper}, \&
  {Quintana}}]{2018ApJS..239....2B}
{Barclay}, T., {Pepper}, J., \& {Quintana}, E.~V. 2018, \apjs, 239, 2

\bibitem[{Barnes(2017)}]{barnes2017}
Barnes, R. 2017, Celestial Mechanics and Dynamical Astronomy, 129, 509

\bibitem[{{Barr} {et~al.}(2018){Barr}, {Dobos}, \&
  {Kiss}}]{2018A&A...613A..37B}
{Barr}, A.~C., {Dobos}, V., \& {Kiss}, L.~L. 2018, \aap, 613, A37

\bibitem[{{Barth} {et~al.}(2021){Barth}, {Carone}, {Barnes}, {Noack},
  {Molli{\`e}re}, \& {Henning}}]{2021AsBio..21.1325B}
{Barth}, P., {Carone}, L., {Barnes}, R., {et~al.} 2021, Astrobiology, 21, 1325

\bibitem[{{Bensby} {et~al.}(2003){Bensby}, {Feltzing}, \&
  {Lundstr{\"o}m}}]{2003AA...410..527B}
{Bensby}, T., {Feltzing}, S., \& {Lundstr{\"o}m}, I. 2003, \aap, 410, 527

\bibitem[{{Blanco-Cuaresma} \& {Bolmont}(2017)}]{blancocuaresma2017}
{Blanco-Cuaresma}, S. \& {Bolmont}, E. 2017, in {IAU Symposium}, Vol. 325,
  {Astroinformatics}, ed. M.~{Brescia}, S.~G. {Djorgovski}, E.~D. {Feigelson},
  G.~{Longo}, \& S.~{Cavuoti}, 341--344

\bibitem[{{Bochanski} {et~al.}(2007){Bochanski}, {West}, {Hawley}, \&
  {Covey}}]{2007AJ....133..531B}
{Bochanski}, J.~J., {West}, A.~A., {Hawley}, S.~L., \& {Covey}, K.~R. 2007,
  \aj, 133, 531

\bibitem[{{Bolmont} {et~al.}(2020){Bolmont}, {Demory}, {Blanco-Cuaresma},
  {Agol}, {Grimm}, {Auclair-Desrotour}, {Selsis}, \& {Leleu}}]{bolmont2020}
{Bolmont}, E., {Demory}, B.~O., {Blanco-Cuaresma}, S., {et~al.} 2020, \aap,
  635, A117

\bibitem[{Bolmont {et~al.}(2014)Bolmont, Raymond, von Paris, Selsis, Hersant,
  Quintana, \& Barclay}]{bolmont2014}
Bolmont, E., Raymond, S.~N., von Paris, P., {et~al.} 2014, The Astrophysical
  Journal, 793, 3

\bibitem[{{Bolmont} {et~al.}(2017){Bolmont}, {Selsis}, {Owen}, {Ribas},
  {Raymond}, {Leconte}, \& {Gillon}}]{2017MNRAS.464.3728B}
{Bolmont}, E., {Selsis}, F., {Owen}, J.~E., {et~al.} 2017, \mnras, 464, 3728

\bibitem[{{Bonfils} {et~al.}(2015){Bonfils}, {Almenara}, {Jocou}, {Wunsche},
  {Kern}, {Delboulb{\'e}}, {Delfosse}, {Feautrier}, {Forveille}, {Gluck},
  {Lafrasse}, {Magnard}, {Maurel}, {Moulin}, {Murgas}, {Rabou}, {Rochat},
  {Roux}, \& {Stadler}}]{Bon2015}
{Bonfils}, X., {Almenara}, J.~M., {Jocou}, L., {et~al.} 2015, in Society of
  Photo-Optical Instrumentation Engineers (SPIE) Conference Series, Vol. 9605,
  Techniques and Instrumentation for Detection of Exoplanets VII, ed.
  S.~{Shaklan}, 96051L

\bibitem[{{Bourrier} {et~al.}(2017){Bourrier}, {de Wit}, {Bolmont},
  {Stamenkovi{\'c}}, {Wheatley}, {Burgasser}, {Delrez}, {Demory}, {Ehrenreich},
  {Gillon}, {Jehin}, {Leconte}, {Lederer}, {Lewis}, {Triaud}, \& {Van
  Grootel}}]{2017AJ....154..121B}
{Bourrier}, V., {de Wit}, J., {Bolmont}, E., {et~al.} 2017, \aj, 154, 121

\bibitem[{{Buder} {et~al.}(2021){Buder}, {Sharma}, {Kos}, {Amarsi},
  {Nordlander}, {Lind}, {Martell}, {Asplund}, {Bland-Hawthorn}, {Casey}, {de
  Silva}, {D'Orazi}, {Freeman}, {Hayden}, {Lewis}, {Lin}, {Schlesinger},
  {Simpson}, {Stello}, {Zucker}, {Zwitter}, {Beeson}, {Buck}, {Casagrande},
  {Clark}, {{\v{C}}otar}, {da Costa}, {de Grijs}, {Feuillet}, {Horner},
  {Kafle}, {Khanna}, {Kobayashi}, {Liu}, {Montet}, {Nandakumar}, {Nataf},
  {Ness}, {Spina}, {Tepper-Garc{\'\i}a}, {Ting}, {Traven},
  {Vogrin{\v{c}}i{\v{c}}}, {Wittenmyer}, {Wyse}, {{\v{Z}}erjal}, \& {GALAH
  Collaboration}}]{2021MNRAS.506..150B}
{Buder}, S., {Sharma}, S., {Kos}, J., {et~al.} 2021, \mnras, 506, 150

\bibitem[{{Burdanov} {et~al.}(2018){Burdanov}, {Delrez}, {Gillon}, \&
  {Jehin}}]{2018haex.bookE.130B}
{Burdanov}, A., {Delrez}, L., {Gillon}, M., \& {Jehin}, E. 2018, {SPECULOOS
  Exoplanet Search and Its Prototype on TRAPPIST}, ed. H.~J. {Deeg} \& J.~A.
  {Belmonte}, 130

\bibitem[{{Burgasser} \& {Mamajek}(2017)}]{2017ApJ...845..110B}
{Burgasser}, A.~J. \& {Mamajek}, E.~E. 2017, \apj, 845, 110

\bibitem[{{Burgasser} \& {Splat Development Team}(2017)}]{splat}
{Burgasser}, A.~J. \& {Splat Development Team}. 2017, in Astronomical Society
  of India Conference Series, Vol.~14, Astronomical Society of India Conference
  Series, 7--12

\bibitem[{{Chambers} {et~al.}(2016){Chambers}, {Magnier}, {Metcalfe},
  {Flewelling}, {Huber}, {Waters}, {Denneau}, {Draper}, {Farrow}, {Finkbeiner},
  {Holmberg}, {Koppenhoefer}, {Price}, {Rest}, {Saglia}, {Schlafly}, {Smartt},
  {Sweeney}, {Wainscoat}, {Burgett}, {Chastel}, {Grav}, {Heasley}, {Hodapp},
  {Jedicke}, {Kaiser}, {Kudritzki}, {Luppino}, {Lupton}, {Monet}, {Morgan},
  {Onaka}, {Shiao}, {Stubbs}, {Tonry}, {White}, {Ba{\~n}ados}, {Bell},
  {Bender}, {Bernard}, {Boegner}, {Boffi}, {Botticella}, {Calamida},
  {Casertano}, {Chen}, {Chen}, {Cole}, {Deacon}, {Frenk}, {Fitzsimmons},
  {Gezari}, {Gibbs}, {Goessl}, {Goggia}, {Gourgue}, {Goldman}, {Grant},
  {Grebel}, {Hambly}, {Hasinger}, {Heavens}, {Heckman}, {Henderson}, {Henning},
  {Holman}, {Hopp}, {Ip}, {Isani}, {Jackson}, {Keyes}, {Koekemoer}, {Kotak},
  {Le}, {Liska}, {Long}, {Lucey}, {Liu}, {Martin}, {Masci}, {McLean}, {Mindel},
  {Misra}, {Morganson}, {Murphy}, {Obaika}, {Narayan}, {Nieto-Santisteban},
  {Norberg}, {Peacock}, {Pier}, {Postman}, {Primak}, {Rae}, {Rai}, {Riess},
  {Riffeser}, {Rix}, {R{\"o}ser}, {Russel}, {Rutz}, {Schilbach}, {Schultz},
  {Scolnic}, {Strolger}, {Szalay}, {Seitz}, {Small}, {Smith}, {Soderblom},
  {Taylor}, {Thomson}, {Taylor}, {Thakar}, {Thiel}, {Thilker}, {Unger},
  {Urata}, {Valenti}, {Wagner}, {Walder}, {Walter}, {Watters}, {Werner},
  {Wood-Vasey}, \& {Wyse}}]{2016arXiv161205560C}
{Chambers}, K.~C., {Magnier}, E.~A., {Metcalfe}, N., {et~al.} 2016, arXiv
  e-prints, arXiv:1612.05560

\bibitem[{{Chen} \& {Kipping}(2017)}]{2017ApJ...834...17C}
{Chen}, J. \& {Kipping}, D. 2017, \apj, 834, 17

\bibitem[{{Ciardi} {et~al.}(2015){Ciardi}, {Beichman}, {Horch}, \&
  {Howell}}]{2015ApJ...805...16C}
{Ciardi}, D.~R., {Beichman}, C.~A., {Horch}, E.~P., \& {Howell}, S.~B. 2015,
  \apj, 805, 16

\bibitem[{{Cincotta} \& {Sim{\'o}}(1999)}]{cincottasimo1999}
{Cincotta}, P. \& {Sim{\'o}}, C. 1999, Celestial Mechanics and Dynamical
  Astronomy, 73, 195

\bibitem[{{Cincotta} {et~al.}(2003){Cincotta}, {Giordano}, \&
  {Sim{\'o}}}]{Cincotta2003}
{Cincotta}, P.~M., {Giordano}, C.~M., \& {Sim{\'o}}, C. 2003, Physica D
  Nonlinear Phenomena, 182, 151

\bibitem[{{Cincotta} \& {Sim{\'o}}(2000)}]{cincottasimo2000}
{Cincotta}, P.~M. \& {Sim{\'o}}, C. 2000, Astronomy and Astrophysics Supplement
  Series, 147, 205

\bibitem[{{Cloutier} {et~al.}(2020){Cloutier}, {Eastman}, {Rodriguez},
  {Astudillo-Defru}, {Bonfils}, {Mortier}, {Watson}, {Stalport}, {Pinamonti},
  {Lienhard}, {Harutyunyan}, {Damasso}, {Latham}, {Collins}, {Massey}, {Irwin},
  {Winters}, {Charbonneau}, {Ziegler}, {Matthews}, {Crossfield}, {Kreidberg},
  {Quinn}, {Ricker}, {Vanderspek}, {Seager}, {Winn}, {Jenkins}, {Vezie},
  {Udry}, {Twicken}, {Tenenbaum}, {Sozzetti}, {S{\'e}gransan}, {Schlieder},
  {Sasselov}, {Santos}, {Rice}, {Rackham}, {Poretti}, {Piotto}, {Phillips},
  {Pepe}, {Molinari}, {Mignon}, {Micela}, {Melo}, {de Medeiros}, {Mayor},
  {Matson}, {Martinez Fiorenzano}, {Mann}, {Magazz{\'u}}, {Lovis},
  {L{\'o}pez-Morales}, {Lopez}, {Lissauer}, {L{\'e}pine}, {Law}, {Kielkopf},
  {Johnson}, {Jensen}, {Howell}, {Gonzales}, {Ghedina}, {Forveille},
  {Figueira}, {Dumusque}, {Dressing}, {Doyon}, {D{\'\i}az}, {Fabrizio},
  {Delfosse}, {Cosentino}, {Conti}, {Collins}, {Cameron}, {Ciardi}, {Caldwell},
  {Burke}, {Buchhave}, {Brice{\~n}o}, {Boyd}, {Bouchy}, {Beichman}, {Artigau},
  \& {Almenara}}]{cloutier2020}
{Cloutier}, R., {Eastman}, J.~D., {Rodriguez}, J.~E., {et~al.} 2020, \aj, 160,
  3

\bibitem[{{Cointepas} {et~al.}(2021){Cointepas}, {Almenara}, {Bonfils},
  {Bouchy}, {Astudillo-Defru}, {Murgas}, {Otegi}, {Wyttenbach}, {Anderson},
  {Artigau}, {Canto Martins}, {Charbonneau}, {Collins}, {Collins}, {Correia},
  {Curaba}, {Delboulb{\'e}}, {Delfosse}, {D{\'\i}az}, {Dorn}, {Doyon},
  {Feautrier}, {Figueira}, {Forveille}, {Gaisne}, {Gan}, {Gluck}, {Helled},
  {Hellier}, {Jocou}, {Kern}, {Lafrasse}, {Law}, {Le{\~a}o}, {Lovis},
  {Magnard}, {Mann}, {Maurel}, {de Medeiros}, {Melo}, {Moulin}, {Pepe},
  {Rabou}, {Rochat}, {Rodriguez}, {Roux}, {Santos}, {S{\'e}gransan}, {Stadler},
  {Ting}, {Twicken}, {Udry}, {Waalkes}, {West}, {W{\"u}nsche}, {Ziegler},
  {Ricker}, {Vanderspek}, {Latham}, {Seager}, {Winn}, \& {Jenkins}}]{Coi2021}
{Cointepas}, M., {Almenara}, J.~M., {Bonfils}, X., {et~al.} 2021, \aap, 650,
  A145

\bibitem[{{Coleman} {et~al.}(2019){Coleman}, {Leleu}, {Alibert}, \&
  {Benz}}]{2019A&A...631A...7C}
{Coleman}, G.~A.~L., {Leleu}, A., {Alibert}, Y., \& {Benz}, W. 2019, \aap, 631,
  A7

\bibitem[{{Crossfield} {et~al.}(2019){Crossfield}, {Waalkes}, {Newton},
  {Narita}, {Muirhead}, {Ment}, {Matthews}, {Kraus}, {Kostov}, {Kosiarek},
  {Kane}, {Isaacson}, {Halverson}, {Gonzales}, {Everett}, {Dragomir},
  {Collins}, {Chontos}, {Berardo}, {Winters}, {Winn}, {Scott}, {Rojas-Ayala},
  {Rizzuto}, {Petigura}, {Peterson}, {Mocnik}, {Mikal-Evans}, {Mehrle},
  {Matson}, {Kuzuhara}, {Irwin}, {Huber}, {Huang}, {Howell}, {Howard},
  {Hirano}, {Fulton}, {Dupuy}, {Dressing}, {Dalba}, {Charbonneau}, {Burt},
  {Berta-Thompson}, {Benneke}, {Watanabe}, {Twicken}, {Tamura}, {Schlieder},
  {Seager}, {Rose}, {Ricker}, {Quintana}, {L{\'e}pine}, {Latham}, {Kotani},
  {Jenkins}, {Hori}, {Colon}, \& {Caldwell}}]{2019ApJ...883L..16C}
{Crossfield}, I. J.~M., {Waalkes}, W., {Newton}, E.~R., {et~al.} 2019, \apjl,
  883, L16

\bibitem[{{Cruz} \& {Reid}(2002)}]{2002AJ....123.2828C}
{Cruz}, K.~L. \& {Reid}, I.~N. 2002, \aj, 123, 2828

\bibitem[{{Cushing} {et~al.}(2004){Cushing}, {Vacca}, \& {Rayner}}]{spextool}
{Cushing}, M.~C., {Vacca}, W.~D., \& {Rayner}, J.~T. 2004, \pasp, 116, 362

\bibitem[{{Cutri} {et~al.}(2021){Cutri}, {Wright}, {Conrow}, {Fowler},
  {Eisenhardt}, {Grillmair}, {Kirkpatrick}, {Masci}, {McCallon}, {Wheelock},
  {Fajardo-Acosta}, {Yan}, {Benford}, {Harbut}, {Jarrett}, {Lake}, {Leisawitz},
  {Ressler}, {Stanford}, {Tsai}, {Liu}, {Helou}, {Mainzer}, {Gettngs},
  {Gonzalez}, {Hoffman}, {Marsh}, {Padgett}, {Skrutskie}, {Beck}, {Papin}, \&
  {Wittman}}]{2014yCat.2328....0C}
{Cutri}, R.~M., {Wright}, E.~L., {Conrow}, T., {et~al.} 2021, VizieR Online
  Data Catalog, II/328

\bibitem[{{de Wit} \& {Seager}(2013)}]{2013Sci...342.1473D}
{de Wit}, J. \& {Seager}, S. 2013, Science, 342, 1473

\bibitem[{{de Wit} {et~al.}(2016){de Wit}, {Wakeford}, {Gillon}, {Lewis},
  {Valenti}, {Demory}, {Burgasser}, {Burdanov}, {Delrez}, {Jehin}, {Lederer},
  {Queloz}, {Triaud}, \& {Van Grootel}}]{2016Natur.537...69D}
{de Wit}, J., {Wakeford}, H.~R., {Gillon}, M., {et~al.} 2016, \nat, 537, 69

\bibitem[{{de Wit} {et~al.}(2018){de Wit}, {Wakeford}, {Lewis}, {Delrez},
  {Gillon}, {Selsis}, {Leconte}, {Demory}, {Bolmont}, {Bourrier}, {Burgasser},
  {Grimm}, {Jehin}, {Lederer}, {Owen}, {Stamenkovi{\'c}}, \&
  {Triaud}}]{2018NatAs...2..214D}
{de Wit}, J., {Wakeford}, H.~R., {Lewis}, N.~K., {et~al.} 2018, Nature
  Astronomy, 2, 214

\bibitem[{{Delrez} {et~al.}(2021){Delrez}, {Ehrenreich}, {Alibert}, {Bonfanti},
  {Borsato}, {Fossati}, {Hooton}, {Hoyer}, {Pozuelos}, {Salmon}, {Sulis},
  {Wilson}, {Adibekyan}, {Bourrier}, {Brandeker}, {Charnoz}, {Deline},
  {Guterman}, {Haldemann}, {Hara}, {Oshagh}, {Sousa}, {Van Grootel}, {Alonso},
  {Anglada-Escud{\'e}}, {B{\'a}rczy}, {Barrado}, {Barros}, {Baumjohann},
  {Beck}, {Bekkelien}, {Benz}, {Billot}, {Bonfils}, {Broeg}, {Cabrera},
  {Collier Cameron}, {Davies}, {Deleuil}, {Delisle}, {Demangeon}, {Demory},
  {Erikson}, {Fortier}, {Fridlund}, {Futyan}, {Gandolfi}, {Garcia Mu{\~n}oz},
  {Gillon}, {Guedel}, {Heng}, {Kiss}, {Laskar}, {Lecavelier des Etangs},
  {Lendl}, {Lovis}, {Maxted}, {Nascimbeni}, {Olofsson}, {Osborn}, {Pagano},
  {Pall{\'e}}, {Piotto}, {Pollacco}, {Queloz}, {Rauer}, {Ragazzoni}, {Ribas},
  {Santos}, {Scandariato}, {S{\'e}gransan}, {Simon}, {Smith}, {Steller},
  {Szab{\'o}}, {Thomas}, {Udry}, \& {Walton}}]{delrez2021}
{Delrez}, L., {Ehrenreich}, D., {Alibert}, Y., {et~al.} 2021, Nature Astronomy,
  5, 775

\bibitem[{{Delrez} {et~al.}(2018){Delrez}, {Gillon}, {Queloz}, {Demory},
  {Almleaky}, {de Wit}, {Jehin}, {Triaud}, {Barkaoui}, {Burdanov}, {Burgasser},
  {Ducrot}, {McCormac}, {Murray}, {Silva Fernandes}, {Sohy}, {Thompson}, {Van
  Grootel}, {Alonso}, {Benkhaldoun}, \& {Rebolo}}]{2018SPIE10700E..1ID}
{Delrez}, L., {Gillon}, M., {Queloz}, D., {et~al.} 2018, in Society of
  Photo-Optical Instrumentation Engineers (SPIE) Conference Series, Vol. 10700,
  Ground-based and Airborne Telescopes VII, ed. H.~K. {Marshall} \&
  J.~{Spyromilio}, 107001I

\bibitem[{{Demory} {et~al.}(2020){Demory}, {Pozuelos}, {G{\'o}mez Maqueo Chew},
  {Sabin}, {Petrucci}, {Schroffenegger}, {Grimm}, {Sestovic}, {Gillon},
  {McCormac}, {Barkaoui}, {Benz}, {Bieryla}, {Bouchy}, {Burdanov}, {Collins},
  {de Wit}, {Dressing}, {Garcia}, {Giacalone}, {Guerra}, {Haldemann}, {Heng},
  {Jehin}, {Jofr{\'e}}, {Kane}, {Lillo-Box}, {Maign{\'e}}, {Mordasini},
  {Morris}, {Niraula}, {Queloz}, {Rackham}, {Savel}, {Soubkiou}, {Srdoc},
  {Stassun}, {Triaud}, {Zambelli}, {Ricker}, {Latham}, {Seager}, {Winn},
  {Jenkins}, {Calvario-Vel{\'a}squez}, {Franco Herrera}, {Colorado}, {Cadena
  Zepeda}, {Figueroa}, {Watson}, {Lugo-Ibarra}, {Carigi}, {Guisa}, {Herrera},
  {Sierra D{\'\i}az}, {Su{\'a}rez}, {Barrado}, {Batalha}, {Benkhaldoun},
  {Chontos}, {Dai}, {Essack}, {Ghachoui}, {Huang}, {Huber}, {Isaacson},
  {Lissauer}, {Morales-Calder{\'o}n}, {Robertson}, {Roy}, {Twicken},
  {Vanderburg}, \& {Weiss}}]{2020A&A...642A..49D}
{Demory}, B.~O., {Pozuelos}, F.~J., {G{\'o}mez Maqueo Chew}, Y., {et~al.} 2020,
  \aap, 642, A49

\bibitem[{{Dhital} {et~al.}(2012){Dhital}, {West}, {Stassun}, {Bochanski},
  {Massey}, \& {Bastien}}]{2012AJ....143...67D}
{Dhital}, S., {West}, A.~A., {Stassun}, K.~G., {et~al.} 2012, \aj, 143, 67

\bibitem[{{Dittmann} {et~al.}(2017){Dittmann}, {Irwin}, {Charbonneau},
  {Bonfils}, {Astudillo-Defru}, {Haywood}, {Berta-Thompson}, {Newton},
  {Rodriguez}, {Winters}, {Tan}, {Almenara}, {Bouchy}, {Delfosse}, {Forveille},
  {Lovis}, {Murgas}, {Pepe}, {Santos}, {Udry}, {W{\"u}nsche}, {Esquerdo},
  {Latham}, \& {Dressing}}]{2017Natur.544..333D}
{Dittmann}, J.~A., {Irwin}, J.~M., {Charbonneau}, D., {et~al.} 2017, \nat, 544,
  333

\bibitem[{{Donati} {et~al.}(2020){Donati}, {Kouach}, {Moutou}, {Doyon},
  {Delfosse}, {Artigau}, {Baratchart}, {Lacombe}, {Barrick}, {H{\'e}brard},
  {Bouchy}, {Saddlemyer}, {Par{\`e}s}, {Rabou}, {Micheau}, {Dolon}, {Reshetov},
  {Challita}, {Carmona}, {Striebig}, {Thibault}, {Martioli}, {Cook},
  {Fouqu{\'e}}, {Vermeulen}, {Wang}, {Arnold}, {Pepe}, {Boisse}, {Figueira},
  {Bouvier}, {Ray}, {Feugeade}, {Morin}, {Alencar}, {Hobson}, {Castilho},
  {Udry}, {Santos}, {Hernandez}, {Benedict}, {Vall{\'e}e}, {Gallou}, {Dupieux},
  {Larrieu}, {Perruchot}, {Sottile}, {Moreau}, {Usher}, {Baril}, {Wildi},
  {Chazelas}, {Malo}, {Bonfils}, {Loop}, {Kerley}, {Wevers}, {Dunn}, {Pazder},
  {Macdonald}, {Dubois}, {Carri{\'e}}, {Valentin}, {Henault}, {Yan}, \&
  {Steinmetz}}]{donati20}
{Donati}, J.~F., {Kouach}, D., {Moutou}, C., {et~al.} 2020, \mnras, 498, 5684

\bibitem[{{Dong} {et~al.}(2018){Dong}, {Jin}, {Lingam}, {Airapetian}, {Ma}, \&
  {van der Holst}}]{2018PNAS..115..260D}
{Dong}, C., {Jin}, M., {Lingam}, M., {et~al.} 2018, Proceedings of the National
  Academy of Science, 115, 260

\bibitem[{{Dorn} {et~al.}(2018){Dorn}, {Mosegaard}, {Grimm}, \&
  {Alibert}}]{2018ApJ...865...20D}
{Dorn}, C., {Mosegaard}, K., {Grimm}, S.~L., \& {Alibert}, Y. 2018, \apj, 865,
  20

\bibitem[{{Dorn} {et~al.}(2014){Dorn}, {Anglada-Escude}, {Baade}, {Bristow},
  {Follert}, {Gojak}, {Grunhut}, {Hatzes}, {Heiter}, {Hilker}, {Ives}, {Jung},
  {K{\"a}ufl}, {Kerber}, {Klein}, {Lizon}, {Lockhart}, {L{\"o}winger},
  {Marquart}, {Oliva}, {Origlia}, {Pasquini}, {Paufique}, {Piskunov}, {Pozna},
  {Reiners}, {Smette}, {Smoker}, {Seemann}, {Stempels}, \& {Valenti}}]{dorn14}
{Dorn}, R.~J., {Anglada-Escude}, G., {Baade}, D., {et~al.} 2014, The Messenger,
  156, 7

\bibitem[{{Douglas} {et~al.}(2014){Douglas}, {Ag{\"u}eros}, {Covey}, {Bowsher},
  {Bochanski}, {Cargile}, {Kraus}, {Law}, {Lemonias}, {Arce}, {Fierroz}, \&
  {Kundert}}]{2014ApJ...795..161D}
{Douglas}, S.~T., {Ag{\"u}eros}, M.~A., {Covey}, K.~R., {et~al.} 2014, \apj,
  795, 161

\bibitem[{{Dreizler} {et~al.}(2020){Dreizler}, {Jeffers}, {Rodr{\'\i}guez},
  {Zechmeister}, {Barnes}, {Haswell}, {Coleman}, {Lalitha}, {Hidalgo Soto},
  {Strachan}, {Hambsch}, {L{\'o}pez-Gonz{\'a}lez}, {Morales}, {Rodr{\'\i}guez
  L{\'o}pez}, {Berdi{\~n}as}, {Ribas}, {Pall{\'e}}, {Reiners}, \&
  {Anglada-Escud{\'e}}}]{2020MNRAS.493..536D}
{Dreizler}, S., {Jeffers}, S.~V., {Rodr{\'\i}guez}, E., {et~al.} 2020, \mnras,
  493, 536

\bibitem[{Dévora-Pajares \& Pozuelos(2022)}]{matrix2022}
Dévora-Pajares, M. \& Pozuelos, F.~J. 2022, {MATRIX: Multi-phAse Transits
  Recovery from Injected eXoplanets}

\bibitem[{Eggleton {et~al.}(1998)Eggleton, Kiseleva, \&
  Hut}]{Eggleton1998TheFriction}
Eggleton, P.~P., Kiseleva, L.~G., \& Hut, P. 1998, The Astrophysical Journal,
  499, 853

\bibitem[{{Eisner} {et~al.}(2020){Eisner}, {Barrag{\'a}n}, {Aigrain},
  {Lintott}, {Miller}, {Zicher}, {Boyajian}, {Brice{\~n}o}, {Bryant},
  {Christiansen}, {Feinstein}, {Flor-Torres}, {Fridlund}, {Gandolfi},
  {Gilbert}, {Guerrero}, {Jenkins}, {Jones}, {Kristiansen}, {Vanderburg},
  {Law}, {L{\'o}pez-S{\'a}nchez}, {Mann}, {Safron}, {Schwamb}, {Stassun},
  {Osborn}, {Wang}, {Zic}, {Ziegler}, {Barnet}, {Bean}, {Bundy}, {Chetnik},
  {Dawson}, {Garstone}, {Stenner}, {Huten}, {Larish}, {Melanson}, {Mitchell},
  {Moore}, {Peltsch}, {Rogers}, {Schuster}, {Smith}, {Simister}, {Tanner},
  {Terentev}, \& {Tsymbal}}]{eisner2020}
{Eisner}, N.~L., {Barrag{\'a}n}, O., {Aigrain}, S., {et~al.} 2020, \mnras, 494,
  750

\bibitem[{{Espinoza} {et~al.}(2019){Espinoza}, {Kossakowski}, \&
  {Brahm}}]{2019MNRAS.490.2262E}
{Espinoza}, N., {Kossakowski}, D., \& {Brahm}, R. 2019, \mnras, 490, 2262

\bibitem[{{Fernandes} {et~al.}(2019){Fernandes}, {Van Grootel}, {Salmon},
  {Aringer}, {Burgasser}, {Scuflaire}, {Brassard}, \&
  {Fontaine}}]{2019ApJ...879...94F}
{Fernandes}, C.~S., {Van Grootel}, V., {Salmon}, S. J.~A.~J., {et~al.} 2019,
  \apj, 879, 94

\bibitem[{{Foreman-Mackey}(2016)}]{2016JOSS....1...24F}
{Foreman-Mackey}, D. 2016, The Journal of Open Source Software, 1, 24

\bibitem[{{Foreman-Mackey} {et~al.}(2014){Foreman-Mackey}, {Hogg}, \&
  {Morton}}]{2014ApJ...795...64F}
{Foreman-Mackey}, D., {Hogg}, D.~W., \& {Morton}, T.~D. 2014, \apj, 795, 64

\bibitem[{{Fukui} {et~al.}(2011){Fukui}, {Narita}, {Tristram}, {Sumi}, {Abe},
  {Itow}, {Sullivan}, {Bond}, {Hirano}, {Tamura}, {Bennett}, {Furusawa},
  {Hayashi}, {Hearnshaw}, {Hosaka}, {Kamiya}, {Kobara}, {Korpela}, {Kilmartin},
  {Lin}, {Ling}, {Makita}, {Masuda}, {Matsubara}, {Miyake}, {Muraki}, {Nagaya},
  {Nishimoto}, {Ohnishi}, {Omori}, {Perrott}, {Rattenbury}, {Saito}, {Skuljan},
  {Suzuki}, {Sweatman}, \& {Wada}}]{2011PASJ...63..287F}
{Fukui}, A., {Narita}, N., {Tristram}, P.~J., {et~al.} 2011, \pasj, 63, 287

\bibitem[{{Fulton} {et~al.}(2018){Fulton}, {Petigura}, {Blunt}, \&
  {Sinukoff}}]{2018PASP..130d4504F}
{Fulton}, B.~J., {Petigura}, E.~A., {Blunt}, S., \& {Sinukoff}, E. 2018, \pasp,
  130, 044504

\bibitem[{{Gaia Collaboration} {et~al.}(2018){Gaia Collaboration}, {Brown},
  {Vallenari}, {Prusti}, {de Bruijne}, {Babusiaux}, {Bailer-Jones}, {Biermann},
  {Evans}, {Eyer}, {Jansen}, {Jordi}, {Klioner}, {Lammers}, {Lindegren},
  {Luri}, {Mignard}, {Panem}, {Pourbaix}, {Randich}, {Sartoretti}, {Siddiqui},
  {Soubiran}, {van Leeuwen}, {Walton}, {Arenou}, {Bastian}, {Cropper},
  {Drimmel}, {Katz}, {Lattanzi}, {Bakker}, {Cacciari}, {Casta{\~n}eda},
  {Chaoul}, {Cheek}, {De Angeli}, {Fabricius}, {Guerra}, {Holl}, {Masana},
  {Messineo}, {Mowlavi}, {Nienartowicz}, {Panuzzo}, {Portell}, {Riello},
  {Seabroke}, {Tanga}, {Th{\'e}venin}, {Gracia-Abril}, {Comoretto},
  {Garcia-Reinaldos}, {Teyssier}, {Altmann}, {Andrae}, {Audard},
  {Bellas-Velidis}, {Benson}, {Berthier}, {Blomme}, {Burgess}, {Busso},
  {Carry}, {Cellino}, {Clementini}, {Clotet}, {Creevey}, {Davidson}, {De
  Ridder}, {Delchambre}, {Dell'Oro}, {Ducourant},
  {Fern{\'a}ndez-Hern{\'a}ndez}, {Fouesneau}, {Fr{\'e}mat}, {Galluccio},
  {Garc{\'\i}a-Torres}, {Gonz{\'a}lez-N{\'u}{\~n}ez}, {Gonz{\'a}lez-Vidal},
  {Gosset}, {Guy}, {Halbwachs}, {Hambly}, {Harrison}, {Hern{\'a}ndez},
  {Hestroffer}, {Hodgkin}, {Hutton}, {Jasniewicz}, {Jean-Antoine-Piccolo},
  {Jordan}, {Korn}, {Krone-Martins}, {Lanzafame}, {Lebzelter}, {L{\"o}ffler},
  {Manteiga}, {Marrese}, {Mart{\'\i}n-Fleitas}, {Moitinho}, {Mora}, {Muinonen},
  {Osinde}, {Pancino}, {Pauwels}, {Petit}, {Recio-Blanco}, {Richards},
  {Rimoldini}, {Robin}, {Sarro}, {Siopis}, {Smith}, {Sozzetti}, {S{\"u}veges},
  {Torra}, {van Reeven}, {Abbas}, {Abreu Aramburu}, {Accart}, {Aerts},
  {Altavilla}, {{\'A}lvarez}, {Alvarez}, {Alves}, {Anderson}, {Andrei},
  {Anglada Varela}, {Antiche}, {Antoja}, {Arcay}, {Astraatmadja}, {Bach},
  {Baker}, {Balaguer-N{\'u}{\~n}ez}, {Balm}, {Barache}, {Barata}, {Barbato},
  {Barblan}, {Barklem}, {Barrado}, {Barros}, {Barstow}, {Bartholom{\'e}
  Mu{\~n}oz}, {Bassilana}, {Becciani}, {Bellazzini}, {Berihuete}, {Bertone},
  {Bianchi}, {Bienaym{\'e}}, {Blanco-Cuaresma}, {Boch}, {Boeche}, {Bombrun},
  {Borrachero}, {Bossini}, {Bouquillon}, {Bourda}, {Bragaglia}, {Bramante},
  {Breddels}, {Bressan}, {Brouillet}, {Br{\"u}semeister}, {Brugaletta},
  {Bucciarelli}, {Burlacu}, {Busonero}, {Butkevich}, {Buzzi}, {Caffau},
  {Cancelliere}, {Cannizzaro}, {Cantat-Gaudin}, {Carballo}, {Carlucci},
  {Carrasco}, {Casamiquela}, {Castellani}, {Castro-Ginard}, {Charlot},
  {Chemin}, {Chiavassa}, {Cocozza}, {Costigan}, {Cowell}, {Crifo}, {Crosta},
  {Crowley}, {Cuypers}, {Dafonte}, {Damerdji}, {Dapergolas}, {David}, {David},
  {de Laverny}, {De Luise}, {De March}, {de Martino}, {de Souza}, {de Torres},
  {Debosscher}, {del Pozo}, {Delbo}, {Delgado}, {Delgado}, {Di Matteo},
  {Diakite}, {Diener}, {Distefano}, {Dolding}, {Drazinos}, {Dur{\'a}n},
  {Edvardsson}, {Enke}, {Eriksson}, {Esquej}, {Eynard Bontemps}, {Fabre},
  {Fabrizio}, {Faigler}, {Falc{\~a}o}, {Farr{\`a}s Casas}, {Federici},
  {Fedorets}, {Fernique}, {Figueras}, {Filippi}, {Findeisen}, {Fonti},
  {Fraile}, {Fraser}, {Fr{\'e}zouls}, {Gai}, {Galleti}, {Garabato},
  {Garc{\'\i}a-Sedano}, {Garofalo}, {Garralda}, {Gavel}, {Gavras}, {Gerssen},
  {Geyer}, {Giacobbe}, {Gilmore}, {Girona}, {Giuffrida}, {Glass}, {Gomes},
  {Granvik}, {Gueguen}, {Guerrier}, {Guiraud}, {Guti{\'e}rrez-S{\'a}nchez},
  {Haigron}, {Hatzidimitriou}, {Hauser}, {Haywood}, {Heiter}, {Helmi}, {Heu},
  {Hilger}, {Hobbs}, {Hofmann}, {Holland}, {Huckle}, {Hypki}, {Icardi},
  {Jan{\ss}en}, {Jevardat de Fombelle}, {Jonker}, {Juh{\'a}sz}, {Julbe},
  {Karampelas}, {Kewley}, {Klar}, {Kochoska}, {Kohley}, {Kolenberg},
  {Kontizas}, {Kontizas}, {Koposov}, {Kordopatis}, {Kostrzewa-Rutkowska},
  {Koubsky}, {Lambert}, {Lanza}, {Lasne}, {Lavigne}, {Le Fustec}, {Le
  Poncin-Lafitte}, {Lebreton}, {Leccia}, {Leclerc}, {Lecoeur-Taibi},
  {Lenhardt}, {Leroux}, {Liao}, {Licata}, {Lindstr{\o}m}, {Lister}, {Livanou},
  {Lobel}, {L{\'o}pez}, {Managau}, {Mann}, {Mantelet}, {Marchal}, {Marchant},
  {Marconi}, {Marinoni}, {Marschalk{\'o}}, {Marshall}, {Martino}, {Marton},
  {Mary}, {Massari}, {Matijevi{\v{c}}}, {Mazeh}, {McMillan}, {Messina},
  {Michalik}, {Millar}, {Molina}, {Molinaro}, {Moln{\'a}r}, {Montegriffo},
  {Mor}, {Morbidelli}, {Morel}, {Morris}, {Mulone}, {Muraveva}, {Musella},
  {Nelemans}, {Nicastro}, {Noval}, {O'Mullane}, {Ord{\'e}novic},
  {Ord{\'o}{\~n}ez-Blanco}, {Osborne}, {Pagani}, {Pagano}, {Pailler},
  {Palacin}, {Palaversa}, {Panahi}, {Pawlak}, {Piersimoni}, {Pineau}, {Plachy},
  {Plum}, {Poggio}, {Poujoulet}, {Pr{\v{s}}a}, {Pulone}, {Racero}, {Ragaini},
  {Rambaux}, {Ramos-Lerate}, {Regibo}, {Reyl{\'e}}, {Riclet}, {Ripepi}, {Riva},
  {Rivard}, {Rixon}, {Roegiers}, {Roelens}, {Romero-G{\'o}mez}, {Rowell},
  {Royer}, {Ruiz-Dern}, {Sadowski}, {Sagrist{\`a} Sell{\'e}s}, {Sahlmann},
  {Salgado}, {Salguero}, {Sanna}, {Santana-Ros}, {Sarasso}, {Savietto},
  {Schultheis}, {Sciacca}, {Segol}, {Segovia}, {S{\'e}gransan}, {Shih},
  {Siltala}, {Silva}, {Smart}, {Smith}, {Solano}, {Solitro}, {Sordo}, {Soria
  Nieto}, {Souchay}, {Spagna}, {Spoto}, {Stampa}, {Steele},
  {Steidelm{\"u}ller}, {Stephenson}, {Stoev}, {Suess}, {Surdej}, {Szabados},
  {Szegedi-Elek}, {Tapiador}, {Taris}, {Tauran}, {Taylor}, {Teixeira},
  {Terrett}, {Teyssandier}, {Thuillot}, {Titarenko}, {Torra Clotet}, {Turon},
  {Ulla}, {Utrilla}, {Uzzi}, {Vaillant}, {Valentini}, {Valette}, {van Elteren},
  {Van Hemelryck}, {van Leeuwen}, {Vaschetto}, {Vecchiato}, {Veljanoski},
  {Viala}, {Vicente}, {Vogt}, {von Essen}, {Voss}, {Votruba}, {Voutsinas},
  {Walmsley}, {Weiler}, {Wertz}, {Wevers}, {Wyrzykowski}, {Yoldas},
  {{\v{Z}}erjal}, {Ziaeepour}, {Zorec}, {Zschocke}, {Zucker}, {Zurbach}, \&
  {Zwitter}}]{2018A&A...616A...1G}
{Gaia Collaboration}, {Brown}, A.~G.~A., {Vallenari}, A., {et~al.} 2018, \aap,
  616, A1

\bibitem[{{Gaia Collaboration} {et~al.}(2021){Gaia Collaboration}, {Brown},
  {Vallenari}, {Prusti}, {de Bruijne}, {Babusiaux}, {Biermann}, {Creevey},
  {Evans}, {Eyer}, {Hutton}, {Jansen}, {Jordi}, {Klioner}, {Lammers},
  {Lindegren}, {Luri}, {Mignard}, {Panem}, {Pourbaix}, {Randich}, {Sartoretti},
  {Soubiran}, {Walton}, {Arenou}, {Bailer-Jones}, {Bastian}, {Cropper},
  {Drimmel}, {Katz}, {Lattanzi}, {van Leeuwen}, {Bakker}, {Cacciari},
  {Casta{\~n}eda}, {De Angeli}, {Ducourant}, {Fabricius}, {Fouesneau},
  {Fr{\'e}mat}, {Guerra}, {Guerrier}, {Guiraud}, {Jean-Antoine Piccolo},
  {Masana}, {Messineo}, {Mowlavi}, {Nicolas}, {Nienartowicz}, {Pailler},
  {Panuzzo}, {Riclet}, {Roux}, {Seabroke}, {Sordo}, {Tanga}, {Th{\'e}venin},
  {Gracia-Abril}, {Portell}, {Teyssier}, {Altmann}, {Andrae}, {Bellas-Velidis},
  {Benson}, {Berthier}, {Blomme}, {Brugaletta}, {Burgess}, {Busso}, {Carry},
  {Cellino}, {Cheek}, {Clementini}, {Damerdji}, {Davidson}, {Delchambre},
  {Dell'Oro}, {Fern{\'a}ndez-Hern{\'a}ndez}, {Galluccio}, {Garc{\'\i}a-Lario},
  {Garcia-Reinaldos}, {Gonz{\'a}lez-N{\'u}{\~n}ez}, {Gosset}, {Haigron},
  {Halbwachs}, {Hambly}, {Harrison}, {Hatzidimitriou}, {Heiter},
  {Hern{\'a}ndez}, {Hestroffer}, {Hodgkin}, {Holl}, {Jan{\ss}en}, {Jevardat de
  Fombelle}, {Jordan}, {Krone-Martins}, {Lanzafame}, {L{\"o}ffler}, {Lorca},
  {Manteiga}, {Marchal}, {Marrese}, {Moitinho}, {Mora}, {Muinonen}, {Osborne},
  {Pancino}, {Pauwels}, {Petit}, {Recio-Blanco}, {Richards}, {Riello},
  {Rimoldini}, {Robin}, {Roegiers}, {Rybizki}, {Sarro}, {Siopis}, {Smith},
  {Sozzetti}, {Ulla}, {Utrilla}, {van Leeuwen}, {van Reeven}, {Abbas}, {Abreu
  Aramburu}, {Accart}, {Aerts}, {Aguado}, {Ajaj}, {Altavilla}, {{\'A}lvarez},
  {{\'A}lvarez Cid-Fuentes}, {Alves}, {Anderson}, {Anglada Varela}, {Antoja},
  {Audard}, {Baines}, {Baker}, {Balaguer-N{\'u}{\~n}ez}, {Balbinot}, {Balog},
  {Barache}, {Barbato}, {Barros}, {Barstow}, {Bartolom{\'e}}, {Bassilana},
  {Bauchet}, {Baudesson-Stella}, {Becciani}, {Bellazzini}, {Bernet}, {Bertone},
  {Bianchi}, {Blanco-Cuaresma}, {Boch}, {Bombrun}, {Bossini}, {Bouquillon},
  {Bragaglia}, {Bramante}, {Breedt}, {Bressan}, {Brouillet}, {Bucciarelli},
  {Burlacu}, {Busonero}, {Butkevich}, {Buzzi}, {Caffau}, {Cancelliere},
  {C{\'a}novas}, {Cantat-Gaudin}, {Carballo}, {Carlucci}, {Carnerero},
  {Carrasco}, {Casamiquela}, {Castellani}, {Castro-Ginard}, {Castro Sampol},
  {Chaoul}, {Charlot}, {Chemin}, {Chiavassa}, {Cioni}, {Comoretto}, {Cooper},
  {Cornez}, {Cowell}, {Crifo}, {Crosta}, {Crowley}, {Dafonte}, {Dapergolas},
  {David}, {David}, {de Laverny}, {De Luise}, {De March}, {De Ridder}, {de
  Souza}, {de Teodoro}, {de Torres}, {del Peloso}, {del Pozo}, {Delbo},
  {Delgado}, {Delgado}, {Delisle}, {Di Matteo}, {Diakite}, {Diener},
  {Distefano}, {Dolding}, {Eappachen}, {Edvardsson}, {Enke}, {Esquej}, {Fabre},
  {Fabrizio}, {Faigler}, {Fedorets}, {Fernique}, {Fienga}, {Figueras},
  {Fouron}, {Fragkoudi}, {Fraile}, {Franke}, {Gai}, {Garabato},
  {Garcia-Gutierrez}, {Garc{\'\i}a-Torres}, {Garofalo}, {Gavras}, {Gerlach},
  {Geyer}, {Giacobbe}, {Gilmore}, {Girona}, {Giuffrida}, {Gomel}, {Gomez},
  {Gonzalez-Santamaria}, {Gonz{\'a}lez-Vidal}, {Granvik},
  {Guti{\'e}rrez-S{\'a}nchez}, {Guy}, {Hauser}, {Haywood}, {Helmi}, {Hidalgo},
  {Hilger}, {H{\l}adczuk}, {Hobbs}, {Holland}, {Huckle}, {Jasniewicz},
  {Jonker}, {Juaristi Campillo}, {Julbe}, {Karbevska}, {Kervella}, {Khanna},
  {Kochoska}, {Kontizas}, {Kordopatis}, {Korn}, {Kostrzewa-Rutkowska},
  {Kruszy{\'n}ska}, {Lambert}, {Lanza}, {Lasne}, {Le Campion}, {Le Fustec},
  {Lebreton}, {Lebzelter}, {Leccia}, {Leclerc}, {Lecoeur-Taibi}, {Liao},
  {Licata}, {Lindstr{\o}m}, {Lister}, {Livanou}, {Lobel}, {Madrero Pardo},
  {Managau}, {Mann}, {Marchant}, {Marconi}, {Marcos Santos}, {Marinoni},
  {Marocco}, {Marshall}, {Martin Polo}, {Mart{\'\i}n-Fleitas}, {Masip},
  {Massari}, {Mastrobuono-Battisti}, {Mazeh}, {McMillan}, {Messina},
  {Michalik}, {Millar}, {Mints}, {Molina}, {Molinaro}, {Moln{\'a}r},
  {Montegriffo}, {Mor}, {Morbidelli}, {Morel}, {Morris}, {Mulone}, {Munoz},
  {Muraveva}, {Murphy}, {Musella}, {Noval}, {Ord{\'e}novic}, {Orr{\`u}},
  {Osinde}, {Pagani}, {Pagano}, {Palaversa}, {Palicio}, {Panahi}, {Pawlak},
  {Pe{\~n}alosa Esteller}, {Penttil{\"a}}, {Piersimoni}, {Pineau}, {Plachy},
  {Plum}, {Poggio}, {Poretti}, {Poujoulet}, {Pr{\v{s}}a}, {Pulone}, {Racero},
  {Ragaini}, {Rainer}, {Raiteri}, {Rambaux}, {Ramos}, {Ramos-Lerate}, {Re
  Fiorentin}, {Regibo}, {Reyl{\'e}}, {Ripepi}, {Riva}, {Rixon}, {Robichon},
  {Robin}, {Roelens}, {Rohrbasser}, {Romero-G{\'o}mez}, {Rowell}, {Royer},
  {Rybicki}, {Sadowski}, {Sagrist{\`a} Sell{\'e}s}, {Sahlmann}, {Salgado},
  {Salguero}, {Samaras}, {Sanchez Gimenez}, {Sanna}, {Santove{\~n}a},
  {Sarasso}, {Schultheis}, {Sciacca}, {Segol}, {Segovia}, {S{\'e}gransan},
  {Semeux}, {Shahaf}, {Siddiqui}, {Siebert}, {Siltala}, {Slezak}, {Smart},
  {Solano}, {Solitro}, {Souami}, {Souchay}, {Spagna}, {Spoto}, {Steele},
  {Steidelm{\"u}ller}, {Stephenson}, {S{\"u}veges}, {Szabados}, {Szegedi-Elek},
  {Taris}, {Tauran}, {Taylor}, {Teixeira}, {Thuillot}, {Tonello}, {Torra},
  {Torra}, {Turon}, {Unger}, {Vaillant}, {van Dillen}, {Vanel}, {Vecchiato},
  {Viala}, {Vicente}, {Voutsinas}, {Weiler}, {Wevers}, {Wyrzykowski}, {Yoldas},
  {Yvard}, {Zhao}, {Zorec}, {Zucker}, {Zurbach}, \& {Zwitter}}]{gaia_edr3}
{Gaia Collaboration}, {Brown}, A.~G.~A., {Vallenari}, A., {et~al.} 2021, \aap,
  650, C3

\bibitem[{{Gan} {et~al.}(2022){Gan}, {Soubkiou}, {Wang}, {Benkhaldoun}, {Mao},
  {Artigau}, {Fouqu{\'e}}, {Arnold}, {Giacalone}, {Theissen}, {Aganze},
  {Burgasser}, {Collins}, {Shporer}, {Barkaoui}, {Ghachoui}, {Howell},
  {Lamman}, {Demangeon}, {Burdanov}, {Cadieux}, {Chouqar}, {Collins}, {Cook},
  {Delrez}, {Demory}, {Doyon}, {Dransfield}, {Dressing}, {Ducrot}, {Fan},
  {Garcia}, {Gill}, {Gillon}, {Gnilka}, {G{\'o}mez Maqueo Chew}, {G{\"u}nther},
  {Henze}, {Huang}, {Jehin}, {Jensen}, {Lin}, {Manset}, {McCormac}, {Murray},
  {Niraula}, {Pedersen}, {Pozuelos}, {Queloz}, {Rackham}, {Savel}, {Schanche},
  {Schwarz}, {Sebastian}, {Thompson}, {Timmermans}, {Triaud}, {Vezie}, {Wells},
  {de Wit}, {Ricker}, {Vanderspek}, {Latham}, {Seager}, {Winn}, \&
  {Jenkins}}]{2022MNRAS.514.4120G}
{Gan}, T., {Soubkiou}, A., {Wang}, S.~X., {et~al.} 2022, \mnras, 514, 4120

\bibitem[{{Garcia} {et~al.}(2022){Garcia}, {Timmermans}, {Pozuelos}, {Ducrot},
  {Gillon}, {Delrez}, {Wells}, \& {Jehin}}]{2022MNRAS.509.4817G}
{Garcia}, L.~J., {Timmermans}, M., {Pozuelos}, F.~J., {et~al.} 2022, \mnras,
  509, 4817

\bibitem[{{Gelman} \& {Rubin}(1992)}]{1992StaSc...7..457G}
{Gelman}, A. \& {Rubin}, D.~B. 1992, Statistical Science, 7, 457

\bibitem[{Geman \& Geman(1984)}]{4767596}
Geman, S. \& Geman, D. 1984, IEEE Transactions on Pattern Analysis and Machine
  Intelligence, PAMI-6, 721

\bibitem[{{Giacalone} \& {Dressing}(2020)}]{2020ascl.soft02004G}
{Giacalone}, S. \& {Dressing}, C.~D. 2020, {triceratops: Candidate exoplanet
  rating tool}

\bibitem[{{Giacalone} {et~al.}(2021){Giacalone}, {Dressing}, {Jensen},
  {Collins}, {Ricker}, {Vanderspek}, {Seager}, {Winn}, {Jenkins}, {Barclay},
  {Barkaoui}, {Cadieux}, {Charbonneau}, {Collins}, {Conti}, {Doyon}, {Evans},
  {Ghachoui}, {Gillon}, {Guerrero}, {Hart}, {Jehin}, {Kielkopf}, {McLean},
  {Murgas}, {Palle}, {Parviainen}, {Pozuelos}, {Relles}, {Shporer}, {Socia},
  {Stockdale}, {Tan}, {Torres}, {Twicken}, {Waalkes}, \&
  {Waite}}]{2021AJ....161...24G}
{Giacalone}, S., {Dressing}, C.~D., {Jensen}, E. L.~N., {et~al.} 2021, \aj,
  161, 24

\bibitem[{{Gillon}(2018)}]{2018NatAs...2..344G}
{Gillon}, M. 2018, Nature Astronomy, 2, 344

\bibitem[{{Gillon} {et~al.}(2014){Gillon}, {Demory}, {Madhusudhan}, {Deming},
  {Seager}, {Zsom}, {Knutson}, {Lanotte}, {Bonfils}, {D{\'e}sert}, {Delrez},
  {Jehin}, {Fraine}, {Magain}, \& {Triaud}}]{2014A&A...563A..21G}
{Gillon}, M., {Demory}, B.~O., {Madhusudhan}, N., {et~al.} 2014, \aap, 563, A21

\bibitem[{{Gillon} {et~al.}(2013){Gillon}, {Jehin}, {Fumel}, {Magain}, \&
  {Queloz}}]{2013EPJWC..4703001G}
{Gillon}, M., {Jehin}, E., {Fumel}, A., {Magain}, P., \& {Queloz}, D. 2013, in
  European Physical Journal Web of Conferences, Vol.~47, European Physical
  Journal Web of Conferences, 03001

\bibitem[{{Gillon} {et~al.}(2016){Gillon}, {Jehin}, {Lederer}, {Delrez}, {de
  Wit}, {Burdanov}, {Van Grootel}, {Burgasser}, {Triaud}, {Opitom}, {Demory},
  {Sahu}, {Bardalez Gagliuffi}, {Magain}, \& {Queloz}}]{2016Natur.533..221G}
{Gillon}, M., {Jehin}, E., {Lederer}, S.~M., {et~al.} 2016, \nat, 533, 221

\bibitem[{{Gillon} {et~al.}(2011){Gillon}, {Jehin}, {Magain}, {Chantry},
  {Hutsem{\'e}kers}, {Manfroid}, {Queloz}, \& {Udry}}]{2011EPJWC..1106002G}
{Gillon}, M., {Jehin}, E., {Magain}, P., {et~al.} 2011, in European Physical
  Journal Web of Conferences, Vol.~11, European Physical Journal Web of
  Conferences, 06002

\bibitem[{{Gillon} {et~al.}(2020){Gillon}, {Meadows}, {Agol}, {Burgasser},
  {Deming}, {Doyon}, {Fortney}, {Kreidberg}, {Owen}, {Selsis}, {de Wit},
  {Lustig-Yaeger}, \& {Rackham}}]{2020arXiv200204798G}
{Gillon}, M., {Meadows}, V., {Agol}, E., {et~al.} 2020, arXiv e-prints,
  arXiv:2002.04798

\bibitem[{{Gillon} {et~al.}(2017){Gillon}, {Triaud}, {Demory}, {Jehin}, {Agol},
  {Deck}, {Lederer}, {de Wit}, {Burdanov}, {Ingalls}, {Bolmont}, {Leconte},
  {Raymond}, {Selsis}, {Turbet}, {Barkaoui}, {Burgasser}, {Burleigh}, {Carey},
  {Chaushev}, {Copperwheat}, {Delrez}, {Fernandes}, {Holdsworth}, {Kotze}, {Van
  Grootel}, {Almleaky}, {Benkhaldoun}, {Magain}, \&
  {Queloz}}]{2017Natur.542..456G}
{Gillon}, M., {Triaud}, A. H.~M.~J., {Demory}, B.-O., {et~al.} 2017, \nat, 542,
  456

\bibitem[{{Gillon} {et~al.}(2012){Gillon}, {Triaud}, {Fortney}, {Demory},
  {Jehin}, {Lendl}, {Magain}, {Kabath}, {Queloz}, {Alonso}, {Anderson},
  {Collier Cameron}, {Fumel}, {Hebb}, {Hellier}, {Lanotte}, {Maxted},
  {Mowlavi}, \& {Smalley}}]{2012A&A...542A...4G}
{Gillon}, M., {Triaud}, A.~H.~M.~J., {Fortney}, J.~J., {et~al.} 2012, \aap,
  542, A4

\bibitem[{{Grimm} {et~al.}(2018){Grimm}, {Demory}, {Gillon}, {Dorn}, {Agol},
  {Burdanov}, {Delrez}, {Sestovic}, {Triaud}, {Turbet}, {Bolmont}, {Caldas},
  {de Wit}, {Jehin}, {Leconte}, {Raymond}, {Van Grootel}, {Burgasser}, {Carey},
  {Fabrycky}, {Heng}, {Hernandez}, {Ingalls}, {Lederer}, {Selsis}, \&
  {Queloz}}]{2018A&A...613A..68G}
{Grimm}, S.~L., {Demory}, B.-O., {Gillon}, M., {et~al.} 2018, \aap, 613, A68

\bibitem[{{Guerrero} {et~al.}(2021){Guerrero}, {Seager}, {Huang}, {Vanderburg},
  {Garcia Soto}, {Mireles}, {Hesse}, {Fong}, {Glidden}, {Shporer}, {Latham},
  {Collins}, {Quinn}, {Burt}, {Dragomir}, {Crossfield}, {Vanderspek},
  {Fausnaugh}, {Burke}, {Ricker}, {Daylan}, {Essack}, {G{\"u}nther}, {Osborn},
  {Pepper}, {Rowden}, {Sha}, {Villanueva}, {Yahalomi}, {Yu}, {Ballard},
  {Batalha}, {Berardo}, {Chontos}, {Dittmann}, {Esquerdo}, {Mikal-Evans},
  {Jayaraman}, {Krishnamurthy}, {Louie}, {Mehrle}, {Niraula}, {Rackham},
  {Rodriguez}, {Rowden}, {Sousa-Silva}, {Watanabe}, {Wong}, {Zhan},
  {Zivanovic}, {Christiansen}, {Ciardi}, {Swain}, {Lund}, {Mullally},
  {Fleming}, {Rodriguez}, {Boyd}, {Quintana}, {Barclay}, {Col{\'o}n},
  {Rinehart}, {Schlieder}, {Clampin}, {Jenkins}, {Twicken}, {Caldwell},
  {Coughlin}, {Henze}, {Lissauer}, {Morris}, {Rose}, {Smith}, {Tenenbaum},
  {Ting}, {Wohler}, {Bakos}, {Bean}, {Berta-Thompson}, {Bieryla}, {Bouma},
  {Buchhave}, {Butler}, {Charbonneau}, {Doty}, {Ge}, {Holman}, {Howard},
  {Kaltenegger}, {Kane}, {Kjeldsen}, {Kreidberg}, {Lin}, {Minsky}, {Narita},
  {Paegert}, {P{\'a}l}, {Palle}, {Sasselov}, {Spencer}, {Sozzetti}, {Stassun},
  {Torres}, {Udry}, \& {Winn}}]{2021ApJS..254...39G}
{Guerrero}, N.~M., {Seager}, S., {Huang}, C.~X., {et~al.} 2021, \apjs, 254, 39

\bibitem[{{G{\"u}nther} {et~al.}(2019){G{\"u}nther}, {Pozuelos}, {Dittmann},
  {Dragomir}, {Kane}, {Daylan}, {Feinstein}, {Huang}, {Morton}, {Bonfanti},
  {Bouma}, {Burt}, {Collins}, {Lissauer}, {Matthews}, {Montet}, {Vanderburg},
  {Wang}, {Winters}, {Ricker}, {Vanderspek}, {Latham}, {Seager}, {Winn},
  {Jenkins}, {Armstrong}, {Barkaoui}, {Batalha}, {Bean}, {Caldwell}, {Ciardi},
  {Collins}, {Crossfield}, {Fausnaugh}, {Furesz}, {Gan}, {Gillon}, {Guerrero},
  {Horne}, {Howell}, {Ireland}, {Isopi}, {Jehin}, {Kielkopf}, {Lepine},
  {Mallia}, {Matson}, {Myers}, {Palle}, {Quinn}, {Relles}, {Rojas-Ayala},
  {Schlieder}, {Sefako}, {Shporer}, {Su{\'a}rez}, {Tan}, {Ting}, {Twicken}, \&
  {Waite}}]{2019NatAs...3.1099G}
{G{\"u}nther}, M.~N., {Pozuelos}, F.~J., {Dittmann}, J.~A., {et~al.} 2019,
  Nature Astronomy, 3, 1099

\bibitem[{{Hadden}(2019)}]{hadden2019}
{Hadden}, S. 2019, {shadden/TTV2Fast2Furious: First release of
  TTV2Fast2Furious}

\bibitem[{Haghighipour \& Kaltenegger(2013)}]{Haghighipour2013}
Haghighipour, N. \& Kaltenegger, L. 2013, The Astrophysical Journal, 777, 166

\bibitem[{{Hamuy} {et~al.}(1994){Hamuy}, {Suntzeff}, {Heathcote}, {Walker},
  {Gigoux}, \& {Phillips}}]{1994PASP..106..566H}
{Hamuy}, M., {Suntzeff}, N.~B., {Heathcote}, S.~R., {et~al.} 1994, \pasp, 106,
  566

\bibitem[{{Hamuy} {et~al.}(1992){Hamuy}, {Walker}, {Suntzeff}, {Gigoux},
  {Heathcote}, \& {Phillips}}]{1992PASP..104..533H}
{Hamuy}, M., {Walker}, A.~R., {Suntzeff}, N.~B., {et~al.} 1992, \pasp, 104, 533

\bibitem[{{Hastings}(1970)}]{1970Bimka..57...97H}
{Hastings}, W.~K. 1970, Biometrika, 57, 97

\bibitem[{{Hinse} {et~al.}(2015){Hinse}, {Haghighipour}, {Kostov}, \&
  {Go{\'z}dziewski}}]{Hinse2015}
{Hinse}, T.~C., {Haghighipour}, N., {Kostov}, V.~B., \& {Go{\'z}dziewski}, K.
  2015, \apj, 799, 88

\bibitem[{{Hippke} {et~al.}(2019){Hippke}, {David}, {Mulders}, \&
  {Heller}}]{wotan2019}
{Hippke}, M., {David}, T.~J., {Mulders}, G.~D., \& {Heller}, R. 2019, \aj, 158,
  143

\bibitem[{{Hippke} \& {Heller}(2019)}]{hippke_TLS_2019}
{Hippke}, M. \& {Heller}, R. 2019, \aap, 623, A39

\bibitem[{{Hirano} {et~al.}(2020){Hirano}, {Kuzuhara}, {Kotani}, {Omiya},
  {Kudo}, {Harakawa}, {Vievard}, {Kurokawa}, {Nishikawa}, {Tamura}, {Hodapp},
  {Ishizuka}, {Jacobson}, {Konishi}, {Serizawa}, {Ueda}, {Gaidos}, \&
  {Sato}}]{2020PASJ...72...93H}
{Hirano}, T., {Kuzuhara}, M., {Kotani}, T., {et~al.} 2020, \pasj, 72, 93

\bibitem[{{Howell} {et~al.}(2016){Howell}, {Everett}, {Horch}, {Winters},
  {Hirsch}, {Nusdeo}, \& {Scott}}]{2016ApJ...829L...2H}
{Howell}, S.~B., {Everett}, M.~E., {Horch}, E.~P., {et~al.} 2016, \apjl, 829,
  L2

\bibitem[{{Howell} {et~al.}(2011){Howell}, {Everett}, {Sherry}, {Horch}, \&
  {Ciardi}}]{2011AJ....142...19H}
{Howell}, S.~B., {Everett}, M.~E., {Sherry}, W., {Horch}, E., \& {Ciardi},
  D.~R. 2011, \aj, 142, 19

\bibitem[{Husser {et~al.}(2013)Husser, {Wende-von Berg}, Dreizler, Homeier,
  Reiners, Barman, \& Hauschildt}]{Husser2013}
Husser, T.-O., {Wende-von Berg}, S., Dreizler, S., {et~al.} 2013, A{\&}A, 553,
  A6

\bibitem[{Hut(1981)}]{Hut1981TidalSystems.}
Hut, P. 1981, Astronomy and Astrophysics, 99, 126

\bibitem[{{Irwin} {et~al.}(2004){Irwin}, {Lewis}, {Hodgkin}, {Bunclark},
  {Evans}, {McMahon}, {Emerson}, {Stewart}, \& {Beard}}]{2004SPIE.5493..411I}
{Irwin}, M.~J., {Lewis}, J., {Hodgkin}, S., {et~al.} 2004, in Society of
  Photo-Optical Instrumentation Engineers (SPIE) Conference Series, Vol. 5493,
  Optimizing Scientific Return for Astronomy through Information Technologies,
  ed. P.~J. {Quinn} \& A.~{Bridger}, 411--422

\bibitem[{{Jehin} {et~al.}(2018){Jehin}, {Gillon}, {Queloz}, {Delrez},
  {Burdanov}, {Murray}, {Sohy}, {Ducrot}, {Sebastian}, {Thompson}, {McCormac},
  {Almleaky}, {Burgasser}, {Demory}, {de Wit}, {Barkaoui}, {Pozuelos},
  {Triaud}, \& {Van Grootel}}]{2018Msngr.174....2J}
{Jehin}, E., {Gillon}, M., {Queloz}, D., {et~al.} 2018, The Messenger, 174, 2

\bibitem[{{Jehin} {et~al.}(2011){Jehin}, {Gillon}, {Queloz}, {Magain},
  {Manfroid}, {Chantry}, {Lendl}, {Hutsem{\'e}kers}, \&
  {Udry}}]{2011Msngr.145....2J}
{Jehin}, E., {Gillon}, M., {Queloz}, D., {et~al.} 2011, The Messenger, 145, 2

\bibitem[{{Jenkins}(2002)}]{2002ApJ...575..493J}
{Jenkins}, J.~M. 2002, \apj, 575, 493

\bibitem[{{Jenkins} {et~al.}(2010){Jenkins}, {Caldwell}, {Chandrasekaran},
  {Twicken}, {Bryson}, {Quintana}, {Clarke}, {Li}, {Allen}, {Tenenbaum}, {Wu},
  {Klaus}, {Middour}, {Cote}, {McCauliff}, {Girouard}, {Gunter}, {Wohler},
  {Sommers}, {Hall}, {Uddin}, {Wu}, {Bhavsar}, {Van Cleve}, {Pletcher},
  {Dotson}, {Haas}, {Gilliland}, {Koch}, \& {Borucki}}]{2010ApJ...713L..87J}
{Jenkins}, J.~M., {Caldwell}, D.~A., {Chandrasekaran}, H., {et~al.} 2010,
  \apjl, 713, L87

\bibitem[{{Jenkins} {et~al.}(2020){Jenkins}, {Tenenbaum}, {Seader}, {Burke},
  {McCauliff}, {Smith}, {Twicken}, \& {Chandrasekaran}}]{Jenkins2020}
{Jenkins}, J.~M., {Tenenbaum}, P., {Seader}, S., {et~al.} 2020, {Kepler Data
  Processing Handbook: Transiting Planet Search}, Kepler Science Document
  KSCI-19081-003

\bibitem[{{Jenkins} {et~al.}(2016){Jenkins}, {Twicken}, {McCauliff},
  {Campbell}, {Sanderfer}, {Lung}, {Mansouri-Samani}, {Girouard}, {Tenenbaum},
  {Klaus}, {Smith}, {Caldwell}, {Chacon}, {Henze}, {Heiges}, {Latham},
  {Morgan}, {Swade}, {Rinehart}, \& {Vanderspek}}]{2016SPIE.9913E..3EJ}
{Jenkins}, J.~M., {Twicken}, J.~D., {McCauliff}, S., {et~al.} 2016, in Society
  of Photo-Optical Instrumentation Engineers (SPIE) Conference Series, Vol.
  9913, Software and Cyberinfrastructure for Astronomy IV, ed. G.~{Chiozzi} \&
  J.~C. {Guzman}, 99133E

\bibitem[{{Jenkins} {et~al.}(2019){Jenkins}, {Pozuelos}, {Tuomi},
  {Berdi{\~n}as}, {D{\'\i}az}, {Vines}, {Su{\'a}rez}, \& {Pe{\~n}a
  Rojas}}]{Jenkins2019}
{Jenkins}, J.~S., {Pozuelos}, F.~J., {Tuomi}, M., {et~al.} 2019, \mnras, 490,
  5585

\bibitem[{Kaltenegger(2017)}]{Kaltenegger2017}
Kaltenegger, L. 2017, Annual Review of Astronomy and Astrophysics, 55, 433

\bibitem[{Kaltenegger \& Haghighipour(2013)}]{Kaltenegger2013}
Kaltenegger, L. \& Haghighipour, N. 2013, The Astrophysical Journal, 777, 165

\bibitem[{{Kaltenegger} \& {Traub}(2009)}]{2009ApJ...698..519K}
{Kaltenegger}, L. \& {Traub}, W.~A. 2009, \apj, 698, 519

\bibitem[{Kane \& Hinkel(2013)}]{Kane2013}
Kane, S.~R. \& Hinkel, N.~R. 2013, The Astrophysical Journal, 762, 7

\bibitem[{Kasting {et~al.}(1993)Kasting, Whitmire, \& Reynolds}]{Kasting1993}
Kasting, J.~F., Whitmire, D.~P., \& Reynolds, R.~T. 1993, Icarus, 101, 108

\bibitem[{{Kipping}(2013)}]{2013MNRAS.435.2152K}
{Kipping}, D.~M. 2013, \mnras, 435, 2152

\bibitem[{{Kirkpatrick} {et~al.}(2010){Kirkpatrick}, {Looper}, {Burgasser},
  {Schurr}, {Cutri}, {Cushing}, {Cruz}, {Sweet}, {Knapp}, {Barman},
  {Bochanski}, {Roellig}, {McLean}, {McGovern}, \& {Rice}}]{Kirkpatrick2010}
{Kirkpatrick}, J.~D., {Looper}, D.~L., {Burgasser}, A.~J., {et~al.} 2010,
  \apjs, 190, 100

\bibitem[{Kopparapu {et~al.}(2016)Kopparapu, Wolf, Haqq-Misra, Yang, Kasting,
  Meadows, Terrien, \& Mahadevan}]{Kopparapu2016}
Kopparapu, Ravi~Kumar, R., Wolf, E.~T., Haqq-Misra, J., {et~al.} 2016, The
  Astrophysical Journal, 819, 84

\bibitem[{{Kopparapu} {et~al.}(2013){Kopparapu}, {Ramirez}, {Kasting}, {Eymet},
  {Robinson}, {Mahadevan}, {Terrien}, {Domagal-Goldman}, {Meadows}, \&
  {Deshpande}}]{2013ApJ...765..131K}
{Kopparapu}, R.~K., {Ramirez}, R., {Kasting}, J.~F., {et~al.} 2013, \apj, 765,
  131

\bibitem[{Kopparapu {et~al.}(2014)Kopparapu, Ramirez, SchottelKotte, Kasting,
  Domagal-Goldman, \& Eymet}]{Kopparapu2014}
Kopparapu, R.~K., Ramirez, R.~M., SchottelKotte, J., {et~al.} 2014, The
  Astrophysical Journal, 787, L29

\bibitem[{{Kostov} {et~al.}(2019){Kostov}, {Schlieder}, {Barclay}, {Quintana},
  {Col{\'o}n}, {Brande}, {Collins}, {Feinstein}, {Hadden}, {Kane}, {Kreidberg},
  {Kruse}, {Lam}, {Matthews}, {Montet}, {Pozuelos}, {Stassun}, {Winters},
  {Ricker}, {Vanderspek}, {Latham}, {Seager}, {Winn}, {Jenkins}, {Afanasev},
  {Armstrong}, {Arney}, {Boyd}, {Barentsen}, {Barkaoui}, {Batalha}, {Beichman},
  {Bayliss}, {Burke}, {Burdanov}, {Cacciapuoti}, {Carson}, {Charbonneau},
  {Christiansen}, {Ciardi}, {Clampin}, {Collins}, {Conti}, {Coughlin},
  {Covone}, {Crossfield}, {Delrez}, {Domagal-Goldman}, {Dressing}, {Ducrot},
  {Essack}, {Everett}, {Fauchez}, {Foreman-Mackey}, {Gan}, {Gilbert}, {Gillon},
  {Gonzales}, {Hamann}, {Hedges}, {Hocutt}, {Hoffman}, {Horch}, {Horne},
  {Howell}, {Hynes}, {Ireland}, {Irwin}, {Isopi}, {Jensen}, {Jehin},
  {Kaltenegger}, {Kielkopf}, {Kopparapu}, {Lewis}, {Lopez}, {Lissauer}, {Mann},
  {Mallia}, {Mandell}, {Matson}, {Mazeh}, {Monsue}, {Moran}, {Moran}, {Morley},
  {Morris}, {Muirhead}, {Mukai}, {Mullally}, {Mullally}, {Murray}, {Narita},
  {Palle}, {Pidhorodetska}, {Quinn}, {Relles}, {Rinehart}, {Ritsko},
  {Rodriguez}, {Rowden}, {Rowe}, {Sebastian}, {Sefako}, {Shahaf}, {Shporer},
  {Ta{\~n}{\'o}n Reyes}, {Tenenbaum}, {Ting}, {Twicken}, {van Belle}, {Vega},
  {Volosin}, {Walkowicz}, \& {Youngblood}}]{2019AJ....158...32K}
{Kostov}, V.~B., {Schlieder}, J.~E., {Barclay}, T., {et~al.} 2019, \aj, 158, 32

\bibitem[{{Kotani} {et~al.}(2018){Kotani}, {Tamura}, {Nishikawa}, {Ueda},
  {Kuzuhara}, {Omiya}, {Hashimoto}, {Ishizuka}, {Hirano}, {Suto}, {Kurokawa},
  {Kokubo}, {Mori}, {Tanaka}, {Kashiwagi}, {Konishi}, {Kudo}, {Sato},
  {Jacobson}, {Hodapp}, {Hall}, {Aoki}, {Usuda}, {Nishiyama}, {Nakajima},
  {Ikeda}, {Yamamuro}, {Morino}, {Baba}, {Hosokawa}, {Ishikawa}, {Narita},
  {Kokubo}, {Hayano}, {Izumiura}, {Kambe}, {Kusakabe}, {Kwon}, {Ikoma}, {Hori},
  {Genda}, {Fukui}, {Fujii}, {Kawahara}, {Olivier}, {Jovanovic}, {Harakawa},
  {Hayashi}, {Hidai}, {Machida}, {Matsuo}, {Nagata}, {Ogihara}, {Takami},
  {Takato}, {Terada}, \& {Oh}}]{2018SPIE10702E..11K}
{Kotani}, T., {Tamura}, M., {Nishikawa}, J., {et~al.} 2018, in Society of
  Photo-Optical Instrumentation Engineers (SPIE) Conference Series, Vol. 10702,
  Ground-based and Airborne Instrumentation for Astronomy VII, ed. C.~J.
  {Evans}, L.~{Simard}, \& H.~{Takami}, 1070211

\bibitem[{{Kov{\'a}cs} {et~al.}(2002){Kov{\'a}cs}, {Zucker}, \&
  {Mazeh}}]{2002A&A...391..369K}
{Kov{\'a}cs}, G., {Zucker}, S., \& {Mazeh}, T. 2002, \aap, 391, 369

\bibitem[{Leconte {et~al.}(2010)Leconte, Chabrier, Baraffe, \&
  Levrard}]{Leconte2010IsEccentricity}
Leconte, J., Chabrier, G., Baraffe, I., \& Levrard, B. 2010, Astronomy and
  Astrophysics, 516, A64

\bibitem[{Leconte {et~al.}(2013)Leconte, Forget, Charnay, Wordsworth, \&
  Pottier}]{Leconte2013}
Leconte, J., Forget, F., Charnay, B., Wordsworth, R., \& Pottier, A. 2013,
  Nature, 504, 268

\bibitem[{{L{\'e}pine} {et~al.}(2013){L{\'e}pine}, {Hilton}, {Mann}, {Wilde},
  {Rojas-Ayala}, {Cruz}, \& {Gaidos}}]{2013AJ....145..102L}
{L{\'e}pine}, S., {Hilton}, E.~J., {Mann}, A.~W., {et~al.} 2013, \aj, 145, 102

\bibitem[{{L{\'e}pine} {et~al.}(2003){L{\'e}pine}, {Rich}, \&
  {Shara}}]{2003AJ....125.1598L}
{L{\'e}pine}, S., {Rich}, R.~M., \& {Shara}, M.~M. 2003, \aj, 125, 1598

\bibitem[{{L{\'e}pine} {et~al.}(2007){L{\'e}pine}, {Rich}, \&
  {Shara}}]{2007ApJ...669.1235L}
{L{\'e}pine}, S., {Rich}, R.~M., \& {Shara}, M.~M. 2007, \apj, 669, 1235

\bibitem[{{Lester} {et~al.}(2021){Lester}, {Matson}, {Howell}, {Furlan},
  {Gnilka}, {Scott}, {Ciardi}, {Everett}, {Hartman}, \&
  {Hirsch}}]{2021AJ....162...75L}
{Lester}, K.~V., {Matson}, R.~A., {Howell}, S.~B., {et~al.} 2021, \aj, 162, 75

\bibitem[{{Li} {et~al.}(2019){Li}, {Tenenbaum}, {Twicken}, {Burke}, {Jenkins},
  {Quintana}, {Rowe}, \& {Seader}}]{2019PASP..131b4506L}
{Li}, J., {Tenenbaum}, P., {Twicken}, J.~D., {et~al.} 2019, \pasp, 131, 024506

\bibitem[{{Lienhard} {et~al.}(2020){Lienhard}, {Queloz}, {Gillon}, {Burdanov},
  {Delrez}, {Ducrot}, {Handley}, {Jehin}, {Murray}, {Triaud}, {Gillen},
  {Mortier}, \& {Rackham}}]{2020MNRAS.497.3790L}
{Lienhard}, F., {Queloz}, D., {Gillon}, M., {et~al.} 2020, \mnras, 497, 3790

\bibitem[{{Lin} {et~al.}(2021){Lin}, {MacDonald}, {Kaltenegger}, \&
  {Wilson}}]{2021MNRAS.505.3562L}
{Lin}, Z., {MacDonald}, R.~J., {Kaltenegger}, L., \& {Wilson}, D.~J. 2021,
  \mnras, 505, 3562

\bibitem[{{Lissauer} {et~al.}(2012){Lissauer}, {Marcy}, {Rowe}, {Bryson},
  {Adams}, {Buchhave}, {Ciardi}, {Cochran}, {Fabrycky}, {Ford}, {Fressin},
  {Geary}, {Gilliland}, {Holman}, {Howell}, {Jenkins}, {Kinemuchi}, {Koch},
  {Morehead}, {Ragozzine}, {Seader}, {Tanenbaum}, {Torres}, \&
  {Twicken}}]{2012ApJ...750..112L}
{Lissauer}, J.~J., {Marcy}, G.~W., {Rowe}, J.~F., {et~al.} 2012, \apj, 750, 112

\bibitem[{{Lithwick} {et~al.}(2012){Lithwick}, {Xie}, \& {Wu}}]{lithwick2012}
{Lithwick}, Y., {Xie}, J., \& {Wu}, Y. 2012, \apj, 761, 122

\bibitem[{{Lomb}(1976)}]{1976Ap&SS..39..447L}
{Lomb}, N.~R. 1976, \apss, 39, 447

\bibitem[{{Luger} {et~al.}(2017){Luger}, {Sestovic}, {Kruse}, {Grimm},
  {Demory}, {Agol}, {Bolmont}, {Fabrycky}, {Fernandes}, {Van Grootel},
  {Burgasser}, {Gillon}, {Ingalls}, {Jehin}, {Raymond}, {Selsis}, {Triaud},
  {Barclay}, {Barentsen}, {Howell}, {Delrez}, {de Wit}, {Foreman-Mackey},
  {Holdsworth}, {Leconte}, {Lederer}, {Turbet}, {Almleaky}, {Benkhaldoun},
  {Magain}, {Morris}, {Heng}, \& {Queloz}}]{2017NatAs...1E.129L}
{Luger}, R., {Sestovic}, M., {Kruse}, E., {et~al.} 2017, Nature Astronomy, 1,
  0129

\bibitem[{{Lustig-Yaeger} {et~al.}(2019){Lustig-Yaeger}, {Meadows}, \&
  {Lincowski}}]{2019AJ....158...27L}
{Lustig-Yaeger}, J., {Meadows}, V.~S., \& {Lincowski}, A.~P. 2019, \aj, 158, 27

\bibitem[{{Luyten}(1979)}]{1979nlcs.book.....L}
{Luyten}, W.~J. 1979, {New Luyten catalogue of stars with proper motions larger
  than two tenths of an arcsecond; and first supplement; NLTT. (Minneapolis
  (1979)); Label 12 = short description; Label 13 = documentation by Warren;
  Label 14 = catalogue}

\bibitem[{{Mandel} \& {Agol}(2002)}]{2002ApJ...580L.171M}
{Mandel}, K. \& {Agol}, E. 2002, \apjl, 580, L171

\bibitem[{{Mann} {et~al.}(2013){Mann}, {Brewer}, {Gaidos}, {L{\'e}pine}, \&
  {Hilton}}]{2013AJ....145...52M}
{Mann}, A.~W., {Brewer}, J.~M., {Gaidos}, E., {L{\'e}pine}, S., \& {Hilton},
  E.~J. 2013, \aj, 145, 52

\bibitem[{{Mann} {et~al.}(2014){Mann}, {Deacon}, {Gaidos}, {Ansdell}, {Brewer},
  {Liu}, {Magnier}, \& {Aller}}]{Mann2014}
{Mann}, A.~W., {Deacon}, N.~R., {Gaidos}, E., {et~al.} 2014, \aj, 147, 160

\bibitem[{{Marconi} {et~al.}(2021){Marconi}, {Abreu}, {Adibekyan}, {Aliverti},
  {Allende Prieto}, {Amado}, {Amate}, {Artigau}, {Augusto}, {Barros},
  {Becerril}, {Benneke}, {Bergin}, {Berio}, {Bezawada}, {Boisse}, {Bonfils},
  {Bouchy}, {Broeg}, {Cabral}, {Calvo-Ortega}, {Canto Martins}, {Chazelas},
  {Chiavassa}, {Christensen}, {Cirami}, {Coretti}, {Covino}, {Cresci},
  {Cristiani}, {Cunha Parro}, {Cupani}, {de Castro Le{\~a}o}, {Renan de
  Medeiros}, {Furlande Souza}, {Di Marcantonio}, {Di Varano}, {D'Odorico},
  {Doyon}, {Drass}, {Figueira}, {Belen Fragoso}, {Uldall Fynbo}, {Gallo},
  {Genoni}, {Gonz{\'a}lez Hern{\'a}ndez}, {Haehnelt}, {Hlavacek-Larrondo},
  {Hughes}, {Huke}, {Humphrey}, {Kjeldsen}, {Korn}, {Kouach}, {Landoni},
  {Liske}, {Lovis}, {Lunney}, {Maiolino}, {Malo}, {Marquart}, {Martins},
  {Mason}, {Molaro}, {Monnier}, {Monteiro}, {Mordasini}, {Morris},
  {Mucciarelli}, {Murray}, {Niedzielski}, {Nunes}, {Oliva}, {Origlia},
  {Pall{\'e}}, {Pariani}, {Parr-Burman}, {Pe{\~n}ate}, {Pepe}, {Pinna},
  {Piskunov}, {Rasilla Pi{\~n}eiro}, {Rebolo}, {Rees}, {Reiners}, {Riva},
  {Romano}, {Rousseau}, {Sanna}, {Santos}, {Sarajlic}, {Shen}, {Sortino},
  {Sosnowska}, {Sousa}, {Stempels}, {Strassmeier}, {Tenegi}, {Tozzi}, {Udry},
  {Valenziano}, {Vanzi}, {Weber}, {Woche}, {Xompero}, {Zackrisson}, \&
  {Zapatero Osorio}}]{Marconi21}
{Marconi}, A., {Abreu}, M., {Adibekyan}, V., {et~al.} 2021, The Messenger, 182,
  27

\bibitem[{{McCormac} {et~al.}(2013){McCormac}, {Pollacco}, {Skillen}, {Faedi},
  {Todd}, \& {Watson}}]{2013PASP..125..548M}
{McCormac}, J., {Pollacco}, D., {Skillen}, I., {et~al.} 2013, \pasp, 125, 548

\bibitem[{{McCully} {et~al.}(2018){McCully}, {Volgenau}, {Harbeck}, {Lister},
  {Saunders}, {Turner}, {Siiverd}, \& {Bowman}}]{2018SPIE10707E..0KM}
{McCully}, C., {Volgenau}, N.~H., {Harbeck}, D.-R., {et~al.} 2018, in Society
  of Photo-Optical Instrumentation Engineers (SPIE) Conference Series, Vol.
  10707, Software and Cyberinfrastructure for Astronomy V, ed. J.~C. {Guzman}
  \& J.~{Ibsen}, 107070K

\bibitem[{{Ment} {et~al.}(2019){Ment}, {Dittmann}, {Astudillo-Defru},
  {Charbonneau}, {Irwin}, {Bonfils}, {Murgas}, {Almenara}, {Forveille}, {Agol},
  {Ballard}, {Berta-Thompson}, {Bouchy}, {Cloutier}, {Delfosse}, {Doyon},
  {Dressing}, {Esquerdo}, {Haywood}, {Kipping}, {Latham}, {Lovis}, {Newton},
  {Pepe}, {Rodriguez}, {Santos}, {Tan}, {Udry}, {Winters}, \&
  {W{\"u}nsche}}]{2019AJ....157...32M}
{Ment}, K., {Dittmann}, J.~A., {Astudillo-Defru}, N., {et~al.} 2019, \aj, 157,
  32

\bibitem[{{Metropolis} {et~al.}(1953){Metropolis}, {Rosenbluth}, {Rosenbluth},
  {Teller}, \& {Teller}}]{1953JChPh..21.1087M}
{Metropolis}, N., {Rosenbluth}, A.~W., {Rosenbluth}, M.~N., {Teller}, A.~H., \&
  {Teller}, E. 1953, \jcp, 21, 1087

\bibitem[{Mignard(1979)}]{Mignard1979TheI}
Mignard, F. 1979, The Moon and the Planets, 20, 301

\bibitem[{{Miguel} {et~al.}(2020){Miguel}, {Cridland}, {Ormel}, {Fortney}, \&
  {Ida}}]{2020MNRAS.491.1998M}
{Miguel}, Y., {Cridland}, A., {Ormel}, C.~W., {Fortney}, J.~J., \& {Ida}, S.
  2020, \mnras, 491, 1998

\bibitem[{{Miller-Ricci} {et~al.}(2009){Miller-Ricci}, {Seager}, \&
  {Sasselov}}]{2009ApJ...690.1056M}
{Miller-Ricci}, E., {Seager}, S., \& {Sasselov}, D. 2009, \apj, 690, 1056

\bibitem[{{Minkowski} \& {Abell}(1963)}]{1963bad..book..481M}
{Minkowski}, R.~L. \& {Abell}, G.~O. 1963, {The National Geographic
  Society-Palomar Observatory Sky Survey}, ed. K.~A. {Strand}, 481

\bibitem[{{Morello} {et~al.}(2022){Morello}, {Parviainen}, {Murgas},
  {Pall{\'e}}, {Oshagh}, {Fukui}, {Hirano}, {Ishikawa}, {Narita}, {Collins},
  {Barkaoui}, {Lewin}, {Cadieux}, {de Leon}, {Soubkiou}, {Crouzet},
  {Esparza-Borges}, {Hori}, {Ikoma}, {Isogai}, {Kagetani}, {Kawauchi},
  {Kimura}, {Kodama}, {Korth}, {Kotani}, {Krishnamurthy}, {Kurita},
  {Livingston}, {Luque}, {Madrigal-Aguado}, {Mori}, {Nishiumi}, {Orell-Miquel},
  {Tamura}, {Terada}, {Watanabe}, {Zou}, {Benkhaldoun}, {Collins}, {Doyon},
  {Garcia}, {Ghachoui}, {Gillon}, {Jehin}, {Jensen}, {Latham}, {Pozuelos},
  {Schwarz}, \& {Timmermans}}]{2022arXiv220113274M}
{Morello}, G., {Parviainen}, H., {Murgas}, F., {et~al.} 2022, arXiv e-prints,
  arXiv:2201.13274

\bibitem[{{Mori} {et~al.}(2022){Mori}, {Livingston}, {Leon}, {Narita},
  {Hirano}, {Fukui}, {Collins}, {Fujita}, {Hori}, {Ishikawa}, {Kawauchi},
  {Stassun}, {Watanabe}, {Giacalone}, {Gore}, {Schroeder}, {Dressing},
  {Bieryla}, {Jensen}, {Massey}, {Shporer}, {Kuzuhara}, {Charbonneau},
  {Ciardi}, {Doty}, {Esparza-Borges}, {Harakawa}, {Hodapp}, {Ikoma}, {Ikuta},
  {Isogai}, {Jenkins}, {Kagetani}, {Kimura}, {Kodama}, {Kotani},
  {Krishnamurthy}, {Kudo}, {Kurita}, {Kurokawa}, {Kusakabe}, {Latham},
  {McLean}, {Murgas}, {Nishikawa}, {Nishiumi}, {Omiya}, {Osborn}, {Palle},
  {Parviainen}, {Ricker}, {Seager}, {Serizawa}, {Teng}, {Terada}, {Twicken},
  {Ueda}, {Vanderspek}, {Vievard}, {Winn}, {Zou}, \&
  {Tamura}}]{2022AJ....163..298M}
{Mori}, M., {Livingston}, J.~H., {Leon}, J.~d., {et~al.} 2022, \aj, 163, 298

\bibitem[{{Muirhead} {et~al.}(2018){Muirhead}, {Dressing}, {Mann},
  {Rojas-Ayala}, {L{\'e}pine}, {Paegert}, {De Lee}, \&
  {Oelkers}}]{2018AJ....155..180M}
{Muirhead}, P.~S., {Dressing}, C.~D., {Mann}, A.~W., {et~al.} 2018, \aj, 155,
  180

\bibitem[{{Mulders} {et~al.}(2021){Mulders}, {Dr{\k{a}}{\.z}kowska}, {van der
  Marel}, {Ciesla}, \& {Pascucci}}]{Mulders2021}
{Mulders}, G.~D., {Dr{\k{a}}{\.z}kowska}, J., {van der Marel}, N., {Ciesla},
  F.~J., \& {Pascucci}, I. 2021, \apjl, 920, L1

\bibitem[{{Murray} {et~al.}(2020){Murray}, {Delrez}, {Pedersen}, {Queloz},
  {Gillon}, {Burdanov}, {Ducrot}, {Garcia}, {Lienhard}, {Demory}, {Jehin},
  {McCormac}, {Sebastian}, {Sohy}, {Thompson}, {Triaud}, {Van Grootel},
  {G{\"u}nther}, \& {Huang}}]{2020MNRAS.495.2446M}
{Murray}, C.~A., {Delrez}, L., {Pedersen}, P.~P., {et~al.} 2020, \mnras, 495,
  2446

\bibitem[{{Murray} {et~al.}(2022){Murray}, {Queloz}, {Gillon}, {Demory},
  {Triaud}, {de Wit}, {Burdanov}, {Chinchilla}, {Delrez}, {Dransfield},
  {Ducrot}, {Garcia}, {G{\'o}mez Maqueo Chew}, {G{\"u}nther}, {Jehin},
  {McCormac}, {Niraula}, {Pedersen}, {Pozuelos}, {Rackham}, {Schanche},
  {Sebastian}, {Thompson}, {Timmermans}, \& {Wells}}]{2022MNRAS.513.2615M}
{Murray}, C.~A., {Queloz}, D., {Gillon}, M., {et~al.} 2022, \mnras, 513, 2615

\bibitem[{{Narita} {et~al.}(2020){Narita}, {Fukui}, {Yamamuro}, {Harbeck},
  {Bowman}, {Elphick}, {Nation}, {Armstrong}, {Han}, {Abe}, {Ikoma}, {Isogai},
  {Kawauchi}, {Kurita}, {Kusakabe}, {de Leon}, {Livingston}, {Mori},
  {Nishiumi}, {Tamura}, {Watanabe}, {Volgenau}, {Heinrich-Josties}, {Foale},
  {Daily}, {McCully}, {Kirby}, {Smith}, {Haworth}, {Conway},
  {Storrie-Lombardi}, {Rosing}, {Chatelain}, {Bachelet}, {Johnson}, \&
  {Rabus}}]{2020SPIE11447E..5KN}
{Narita}, N., {Fukui}, A., {Yamamuro}, T., {et~al.} 2020, in Society of
  Photo-Optical Instrumentation Engineers (SPIE) Conference Series, Vol. 11447,
  Society of Photo-Optical Instrumentation Engineers (SPIE) Conference Series,
  114475K

\bibitem[{{Neron de Surgy} \& {Laskar}(1997)}]{neron1997}
{Neron de Surgy}, O. \& {Laskar}, J. 1997, \aap, 318, 975

\bibitem[{{Newton} {et~al.}(2017){Newton}, {Irwin}, {Charbonneau}, {Berlind},
  {Calkins}, \& {Mink}}]{2017ApJ...834...85N}
{Newton}, E.~R., {Irwin}, J., {Charbonneau}, D., {et~al.} 2017, \apj, 834, 85

\bibitem[{{Niraula} {et~al.}(2020){Niraula}, {Wit}, {Rackham}, {Ducrot},
  {Burdanov}, {Crossfield}, {Van Grootel}, {Murray}, {Garcia}, {Alonso},
  {Beard}, {Maqueo Chew}, {Delrez}, {Demory}, {Fulton}, {Gillon},
  {G{\"u}nther}, {Howard}, {Issacson}, {Jehin}, {Pedersen}, {Pozuelos},
  {Queloz}, {Rebolo-L{\'o}pez}, {Lalitha}, {Sebastian}, {Thompson}, \&
  {Triaud}}]{2020AJ....160..172N}
{Niraula}, P., {Wit}, J.~d., {Rackham}, B.~V., {et~al.} 2020, \aj, 160, 172

\bibitem[{{O'Malley-James} \& {Kaltenegger}(2017)}]{2017MNRAS.469L..26O}
{O'Malley-James}, J.~T. \& {Kaltenegger}, L. 2017, \mnras, 469, L26

\bibitem[{{Ormel} {et~al.}(2017){Ormel}, {Liu}, \&
  {Schoonenberg}}]{2017A&A...604A...1O}
{Ormel}, C.~W., {Liu}, B., \& {Schoonenberg}, D. 2017, \aap, 604, A1

\bibitem[{{Parviainen}(2015)}]{2015MNRAS.450.3233P}
{Parviainen}, H. 2015, \mnras, 450, 3233

\bibitem[{Parviainen \& Aigrain(2015)}]{Parviainen2015}
Parviainen, H. \& Aigrain, S. 2015, MNRAS, 453, 3821

\bibitem[{{Pepe} {et~al.}(2021){Pepe}, {Cristiani}, {Rebolo}, {Santos},
  {Dekker}, {Cabral}, {Di Marcantonio}, {Figueira}, {Lo Curto}, {Lovis},
  {Mayor}, {M{\'e}gevand}, {Molaro}, {Riva}, {Zapatero Osorio}, {Amate},
  {Manescau}, {Pasquini}, {Zerbi}, {Adibekyan}, {Abreu}, {Affolter}, {Alibert},
  {Aliverti}, {Allart}, {Allende Prieto}, {{\'A}lvarez}, {Alves}, {Avila},
  {Baldini}, {Bandy}, {Barros}, {Benz}, {Bianco}, {Borsa}, {Bourrier},
  {Bouchy}, {Broeg}, {Calderone}, {Cirami}, {Coelho}, {Conconi}, {Coretti},
  {Cumani}, {Cupani}, {D'Odorico}, {Damasso}, {Deiries}, {Delabre},
  {Demangeon}, {Dumusque}, {Ehrenreich}, {Faria}, {Fragoso}, {Genolet},
  {Genoni}, {G{\'e}nova Santos}, {Gonz{\'a}lez Hern{\'a}ndez}, {Hughes},
  {Iwert}, {Kerber}, {Knudstrup}, {Landoni}, {Lavie}, {Lillo-Box}, {Lizon},
  {Maire}, {Martins}, {Mehner}, {Micela}, {Modigliani}, {Monteiro}, {Monteiro},
  {Moschetti}, {Murphy}, {Nunes}, {Oggioni}, {Oliveira}, {Oshagh}, {Pall{\'e}},
  {Pariani}, {Poretti}, {Rasilla}, {Rebord{\~a}o}, {Redaelli}, {Santana
  Tschudi}, {Santin}, {Santos}, {S{\'e}gransan}, {Schmidt}, {Segovia},
  {Sosnowska}, {Sozzetti}, {Sousa}, {Span{\`o}}, {Su{\'a}rez Mascare{\~n}o},
  {Tabernero}, {Tenegi}, {Udry}, \& {Zanutta}}]{pepe21}
{Pepe}, F., {Cristiani}, S., {Rebolo}, R., {et~al.} 2021, \aap, 645, A96

\bibitem[{{Pidhorodetska} {et~al.}(2020){Pidhorodetska}, {Fauchez},
  {Villanueva}, {Domagal-Goldman}, \& {Kopparapu}}]{2020ApJ...898L..33P}
{Pidhorodetska}, D., {Fauchez}, T.~J., {Villanueva}, G.~L., {Domagal-Goldman},
  S.~D., \& {Kopparapu}, R.~K. 2020, \apjl, 898, L33

\bibitem[{Pierrehumbert \& Gaidos(2011)}]{Pierrehumbert2011}
Pierrehumbert, R. \& Gaidos, E. 2011, The Astrophysical Journal, 734, L13

\bibitem[{{Pozuelos} {et~al.}(2020){Pozuelos}, {Su{\'a}rez}, {de El{\'\i}a},
  {Berdi{\~n}as}, {Bonfanti}, {Dugaro}, {Gillon}, {Jehin}, {G{\"u}nther}, {Van
  Grootel}, {Garcia}, {Thuillier}, {Delrez}, \& {Rod{\'o}n}}]{pozuelos2020}
{Pozuelos}, F.~J., {Su{\'a}rez}, J.~C., {de El{\'\i}a}, G.~C., {et~al.} 2020,
  \aap, 641, A23

\bibitem[{{Press} {et~al.}(1992){Press}, {Teukolsky}, {Vetterling}, \&
  {Flannery}}]{1992nrfa.book.....P}
{Press}, W.~H., {Teukolsky}, S.~A., {Vetterling}, W.~T., \& {Flannery}, B.~P.
  1992, {Numerical recipes in FORTRAN. The art of scientific computing}

\bibitem[{Ramirez \& Kaltenegger(2016)}]{Ramirez2016}
Ramirez, R.~M. \& Kaltenegger, L. 2016, The Astrophysical Journal, 823, 6

\bibitem[{Ramirez \& Kaltenegger(2017)}]{Ramirez2017}
Ramirez, R.~M. \& Kaltenegger, L. 2017, The Astrophysical Journal, 837, L4

\bibitem[{Ramirez \& Kaltenegger(2018)}]{Ramirez2018}
Ramirez, R.~M. \& Kaltenegger, L. 2018, The Astrophysical Journal, 858, 72

\bibitem[{{Rasmussen} \& {Williams}(2006)}]{2006gpml.book.....R}
{Rasmussen}, C.~E. \& {Williams}, C. K.~I. 2006, {Gaussian Processes for
  Machine Learning}

\bibitem[{{Raymond} {et~al.}(2022){Raymond}, {Izidoro}, {Bolmont}, {Dorn},
  {Selsis}, {Turbet}, {Agol}, {Barth}, {Carone}, {Dasgupta}, {Gillon}, \&
  {Grimm}}]{2022NatAs...6...80R}
{Raymond}, S.~N., {Izidoro}, A., {Bolmont}, E., {et~al.} 2022, Nature
  Astronomy, 6, 80

\bibitem[{{Rayner} {et~al.}(2003){Rayner}, {Toomey}, {Onaka}, {Denault},
  {Stahlberger}, {Vacca}, {Cushing}, \& {Wang}}]{spex}
{Rayner}, J.~T., {Toomey}, D.~W., {Onaka}, P.~M., {et~al.} 2003, \pasp, 115,
  362

\bibitem[{{Reid} {et~al.}(1991){Reid}, {Brewer}, {Brucato}, {McKinley},
  {Maury}, {Mendenhall}, {Mould}, {Mueller}, {Neugebauer}, {Phinney},
  {Sargent}, {Schombert}, \& {Thicksten}}]{1991PASP..103..661R}
{Reid}, I.~N., {Brewer}, C., {Brucato}, R.~J., {et~al.} 1991, \pasp, 103, 661

\bibitem[{{Rein} \& {Liu}(2012)}]{rein2012}
{Rein}, H. \& {Liu}, S.-F. 2012, 537, A128

\bibitem[{{Rein} \& {Tamayo}(2015)}]{rein2015}
{Rein}, H. \& {Tamayo}, D. 2015, 452, 376

\bibitem[{{Ricker} {et~al.}(2015){Ricker}, {Winn}, {Vanderspek}, {Latham},
  {Bakos}, {Bean}, {Berta-Thompson}, {Brown}, {Buchhave}, {Butler}, {Butler},
  {Chaplin}, {Charbonneau}, {Christensen-Dalsgaard}, {Clampin}, {Deming},
  {Doty}, {De Lee}, {Dressing}, {Dunham}, {Endl}, {Fressin}, {Ge}, {Henning},
  {Holman}, {Howard}, {Ida}, {Jenkins}, {Jernigan}, {Johnson}, {Kaltenegger},
  {Kawai}, {Kjeldsen}, {Laughlin}, {Levine}, {Lin}, {Lissauer}, {MacQueen},
  {Marcy}, {McCullough}, {Morton}, {Narita}, {Paegert}, {Palle}, {Pepe},
  {Pepper}, {Quirrenbach}, {Rinehart}, {Sasselov}, {Sato}, {Seager},
  {Sozzetti}, {Stassun}, {Sullivan}, {Szentgyorgyi}, {Torres}, {Udry}, \&
  {Villasenor}}]{2015JATIS...1a4003R}
{Ricker}, G.~R., {Winn}, J.~N., {Vanderspek}, R., {et~al.} 2015, Journal of
  Astronomical Telescopes, Instruments, and Systems, 1, 014003

\bibitem[{{Riddick} {et~al.}(2007){Riddick}, {Roche}, \&
  {Lucas}}]{2007MNRAS.381.1067R}
{Riddick}, F.~C., {Roche}, P.~F., \& {Lucas}, P.~W. 2007, \mnras, 381, 1067

\bibitem[{{Rojas-Ayala} {et~al.}(2012){Rojas-Ayala}, {Covey}, {Muirhead}, \&
  {Lloyd}}]{Rojas-Ayala2012}
{Rojas-Ayala}, B., {Covey}, K.~R., {Muirhead}, P.~S., \& {Lloyd}, J.~P. 2012,
  \apj, 748, 93

\bibitem[{{Sabin} {et~al.}(2018){Sabin}, {G{\'o}mez Maqueo Chew}, {Demory},
  {Petrucci}, \& {Saint-Ex Consortium}}]{2018csss.confE..59S}
{Sabin}, L., {G{\'o}mez Maqueo Chew}, Y., {Demory}, B.-O., {Petrucci}, R., \&
  {Saint-Ex Consortium}. 2018, in 20th Cambridge Workshop on Cool Stars,
  Stellar Systems and the Sun, Cambridge Workshop on Cool Stars, Stellar
  Systems, and the Sun, 59

\bibitem[{{Savitzky} \& {Golay}(1964)}]{savgol}
{Savitzky}, A. \& {Golay}, M.~J.~E. 1964, Analytical Chemistry, 36, 1627

\bibitem[{{Scargle}(1982)}]{1982ApJ...263..835S}
{Scargle}, J.~D. 1982, \apj, 263, 835

\bibitem[{{Schanche} {et~al.}(2022){Schanche}, {Pozuelos}, {G{\"u}nther},
  {Wells}, {Burgasser}, {Chinchilla}, {Delrez}, {Ducrot}, {Garcia}, {G{\'o}mez
  Maqueo Chew}, {Jofr{\'e}}, {Rackham}, {Sebastian}, {Stassun}, {Stern},
  {Timmermans}, {Barkaoui}, {Belinski}, {Benkhaldoun}, {Benz}, {Bieryla},
  {Bouchy}, {Burdanov}, {Charbonneau}, {Christiansen}, {Collins}, {Demory},
  {D{\'e}vora-Pajares}, {de Wit}, {Dragomir}, {Dransfield}, {Furlan},
  {Ghachoui}, {Gillon}, {Gnilka}, {G{\'o}mez-Mu{\~n}oz}, {Guerrero}, {Harris},
  {Heng}, {Henze}, {Hesse}, {Howell}, {Jehin}, {Jenkins}, {Jensen}, {Kunimoto},
  {Latham}, {Lester}, {McLeod}, {Mireles}, {Murray}, {Niraula}, {Pedersen},
  {Queloz}, {Quintana}, {Ricker}, {Rudat}, {Sabin}, {Safonov},
  {Schroffenegger}, {Scott}, {Seager}, {Strakhov}, {Triaud}, {Vanderspek},
  {Vezie}, \& {Winn}}]{nicole2022}
{Schanche}, N., {Pozuelos}, F.~J., {G{\"u}nther}, M.~N., {et~al.} 2022, \aap,
  657, A45

\bibitem[{{Schwarz}(1978)}]{1978AnSta...6..461S}
{Schwarz}, G. 1978, Annals of Statistics, 6, 461

\bibitem[{{Scott} {et~al.}(2021){Scott}, {Howell}, {Gnilka}, {Stephens},
  {Salinas}, {Matson}, {Furlan}, {Horch}, {Everett}, {Ciardi}, {Mills}, \&
  {Quigley}}]{2021FrASS...8..138S}
{Scott}, N.~J., {Howell}, S.~B., {Gnilka}, C.~L., {et~al.} 2021, Frontiers in
  Astronomy and Space Sciences, 8, 138

\bibitem[{{Seager} \& {Mall{\'e}n-Ornelas}(2003)}]{2003ApJ...585.1038S}
{Seager}, S. \& {Mall{\'e}n-Ornelas}, G. 2003, \apj, 585, 1038

\bibitem[{{Sebastian} {et~al.}(2021){Sebastian}, {Gillon}, {Ducrot},
  {Pozuelos}, {Garcia}, {G{\"u}nther}, {Delrez}, {Queloz}, {Demory}, {Triaud},
  {Burgasser}, {de Wit}, {Burdanov}, {Dransfield}, {Jehin}, {McCormac},
  {Murray}, {Niraula}, {Pedersen}, {Rackham}, {Sohy}, {Thompson}, \& {Van
  Grootel}}]{2021A&A...645A.100S}
{Sebastian}, D., {Gillon}, M., {Ducrot}, E., {et~al.} 2021, \aap, 645, A100

\bibitem[{{Seifahrt} {et~al.}(2020){Seifahrt}, {Bean}, {St{\"u}rmer}, {Kasper},
  {Gers}, {Schwab}, {Zechmeister}, {Stef{\'a}nsson}, {Montet}, {Dos Santos},
  {Peck}, {White}, \& {Tapia}}]{MAROON-X2021}
{Seifahrt}, A., {Bean}, J.~L., {St{\"u}rmer}, J., {et~al.} 2020, in
  Ground-based and Airborne Instrumentation for Astronomy VIII, ed. C.~J.
  Evans, J.~J. Bryant, \& K.~Motohara, Vol. 11447, International Society for
  Optics and Photonics (SPIE), 305 -- 325

\bibitem[{{Sergeev} {et~al.}(2021){Sergeev}, {Fauchez}, {Turbet}, {Boutle},
  {Tsigaridis}, {Way}, {Wolf}, {Domagal-Goldman}, {Forget}, {Haqq-Misra},
  {Kopparapu}, {Lambert}, {Manners}, \& {Mayne}}]{2021arXiv210911459S}
{Sergeev}, D.~E., {Fauchez}, T.~J., {Turbet}, M., {et~al.} 2021, arXiv
  e-prints, arXiv:2109.11459

\bibitem[{{Sharma} {et~al.}(2018){Sharma}, {Stello}, {Buder}, {Kos},
  {Bland-Hawthorn}, {Asplund}, {Duong}, {Lin}, {Lind}, {Ness}, {Huber},
  {Zwitter}, {Traven}, {Hon}, {Kafle}, {Khanna}, {Saddon}, {Anguiano}, {Casey},
  {Freeman}, {Martell}, {De Silva}, {Simpson}, {Wittenmyer}, \&
  {Zucker}}]{2018MNRAS.473.2004S}
{Sharma}, S., {Stello}, D., {Buder}, S., {et~al.} 2018, \mnras, 473, 2004

\bibitem[{{Skrutskie} {et~al.}(2006){Skrutskie}, {Cutri}, {Stiening},
  {Weinberg}, {Schneider}, {Carpenter}, {Beichman}, {Capps}, {Chester},
  {Elias}, {Huchra}, {Liebert}, {Lonsdale}, {Monet}, {Price}, {Seitzer},
  {Jarrett}, {Kirkpatrick}, {Gizis}, {Howard}, {Evans}, {Fowler}, {Fullmer},
  {Hurt}, {Light}, {Kopan}, {Marsh}, {McCallon}, {Tam}, {Van Dyk}, \&
  {Wheelock}}]{2006AJ....131.1163S}
{Skrutskie}, M.~F., {Cutri}, R.~M., {Stiening}, R., {et~al.} 2006, \aj, 131,
  1163

\bibitem[{{Smith} {et~al.}(2012){Smith}, {Stumpe}, {Van Cleve}, {Jenkins},
  {Barclay}, {Fanelli}, {Girouard}, {Kolodziejczak}, {McCauliff}, {Morris}, \&
  {Twicken}}]{2012PASP..124.1000S}
{Smith}, J.~C., {Stumpe}, M.~C., {Van Cleve}, J.~E., {et~al.} 2012, \pasp, 124,
  1000

\bibitem[{{Speagle}(2020)}]{2020MNRAS.493.3132S}
{Speagle}, J.~S. 2020, \mnras, 493, 3132

\bibitem[{{Stassun} {et~al.}(2017){Stassun}, {Collins}, \&
  {Gaudi}}]{Stassun:2017}
{Stassun}, K.~G., {Collins}, K.~A., \& {Gaudi}, B.~S. 2017, \aj, 153, 136

\bibitem[{{Stassun} {et~al.}(2018{\natexlab{a}}){Stassun}, {Corsaro}, {Pepper},
  \& {Gaudi}}]{Stassun:2018}
{Stassun}, K.~G., {Corsaro}, E., {Pepper}, J.~A., \& {Gaudi}, B.~S.
  2018{\natexlab{a}}, \aj, 155, 22

\bibitem[{{Stassun} {et~al.}(2018{\natexlab{b}}){Stassun}, {Oelkers}, {Pepper},
  {Paegert}, {De Lee}, {Torres}, {Latham}, {Charpinet}, {Dressing}, {Huber},
  {Kane}, {L{\'e}pine}, {Mann}, {Muirhead}, {Rojas-Ayala}, {Silvotti},
  {Fleming}, {Levine}, \& {Plavchan}}]{2018AJ....156..102S}
{Stassun}, K.~G., {Oelkers}, R.~J., {Pepper}, J., {et~al.} 2018{\natexlab{b}},
  \aj, 156, 102

\bibitem[{{Stassun} \& {Torres}(2016)}]{Stassun:2016}
{Stassun}, K.~G. \& {Torres}, G. 2016, \aj, 152, 180

\bibitem[{{Stassun} \& {Torres}(2021)}]{StassunTorres:2021}
{Stassun}, K.~G. \& {Torres}, G. 2021, \apjl, 907, L33

\bibitem[{{Stumpe} {et~al.}(2014){Stumpe}, {Smith}, {Catanzarite}, {Van Cleve},
  {Jenkins}, {Twicken}, \& {Girouard}}]{2014PASP..126..100S}
{Stumpe}, M.~C., {Smith}, J.~C., {Catanzarite}, J.~H., {et~al.} 2014, \pasp,
  126, 100

\bibitem[{{Stumpe} {et~al.}(2012){Stumpe}, {Smith}, {Van Cleve}, {Twicken},
  {Barclay}, {Fanelli}, {Girouard}, {Jenkins}, {Kolodziejczak}, {McCauliff}, \&
  {Morris}}]{2012PASP..124..985S}
{Stumpe}, M.~C., {Smith}, J.~C., {Van Cleve}, J.~E., {et~al.} 2012, \pasp, 124,
  985

\bibitem[{{Sullivan} {et~al.}(2015){Sullivan}, {Winn}, {Berta-Thompson},
  {Charbonneau}, {Deming}, {Dressing}, {Latham}, {Levine}, {McCullough},
  {Morton}, {Ricker}, {Vanderspek}, \& {Woods}}]{2015ApJ...809...77S}
{Sullivan}, P.~W., {Winn}, J.~N., {Berta-Thompson}, Z.~K., {et~al.} 2015, \apj,
  809, 77

\bibitem[{{Tamura} {et~al.}(2012){Tamura}, {Suto}, {Nishikawa}, {Kotani},
  {Sato}, {Aoki}, {Usuda}, {Kurokawa}, {Kashiwagi}, {Nishiyama}, {Ikeda},
  {Hall}, {Hodapp}, {Hashimoto}, {Morino}, {Inoue}, {Mizuno}, {Washizaki},
  {Tanaka}, {Suzuki}, {Kwon}, {Suenaga}, {Oh}, {Narita}, {Kokubo}, {Hayano},
  {Izumiura}, {Kambe}, {Kudo}, {Kusakabe}, {Ikoma}, {Hori}, {Omiya}, {Genda},
  {Fukui}, {Fujii}, {Guyon}, {Harakawa}, {Hayashi}, {Hidai}, {Hirano},
  {Kuzuhara}, {Machida}, {Matsuo}, {Nagata}, {Ohnuki}, {Ogihara}, {Oshino},
  {Suzuki}, {Takami}, {Takato}, {Takahashi}, {Tachinami}, \&
  {Terada}}]{2012SPIE.8446E..1TT}
{Tamura}, M., {Suto}, H., {Nishikawa}, J., {et~al.} 2012, in Society of
  Photo-Optical Instrumentation Engineers (SPIE) Conference Series, Vol. 8446,
  Ground-based and Airborne Instrumentation for Astronomy IV, ed. I.~S.
  {McLean}, S.~K. {Ramsay}, \& H.~{Takami}, 84461T

\bibitem[{{Teyssandier} {et~al.}(2022){Teyssandier}, {Libert}, \&
  {Agol}}]{2022A&A...658A.170T}
{Teyssandier}, J., {Libert}, A.~S., \& {Agol}, E. 2022, \aap, 658, A170

\bibitem[{{Turbet} {et~al.}(2018){Turbet}, {Bolmont}, {Leconte}, {Forget},
  {Selsis}, {Tobie}, {Caldas}, {Naar}, \& {Gillon}}]{turbet2018}
{Turbet}, M., {Bolmont}, E., {Leconte}, J., {et~al.} 2018, \aap, 612, A86

\bibitem[{{Turbet} {et~al.}(2021){Turbet}, {Fauchez}, {Sergeev}, {Boutle},
  {Tsigaridis}, {Way}, {Wolf}, {Domagal-Goldman}, {Forget}, {Haqq-Misra},
  {Kopparapu}, {Lambert}, {Manners}, {Mayne}, \& {Sohl}}]{2021arXiv210911457T}
{Turbet}, M., {Fauchez}, T.~J., {Sergeev}, D.~E., {et~al.} 2021, arXiv
  e-prints, arXiv:2109.11457

\bibitem[{{Twicken} {et~al.}(2018){Twicken}, {Catanzarite}, {Clarke},
  {Girouard}, {Jenkins}, {Klaus}, {Li}, {McCauliff}, {Seader}, {Tenenbaum},
  {Wohler}, {Bryson}, {Burke}, {Caldwell}, {Haas}, {Henze}, \&
  {Sanderfer}}]{2018PASP..130f4502T}
{Twicken}, J.~D., {Catanzarite}, J.~H., {Clarke}, B.~D., {et~al.} 2018, \pasp,
  130, 064502

\bibitem[{{Van Grootel} {et~al.}(2018){Van Grootel}, {Fernandes}, {Gillon},
  {Jehin}, {Manfroid}, {Scuflaire}, {Burgasser}, {Barkaoui}, {Benkhaldoun},
  {Burdanov}, {Delrez}, {Demory}, {de Wit}, {Queloz}, \&
  {Triaud}}]{2018ApJ...853...30V}
{Van Grootel}, V., {Fernandes}, C.~S., {Gillon}, M., {et~al.} 2018, \apj, 853,
  30

\bibitem[{{Vanderspek} {et~al.}(2019){Vanderspek}, {Huang}, {Vanderburg},
  {Ricker}, {Latham}, {Seager}, {Winn}, {Jenkins}, {Burt}, {Dittmann},
  {Newton}, {Quinn}, {Shporer}, {Charbonneau}, {Irwin}, {Ment}, {Winters},
  {Collins}, {Evans}, {Gan}, {Hart}, {Jensen}, {Kielkopf}, {Mao}, {Waalkes},
  {Bouchy}, {Marmier}, {Nielsen}, {Ottoni}, {Pepe}, {S{\'e}gransan}, {Udry},
  {Henry}, {Paredes}, {James}, {Hinojosa}, {Silverstein}, {Palle},
  {Berta-Thompson}, {Crossfield}, {Davies}, {Dragomir}, {Fausnaugh}, {Glidden},
  {Pepper}, {Morgan}, {Rose}, {Twicken}, {Villase{\~n}or}, {Yu}, {Bakos},
  {Bean}, {Buchhave}, {Christensen-Dalsgaard}, {Christiansen}, {Ciardi},
  {Clampin}, {De Lee}, {Deming}, {Doty}, {Jernigan}, {Kaltenegger}, {Lissauer},
  {McCullough}, {Narita}, {Paegert}, {Pal}, {Rinehart}, {Sasselov}, {Sato},
  {Sozzetti}, {Stassun}, \& {Torres}}]{2019ApJ...871L..24V}
{Vanderspek}, R., {Huang}, C.~X., {Vanderburg}, A., {et~al.} 2019, \apjl, 871,
  L24

\bibitem[{{Wells} {et~al.}(2021){Wells}, {Rackham}, {Schanche}, {Petrucci},
  {G{\'o}mez Maqueo Chew}, {Demory}, {Burgasser}, {Burn}, {Pozuelos},
  {G{\"u}nther}, {Sabin}, {Schroffenegger}, {G{\'o}mez-Mu{\~n}oz}, {Stassun},
  {Van Grootel}, {Howell}, {Sebastian}, {Triaud}, {Apai}, {Plauchu-Frayn},
  {Guerrero}, {Guill{\'e}n}, {Landa}, {Melgoza}, {Montalvo}, {Serrano},
  {Riesgo}, {Barkaoui}, {Bixel}, {Burdanov}, {Chen}, {Chinchilla}, {Collins},
  {Daylan}, {de Wit}, {Delrez}, {D{\'e}vora-Pajares}, {Dietrich}, {Dransfield},
  {Ducrot}, {Fausnaugh}, {Furlan}, {Gabor}, {Gan}, {Garcia}, {Ghachoui},
  {Giacalone}, {Gibbs}, {Gillon}, {Gnilka}, {Gore}, {Guerrero}, {Henning},
  {Hesse}, {Jehin}, {Jenkins}, {Latham}, {Lester}, {McCormac}, {Murray},
  {Niraula}, {Pedersen}, {Queloz}, {Ricker}, {Rodriguez}, {Schroeder},
  {Schwarz}, {Scott}, {Seager}, {Theissen}, {Thompson}, {Timmermans},
  {Twicken}, \& {Winn}}]{wells2021}
{Wells}, R.~D., {Rackham}, B.~V., {Schanche}, N., {et~al.} 2021, \aap, 653, A97

\bibitem[{{Wildi} {et~al.}(2017){Wildi}, {Blind}, {Reshetov}, {Hernandez},
  {Genolet}, {Conod}, {Sordet}, {Segovilla}, {Rasilla}, {Brousseau},
  {Thibault}, {Delabre}, {Bandy}, {Sarajlic}, {Cabral}, {Bovay}, {Vall{\'e}e},
  {Bouchy}, {Doyon}, {Artigau}, {Pepe}, {Hagelberg}, {Melo}, {Delfosse},
  {Figueira}, {Santos}, {Gonz{\'a}lez Hern{\'a}ndez}, {de Medeiros}, {Rebolo},
  {Broeg}, {Benz}, {Boisse}, {Malo}, {K{\"a}ufl}, \& {Saddlemyer}}]{wildi17}
{Wildi}, F., {Blind}, N., {Reshetov}, V., {et~al.} 2017, in Society of
  Photo-Optical Instrumentation Engineers (SPIE) Conference Series, Vol. 10400,
  Society of Photo-Optical Instrumentation Engineers (SPIE) Conference Series,
  1040018

\bibitem[{Wolf \& Toon(2014)}]{Wolf2014}
Wolf, E.~T. \& Toon, O.~B. 2014, Geophysical Research Letters, 41, 167

\bibitem[{Yang {et~al.}(2013)Yang, Cowan, \& Abbot}]{Yang2013}
Yang, J., Cowan, N.~B., \& Abbot, D.~S. 2013, The Astrophysical Journal, 771,
  L45

\bibitem[{{Zechmeister} {et~al.}(2019){Zechmeister}, {Dreizler}, {Ribas},
  {Reiners}, {Caballero}, {Bauer}, {B{\'e}jar}, {Gonz{\'a}lez-Cuesta},
  {Herrero}, {Lalitha}, {L{\'o}pez-Gonz{\'a}lez}, {Luque}, {Morales},
  {Pall{\'e}}, {Rodr{\'\i}guez}, {Rodr{\'\i}guez L{\'o}pez}, {Tal-Or},
  {Anglada-Escud{\'e}}, {Quirrenbach}, {Amado}, {Abril}, {Aceituno},
  {Aceituno}, {Alonso-Floriano}, {Ammler-von Eiff}, {Antona Jim{\'e}nez},
  {Anwand-Heerwart}, {Arroyo-Torres}, {Azzaro}, {Baroch}, {Barrado},
  {Becerril}, {Ben{\'\i}tez}, {Berdi{\~n}as}, {Bergond}, {Bluhm},
  {Brinkm{\"o}ller}, {del Burgo}, {Calvo Ortega}, {Cano}, {Cardona
  Guill{\'e}n}, {Carro}, {C{\'a}rdenas V{\'a}zquez}, {Casal},
  {Casasayas-Barris}, {Casanova}, {Chaturvedi}, {Cifuentes}, {Claret},
  {Colom{\'e}}, {Cort{\'e}s-Contreras}, {Czesla}, {D{\'\i}ez-Alonso}, {Dorda},
  {Fern{\'a}ndez}, {Fern{\'a}ndez-Mart{\'\i}n}, {Fuhrmeister}, {Fukui},
  {Galad{\'\i}-Enr{\'\i}quez}, {Gallardo Cava}, {Garcia de la Fuente},
  {Garcia-Piquer}, {Garc{\'\i}a Vargas}, {Gesa}, {G{\'o}ngora Rueda},
  {Gonz{\'a}lez-{\'A}lvarez}, {Gonz{\'a}lez Hern{\'a}ndez},
  {Gonz{\'a}lez-Peinado}, {Gr{\"o}zinger}, {Gu{\`a}rdia}, {Guijarro}, {de
  Guindos}, {Hatzes}, {Hauschildt}, {Hedrosa}, {Helmling}, {Henning},
  {Hermelo}, {Hern{\'a}ndez Arabi}, {Hern{\'a}ndez Casta{\~n}o}, {Hern{\'a}ndez
  Otero}, {Hintz}, {Huke}, {Huber}, {Jeffers}, {Johnson}, {de Juan},
  {Kaminski}, {Kemmer}, {Kim}, {Klahr}, {Klein}, {Kl{\"u}ter}, {Klutsch},
  {Kossakowski}, {K{\"u}rster}, {Labarga}, {Lafarga}, {Llamas}, {Lamp{\'o}n},
  {Lara}, {Launhardt}, {L{\'a}zaro}, {Lodieu}, {L{\'o}pez del Fresno},
  {L{\'o}pez-Puertas}, {L{\'o}pez Salas}, {L{\'o}pez-Santiago}, {Mag{\'a}n
  Madinabeitia}, {Mall}, {Mancini}, {Mandel}, {Marfil}, {Mar{\'\i}n Molina},
  {Maroto Fern{\'a}ndez}, {Mart{\'\i}n}, {Mart{\'\i}n-Fern{\'a}ndez},
  {Mart{\'\i}n-Ruiz}, {Marvin}, {Mirabet}, {Monta{\~n}{\'e}s-Rodr{\'\i}guez},
  {Montes}, {Moreno-Raya}, {Nagel}, {Naranjo}, {Narita}, {Nortmann}, {Nowak},
  {Ofir}, {Oshagh}, {Panduro}, {Parviainen}, {Pascual}, {Passegger}, {Pavlov},
  {Pedraz}, {P{\'e}rez-Calpena}, {P{\'e}rez Medialdea}, {Perger}, {Perryman},
  {Rabaza}, {Ram{\'o}n Ballesta}, {Rebolo}, {Redondo}, {Reffert}, {Reinhardt},
  {Rhode}, {Rix}, {Rodler}, {Rodr{\'\i}guez Trinidad}, {Rosich}, {Sadegi},
  {S{\'a}nchez-Blanco}, {S{\'a}nchez Carrasco}, {S{\'a}nchez-L{\'o}pez},
  {Sanz-Forcada}, {Sarkis}, {Sarmiento}, {Sch{\"a}fer}, {Schmitt},
  {Sch{\"o}fer}, {Schweitzer}, {Seifert}, {Shulyak}, {Solano}, {Sota}, {Stahl},
  {Stock}, {Strachan}, {Stuber}, {St{\"u}rmer}, {Su{\'a}rez}, {Tabernero},
  {Tala Pinto}, {Trifonov}, {Veredas}, {Vico Linares}, {Vilardell}, {Wagner},
  {Wolthoff}, {Xu}, {Yan}, \& {Zapatero Osorio}}]{2019A&A...627A..49Z}
{Zechmeister}, M., {Dreizler}, S., {Ribas}, I., {et~al.} 2019, \aap, 627, A49

\bibitem[{{Zhang} {et~al.}(2019){Zhang}, {Luo}, {Comte}, {Gizis}, {Wang}, {Li},
  {Qin}, {Kong}, {Bai}, \& {Yi}}]{2019ApJS..240...31Z}
{Zhang}, S., {Luo}, A.~L., {Comte}, G., {et~al.} 2019, \apjs, 240, 31

\end{thebibliography}

\onecolumn
\begin{appendix}

\section{Radial velocities}

\begin{table}[hbt!]
\centering
\begin{tabular}{lcc}
\toprule
\toprule
$\mathrm{BJD_{TDB}} - 2\,450\,000$ & RV (km/s) & $\sigma_{\mathrm{RV}}$ (km/s) \\
\hline
$YJ$-band & & \\
9467.104595 & 8.975819 & 0.01608 \\
9467.125868 & 8.926950 & 0.01971 \\
9488.104150 & 8.864151 & 0.00970 \\
9508.011182 & 8.920489 & 0.01211 \\
9509.985105 & 8.945039 & 0.01015 \\
9510.006380 & 8.924036 & 0.01078 \\
9515.018797 & 8.906155 & 0.01468 \\
9515.040059 & 8.958196 & 0.01124 \\
9515.061315 & 8.898919 & 0.01441 \\
9530.936633 & 8.911857 & 0.01190 \\
9530.957919 & 8.834334 & 0.01502 \\
\hline
$H$-band & & \\
9467.104595 & 3.504429 & 0.01054 \\
9467.125868 & 3.520200 & 0.01299 \\
9488.104150 & 3.452051 & 0.00735 \\
9508.011182 & 3.450579 & 0.00872 \\
9509.985105 & 3.524839 & 0.00767 \\
9510.006380 & 3.500336 & 0.00786 \\
9515.018797 & 3.458665 & 0.01036 \\
9515.040059 & 3.508536 & 0.00820 \\
9515.061315 & 3.492029 & 0.00974 \\
9530.936633 & 3.489667 & 0.00914 \\
9530.957919 & 3.464614 & 0.01166 \\
9588.854504 & 3.497201 & 0.00797 \\
9588.875855 & 3.508872 & 0.00809 \\
9588.897185 & 3.495941 & 0.00792 \\
\bottomrule
\bottomrule
\end{tabular}
\caption{Subaru/IRD radial velocity measurements. We recommend treating the $YJ$-band and $H$-band RVs as independent datasets, as done in the analysis presented in Sect. \ref{sec:rv_analysis}.}
\label{tab:RVs}
\end{table}

\clearpage
\section{Photometric baseline models and error scaling factors}

\begin{longtable}{l l c c c c c c}
\hline
\hline
Planet & Facility & Bandpass & $T_{0}$ ($\mathrm{BJD_{TDB}}$ & Baseline model & Residual RMS & $\beta_w$ & $\beta_r$ \\
 & & & $-$ 2 450 000) & & (exp. time) & & \\
\hline
\endfirsthead
\multicolumn{8}{c}{\tablename\ \thetable{} -- \textit{continued from previous page}} \\
\hline
Planet & Facility & Bandpass & $T_{0}$ ($\mathrm{BJD_{TDB}}$ & Baseline model & Residual RMS & $\beta_w$ & $\beta_r$ \\
 & & & $-$ 2 450 000) & & (exp. time) & & \\
\hline
\endhead
\hline \multicolumn{8}{c}{\textit{Continued on next page}} \\ 
\endfoot
\hline \hline
\caption{Photometric baseline models, residual RMS, and error scaling factors $\beta_w$ and $\beta_r$ for each transit light curve used in our global transit analysis (see Sect. \ref{sec:glob_an}). For the baseline function, p($\alpha^n$) denotes a $n$-order polynomial function of the parameter $\alpha$, with $\alpha$ that can be $t=$ 
time, $a=$ airmass, $x$ and $y=$ $x$- and $y$- position of the target on the detector, or $f=$ full width at half maximum of the point spread function.}
\label{tab:baselines}\\
\endlastfoot
  b & TESS Sector 4 & TESS & 8413.1930 & offset & 0.0197 (120 s) & 0.85 & 1.25 \\
    & TESS Sector 4 & TESS & 8415.9229 & offset & 0.0218 (120 s) & 0.94 & 1.46 \\
    & TESS Sector 4 & TESS & 8426.8425 & offset & 0.0231 (120 s) & 1.00 & 1.00 \\
    & TESS Sector 4 & TESS & 8429.5724 & offset & 0.0217 (120 s) & 0.95 & 1.23 \\  
    & TESS Sector 4 & TESS & 8432.3023 & offset & 0.0219 (120 s) & 0.95 & 1.00 \\
    & TESS Sector 4 & TESS & 8435.0322 & offset & 0.0200 (120 s) & 0.87 & 1.30 \\ 
    & TESS Sector 5 & TESS & 8440.4920 & offset & 0.0160 (120 s) & 0.90 & 1.67 \\
    & TESS Sector 5 & TESS & 8443.2219 & offset & 0.0145 (120 s) & 0.82 & 1.44 \\
    & TESS Sector 5 & TESS & 8445.9518 & offset & 0.0151 (120 s) & 0.86 & 1.06 \\ 
    & TESS Sector 5 & TESS & 8448.6817 & offset & 0.0158 (120 s) & 0.89 & 1.30 \\  
    & TESS Sector 5 & TESS & 8454.1415 & offset & 0.0170 (120 s) & 0.96 & 1.12 \\ 
    & TESS Sector 5 & TESS & 8456.8714 & offset & 0.0156 (120 s) & 0.89 & 1.11 \\   
    & TESS Sector 5 & TESS & 8459.6013 & offset & 0.0162 (120 s) & 0.92 & 1.29 \\ 
    & TESS Sector 5 & TESS & 8462.3312 & p($t^1$) & 0.0160 (120 s) & 0.91 & 1.10 \\
    & TESS Sector 31 & TESS & 9144.8071 & offset & 0.0194 (120 s) & 1.03 & 1.18 \\
    & TESS Sector 31 & TESS & 9147.5370 & offset & 0.0181 (120 s) & 0.97 & 1.29 \\ 
    & TESS Sector 31 & TESS & 9150.2669 & offset & 0.0178 (120 s) & 0.95 & 1.17 \\ 
    & TESS Sector 31 & TESS & 9152.9968 & offset & 0.0157 (120 s) & 0.84 & 1.26 \\    
    & TESS Sector 31 & TESS & 9155.7267 & p($t^1$) & 0.0180 (120 s) & 0.93 & 1.62 \\  
    & TESS Sector 31 & TESS & 9161.1865 & offset & 0.0158 (120 s) & 0.85 & 1.59 \\  
    & TESS Sector 31 & TESS & 9163.9164 & offset & 0.0167 (120 s) & 0.90 & 1.78 \\ 
    & TESS Sector 31 & TESS & 9166.6463 & offset & 0.0189 (120 s) & 1.01 & 1.12 \\   
    & TESS Sector 31 & TESS & 9169.3762 & offset & 0.0170 (120 s) & 0.91 & 1.16 \\ 
    & TESS Sector 32 & TESS & 9174.8360 & offset & 0.0217 (120 s) & 1.10 & 1.00 \\ 
    & TESS Sector 32 & TESS & 9177.5659 & offset & 0.0178 (120 s) & 0.90 & 1.00 \\   
    & TESS Sector 32 & TESS & 9180.2958 & offset & 0.0183 (120 s) & 0.93 & 1.15 \\ 
    & TESS Sector 32 & TESS & 9183.0257 & offset & 0.0179 (120 s) & 0.91 & 1.12 \\ 
    & TESS Sector 32 & TESS & 9188.4855 & offset & 0.0184 (120 s) & 0.93 & 1.16 \\ 
    & TESS Sector 32 & TESS & 9191.2154 & offset & 0.0170 (120 s) & 0.86 & 1.56 \\
    & TESS Sector 32 & TESS & 9193.9453 & offset & 0.0180 (120 s) & 0.92 & 1.42 \\ 
    & TESS Sector 32 & TESS & 9196.6752 & offset & 0.0168 (120 s) & 0.86 & 1.13 \\
    & TESS Sector 32 & TESS & 9199.4051 & offset & 0.0173 (120 s) & 0.88 & 1.43 \\
    & SSO/Europa & $I+z$ & 9436.9067 & p($y^1$) & 0.00238 (35 s) & 0.77 & 1.10 \\
    & TRAPPIST-South & blue-blocking & 9436.9067 & p($a^1$) & 0.00329 (120 s) & 0.62 & 1.83 \\
    & SSO/Callisto & $i'$ & 9447.8263 & p($a^1$) & 0.00236 (92 s) & 0.83 & 1.15 \\
    & SSO/Io & $I+z$ & 9447.8263 & p($a^1$) & 0.00236 (39 s) & 0.64 & 1.29 \\
    & SSO/Europa & $i'$ & 9458.7460 & p($t^1$) & 0.00255 (92 s) & 0.58 & 1.78 \\
    & SSO/Io & $z'$ & 9458.7460 & p($a^1$, $y^1$) & 0.00252 (49 s) & 0.76 & 1.10 \\
    & SSO/Ganymede & $r'$ & 9477.8553 & p($a^1$) & 0.00616 (180 s) & 1.12 & 1.11 \\
    & SSO/Io & $r'$ & 9477.8553 & p($t^1$) & 0.00584 (180 s) & 0.94 & 1.23 \\
    & SSO/Europa & $r'$ & 9477.8553 & p($a^1$) & 0.00630 (180 s) & 1.10 & 1.00 \\
    & ExTrA1 & 0.85$-$1.55 $\mu$m & 9477.8553 & p($t^4$) & 0.00540 (60 s) & 1.02 & 1.43 \\
    & ExTrA2 & 0.85$-$1.55 $\mu$m & 9477.8553 & p($t^4$) & 0.00593 (60 s) & 0.95 & 1.00 \\
    & MuSCAT3 & $g'$ & 9486.0450 & p($t^1$) & 0.0139 (120 s) & 0.89 & 1.22 \\
    & MuSCAT3 & $r'$ & 9486.0450 & offset & 0.00430 (120 s) & 0.90 & 1.15 \\
    & MuSCAT3 & $i'$ & 9486.0450 & p($y^2$) & 0.00214 (110 s) & 1.14 & 1.50 \\
    & MuSCAT3 & $z'$ & 9486.0450 & p($a^1$, $y^1$) & 0.00152 (100 s) & 1.14 & 1.23 \\
    & SSO/Europa & $I+z$ & 9488.7749 & offset & 0.00256 (39 s) & 0.56 & 3.46 \\
    & SSO/Europa & $I+z$ & 9510.6141 & p($f^1$) & 0.00246 (39 s) & 0.79 & 1.42 \\
    & MuSCAT3 & $g'$ & 9516.0739 & offset & 0.0158 (120 s) & 0.86 & 1.00 \\
    & MuSCAT3 & $r'$ & 9516.0739 & p($a^1$) & 0.00558 (120 s) & 0.98 & 1.35 \\
    & MuSCAT3 & $i'$ & 9516.0739 & offset & 0.00338 (55 s) & 1.09 & 1.35 \\
    & MuSCAT3 & $z'$ & 9516.0739 & p($f^1$) & 0.00303 (29 s) & 0.92 & 1.55 \\
    & MuSCAT3 & $g'$ & 9526.9935 & p($t^1$) & 0.0120 (120 s) & 1.10 & 2.00 \\
    & MuSCAT3 & $r'$ & 9526.9935 & p($t^1$) & 0.00422 (120 s) & 1.08 & 1.72 \\
    & MuSCAT3 & $i'$ & 9526.9935 & p($t^1$) & 0.00246 (55 s) & 0.96 & 1.08 \\
    & MuSCAT3 & $z'$ & 9526.9935 & offset & 0.00259 (29 s) & 0.97 & 1.22 \\ 
    & SSO/Europa & $z'$ & 9529.7235 & p($f^1$) & 0.00259 (49 s) & 0.77 & 1.63 \\ 
    & SSO/Europa & $I+z$ & 9548.8328 & p($t^1$) & 0.00283 (39 s) & 0.47 & 1.91 \\
    & SAINT-EX & $I+z$ & 9548.8328 & p($t^2$,$y^1$) & 0.00214 (128 s) & 1.66 & 1.14 \\
    & SSO/Europa & $I+z$ & 9551.5627 & p($t^2$) & 0.00253 (39 s) & 0.50 & 1.36 \\
    & SSO/Europa & $I+z$ & 9581.5916 & p($t^2$,$y^1$) & 0.00294 (39 s) & 0.66 & 3.01 \\
    \hline
  c & TESS Sector 4 & TESS & 8412.7784 & offset & 0.0213 (120 s) & 0.93 & 1.08 \\ 
    & TESS Sector 4 & TESS & 8429.6934 & offset & 0.0217 (120 s) & 0.94 & 1.43 \\ 
    & TESS Sector 5 & TESS & 8438.1508 & offset & 0.0186 (120 s) & 0.94 & 1.89 \\  
    & TESS Sector 5 & TESS & 8446.6083 & offset & 0.0160 (120 s) & 0.91 & 1.00 \\ 
    & TESS Sector 5 & TESS & 8455.0658 & offset & 0.0161 (120 s) & 0.92 & 1.17 \\
    & TESS Sector 5 & TESS & 8463.5232 & offset & 0.0161 (120 s) & 0.90 & 1.57 \\
    & TESS Sector 31 & TESS & 9148.5776 & offset & 0.0164 (120 s) & 0.87 & 1.43 \\ 
    & TESS Sector 31 & TESS & 9165.4926 & offset & 0.0168 (120 s) & 0.90 & 1.29 \\ 
    & TESS Sector 32 & TESS & 9182.4075 & offset & 0.0187 (120 s) & 0.95 & 1.22 \\
    & TESS Sector 32 & TESS & 9190.8650 & offset & 0.0180 (120 s) & 0.92 & 1.45 \\           
    & TESS Sector 32 & TESS & 9199.3224 & offset & 0.0173 (120 s) & 0.88 & 1.43 \\ 
    & SSO/Io & $I+z$ & 9503.7910 & p($a^1$, $f^1$) & 0.00231 (39 s) & 0.83 & 1.47 \\
    & SSO/Europa & $I+z$ & 9520.7060 & p($a^1$) & 0.00232 (39 s) & 0.74 & 1.13 \\
    & SSO/Europa & $I+z$ & 9537.6209 & p($a^1$) & 0.00287 (39 s) & 0.43 & 1.70 \\
    & MuSCAT3 & $r'$ & 9596.8231 & p($a^1$, $x^1$) & 0.00682 (120 s) & 1.03 & 1.30 \\
    & MuSCAT3 & $i'$ & 9596.8231 & p($t^1$) & 0.00311 (55 s) & 1.06 & 1.56 \\
    & MuSCAT3 & $z'$ & 9596.8231 & p($t^1$) & 0.00268 (29 s) & 0.96 & 1.38 \\
\end{longtable}

\clearpage

\section{Transit fit posterior distributions}

\begin{figure*}[hbt!]
    \centering
    \includegraphics[width=0.9\textwidth]{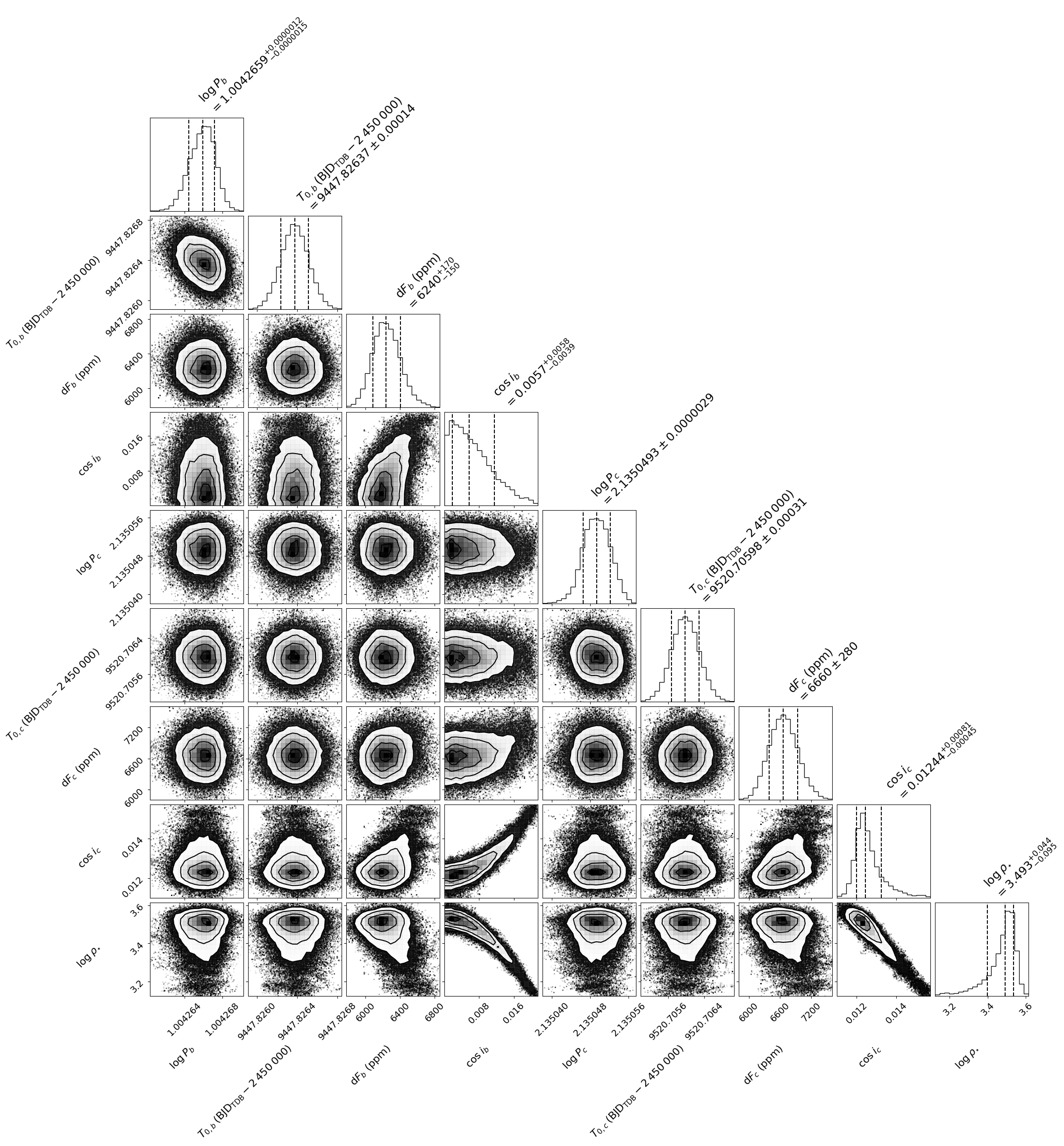}
    \caption{MCMC posterior distributions for the main fitted transit parameters (see Sect. \ref{sec:glob_an}): the log of the orbital period (log $P$), the mid-transit time ($T_0$), the transit depth ($\mathrm{d}F$), and the cosine of the orbital inclination (\hbox{cos $i_{\rm{p}}$}) for each of the two planets; and the log of the stellar density \hbox{(log $\rho_\star$)}. This plot was made using the \texttt{corner} python package \citep{2016JOSS....1...24F}.}
    \label{fig:posteriors}
\end{figure*}

\end{appendix}


\end{document}